\documentclass[]{aa}

\usepackage{txfonts}
\usepackage{graphicx}

\begin{document}

\title{Accuracy of core mass estimates in simulated observations of dust emission}

%\subtitle{}

\author{
J. Malinen \inst{1} \and
M. Juvela \inst{1} \and
D.C. Collins \inst{2} \and
T. Lunttila \inst{1} \and
P. Padoan \inst{2,3}
}

\institute{
Department of Physics, University of Helsinki, P.O. Box 64, FI-00014 Helsinki, Finland; johanna.malinen@helsinki.fi
\and
Center for Astrophysics and Space Sciences,
University of California, San Diego, 9500 Gilman Drive, La Jolla, CA 92093, USA
\and
ICREA \& ICC, University of Barcelona, Marti i Franqu\`{e}s 1, E-08028 Barcelona, Spain
}

%\date{Received 1 January 2010 / Accepted 2 January 2010}

\abstract
% context
{}
% aims
{
We study the reliability of the mass estimates obtained for
molecular cloud cores using sub-millimetre and infrared dust emission.
}
% methods
{
We use magnetohydrodynamic simulations and radiative transfer
to produce synthetic observations with spatial resolution
and noise levels typical of Herschel surveys.
We estimate dust colour temperatures using different pairs of intensities,
calculate column densities with opacity at one wavelength, and 
compare the estimated masses with the true values.
We compare these results to the case when all five Herschel wavelengths are available. 
We investigate the effects of spatial variations of dust properties and
the influence of embedded heating sources.
}
% results
{
Wrong assumptions of dust opacity and its spectral index $\beta$
can cause significant systematic errors in mass estimates. These
are mainly multiplicative and leave the slope of the mass spectrum
intact, unless cores with very high optical depth are included.
Temperature variations bias the colour temperature estimates and, in
quiescent cores with optical depths higher than for normal stable
cores, masses can be underestimated by up to one order of magnitude.
When heated by internal radiation sources, the dust in the core centre
becomes visible and the observations recover the true mass spectra.
}
% conclusions
{
The shape, although not the position, of the mass spectrum is
reliable against observational errors and biases introduced in the
analysis. This changes only if the cores have optical depths much
higher than expected for basic hydrostatic equilibrium conditions.
Observations underestimate the value of $\beta$ whenever there are
temperature variations along the line of sight. A bias can also be
observed when the true $\beta$ varies with wavelength. Internal
heating sources produce an inverse correlation between colour
temperature and $\beta$ that may be difficult to separate from any
intrinsic $\beta(T)$ relation of the dust grains. This suggests
caution when interpreting the observed mass spectra and the
spectral indices.
}


\maketitle

\section{Introduction} \label{sect:intro}

The initial conditions in cold molecular cloud cores determine many
fundamental aspects of star formation: the stellar mass distribution,
the star formation efficiencies, the main mode of clustered vs.
isolated star formation, the evolution time scales etc. In particular,
the stellar initial mass function (IMF) appears to be directly linked
to the mass function of pre-stellar cores. In order to understand the
star formation process, especially in its earliest phases, we must
study the cold, pre-stellar cloud cores. In the cores many of the
common molecular tracers have frozen onto dust grains. This makes it
difficult to determine precise core properties from molecular line
data and one has to resort to observations of the infrared and
sub-millimeter dust emission. However, also the analysis of these data
is not free from errors because temperature gradients and changes in
dust emissivity may distort the column density estimates.

The core mass function (CMF) represents an intermediate stage between
large scale cloud structure and turbulence and the final stellar
population. Studies appear to confirm the similarity between the CMF and
the IMF (Motte et al. \cite{motte98},~\cite{motte01};
\cite{johnstone00}; \cite{enoch08}; Andr\'e et al. \cite{Andre2010})
but the observed mass spectra can be affected by several sources of
uncertainty. The analysis may assume a constant dust temperature and,
when several wavelengths are observed, the colour temperatures will be
biased towards the highest actual dust temperatures along the
line-of-sight. These systematic errors will have some effect also on
the derived mass spectra. When protostars are formed within the cores,
the internal heating will create even stronger temperature gradients
that must be reflected in the mass estimates in a way that could be
visible even in the details of the observed CMF.
Also the used algorithms and their parameter values, or image acquisition techniques, can affect the derived CMF
(e.g. Smith et al.~\cite{smith08}; Pineda et al.~\cite{pineda09}; Reid et al.~\cite{reid2010}).

Several authors have investigated different aspects of star
formation and the related observations with simulations (see e.g. Smith
et al.~(\cite{smith08}); Stamatellos et al. (\cite{stamatellos07},
\cite{stamatellos10}); Urban et al. (\cite{urban09}); Shetty et al.
(\cite{shetty09a}, \cite{shetty09b}, \cite{shetty10})).
In particular, Shetty et al. (\cite{shetty09b}) studied the effects of
observational noise and temperature variations on the reliability of
the derived dust properties. As expected, in the presence of
temperature variations, the observationally derived dust spectral index $\beta$
underestimates the real spectral index of the dust grains. More
surprisingly, the estimated colour temperatures are not only biased
towards the warm regions but, in their two component models, were even
higher than the physical temperature of either temperature component.
The reliability of the spectral index determination is perhaps not as
important for the mass estimation -- although a wrong assumption of
$\beta$ will automatically lead to wrong temperature estimates.
However, the spectral index and its temperature dependence have
received a lot of attention because they may provide additional
information on the chemical composition, structure, and the size
distribution of interstellar dust grains (e.g., Ossenkopf \& Henning
\cite{ossenkopf94}; Kr\"ugel \& Siebenmorgen \cite{Krugel1994}; Mennella
et al. \cite{Mennella1998}; M\'eny et al. \cite{Meny2007}; Compi\`egne
et al. \cite{Compiegne2011}). 
The observations consistently show an inverse $T-\beta$ relation (e.g.
Dupac~\cite{Dupac2001}, \cite{Dupac2003}; Hill et al.~\cite{Hill2006};
Veneziani et al.~\cite{Veneziani2010} and the initial Herschel
results). The reliability of this relation is difficult to
estimate because any noise present in the measurements also tends to
produce a similar anticorrelation (see Shetty~\cite{shetty09a}). The
absolute value of the dust opacity, $\kappa$, is also
expected to vary because of changes in the abundance of different dust
populations and changes in the grain size distribution. For the
sub-millimeter observations, the changes should be strongest in dense
and cold regions where the grains acquire ice mantles and may
coagulate to form larger grains with much higher $\kappa$ values
(e.g., Ossenkopf \& Henning \cite{ossenkopf94}, Krugel \&
Siebenmorgen \cite{Krugel1994}, Stepnik et al. \cite{Stepnik2003}).
The uncertainty of $\kappa$ affects directly any mass estimates. On
the other hand, the absolute value of $\kappa$ is not easy to measure
because it requires an independent column density estimate that is
reliable over the same $A_{\rm V}$ range for which sub-millimetre
observations exist. The total errors resulting from noise, temperature
inhomogeneities, and dust variations are best estimated with 
modelling.

Magnetohydrodynamic (MHD) simulations are found to provide a good
description for the general structure of interstellar clouds (Padoan et al.~\cite{padoan2004,padoan2006}). When
self-gravity is included, predictions can be made for the frequency
and internal structure of dense cloud cores. Some of the objects that
in continuum observations are categorized as cores may be transient
while others will be held together by gravity and can potentially
collapse forming stars. The models need to be compared with
observations to see if the IMF can truly be explained as a direct
consequence of the turbulent cascade (Padoan~\cite{padoan95}; Padoan
\& Nordlund~\cite{padoan2002}). Nevertheless, the MHD runs already
provide the best starting point for the realistic modelling of dust
emission from dense clouds.

In this paper, we investigate the reliability of mass estimates by
combining MHD simulations with radiative
transfer modelling of dust emission. We start with some spherically
symmetric (1D) models that help us to interpret the results of the
more complex 3D clouds. We continue with low density contrast MHD
models where we also test the consequences of spatially
varying dust properties. We then present the first results from
studies where we use cloud models that are defined with the help of
hierarchical grids. With adaptive mesh refinement (AMR) one can
follow the evolution of individual cores much further in the MHD
calculations (see Collins et al.~\cite{Collins10}) and, because of the
stronger density and temperature variations, also the errors in the
observationally derived core masses can be expected to be larger. Our main interest
is not the shape of the CMF itself (as a result of the MHD modelling)
but how the observational effects change the core mass estimates and
to what extent these are visible in the observationally derived CMF. With the help of
simulated observations, we can also draw some conclusions on the
observability of dust spectral index variations.

The present study is relevant to many Galactic studies that are being
conducted with the Planck (Tauber et al. \cite{Tauber2010}) and
Herschel (Pilbratt et al. \cite{Pilbratt2010}) satellites. Herschel
has already provided hundreds of new detections of both starless and
protostellar cores (Andr\'e et al. \cite{Andre2010}; Bontemps et al.
\cite{Bontemps2010}; K\"onyves et al. \cite{Konyves2010}; Molinari et
al. \cite{Molinari2010}; Ward-Thompson et al. \cite{WardThompson2010})
and detailed studies of these objects are now under way. In this
context, it is crucial to know the uncertainties and the possible
biases that might exist in the core and dust parameters that are
derived from such sub-millimetre observations.
The principal authors of this paper are also involved in the survey of
Galactic cold cores that is being carried out with data from the
Planck and Herschel satellites. The project will present an overview
of the future galactic star formation areas by locating pre-stellar cores
from the Planck all-sky survey and by following up a representative
set of cores with higher resolution Herschel observations (see Juvela
et al. \cite{Juvela2010}).

The methods used to create model clouds, to predict their
sub-millimetre emission, and to analyse the resulting surface
brightness maps are presented in Sect.~\ref{sect:methods}. The results
from the analysis are presented in Sect.~\ref{sect:results}, first for
simple spherically symmetric cores (Sect.~\ref{sect:results_1d}) and
then for the cloud models based on MHD simulations
(Sect.~\ref{sect:results_3d}). Results of the estimation of dust spectral indices are presented in Sect.~\ref{sect:specind}.
We discuss the results in
Sect.~\ref{sect:discussion} before presenting our conclusions in
Sect.~\ref{sect:conclusions}.

\begin{figure*}
\centering
\includegraphics[width=6cm]{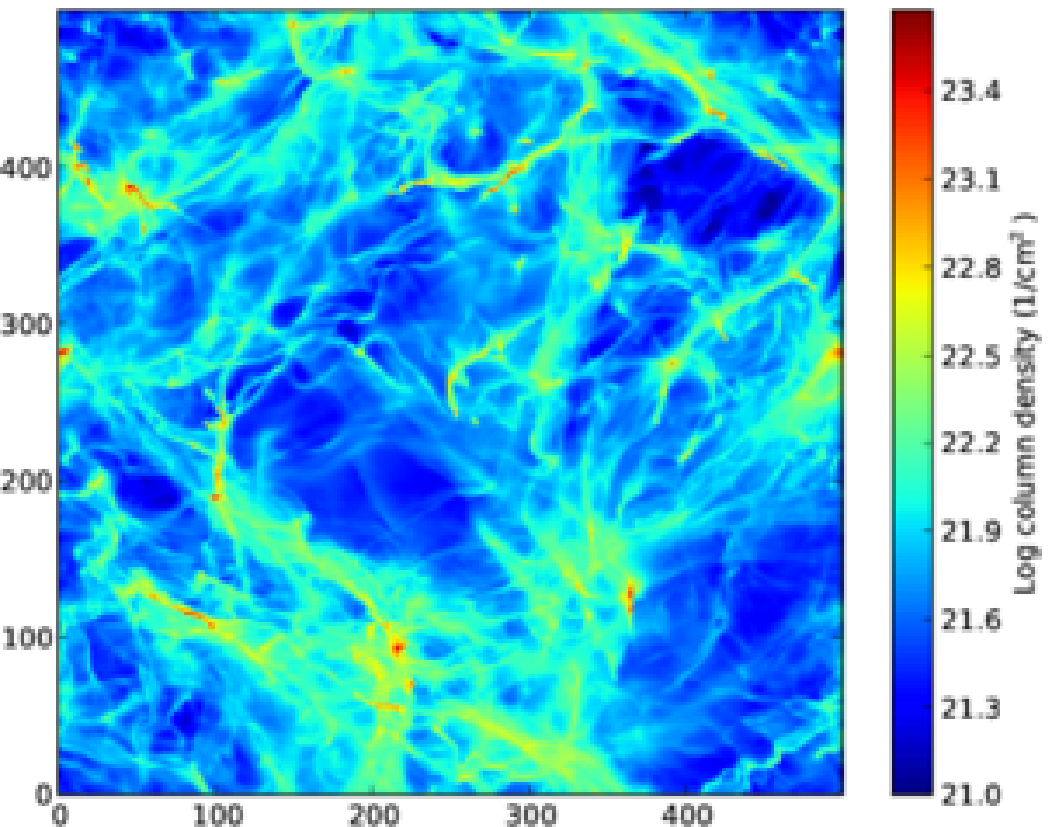}
\includegraphics[width=6cm]{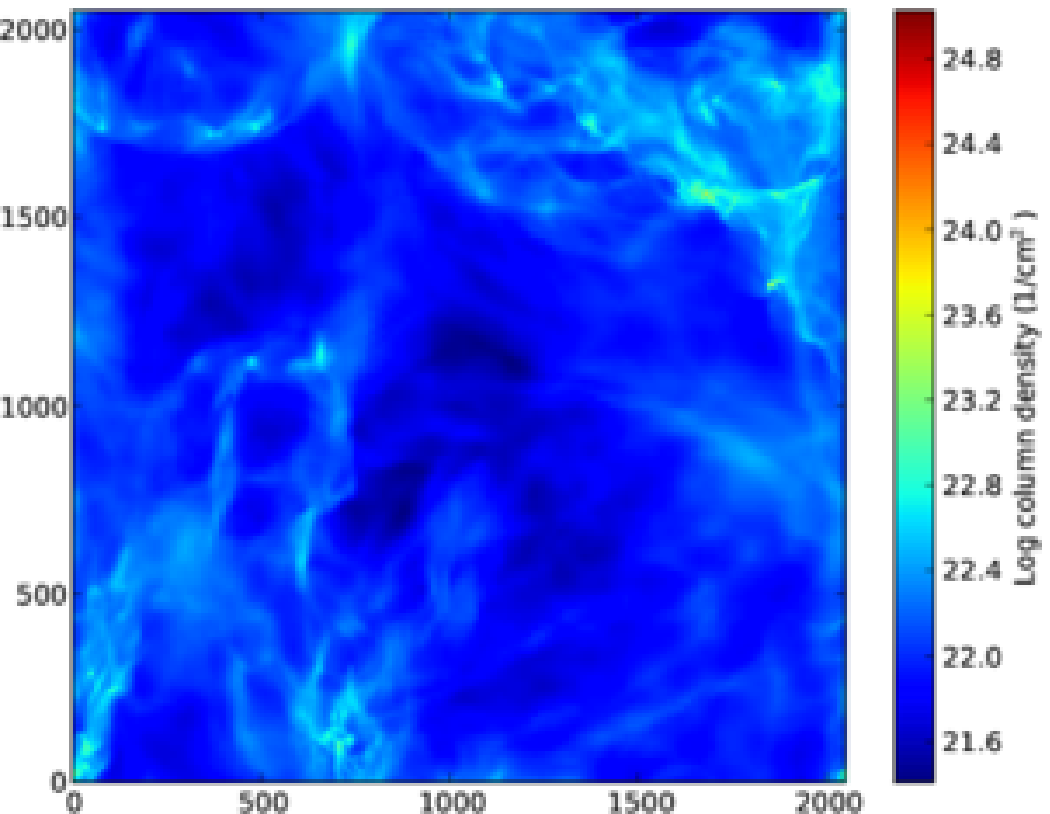}
\includegraphics[width=6cm]{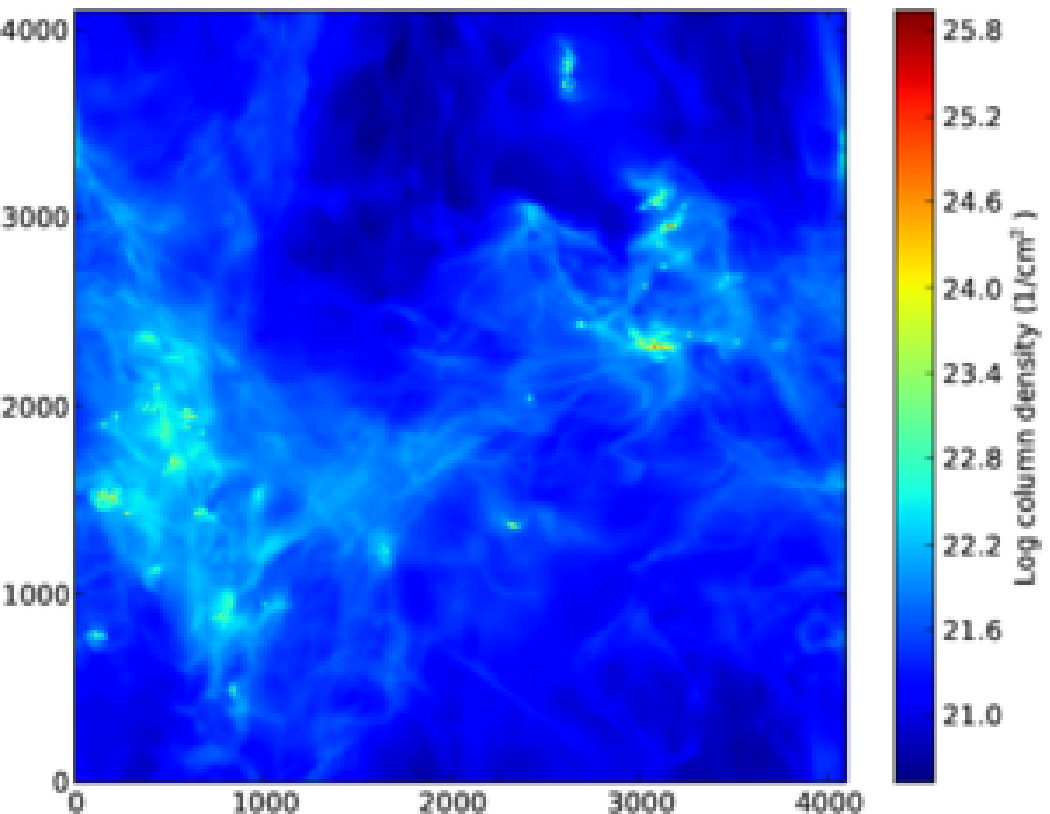}
\includegraphics[width=6cm]{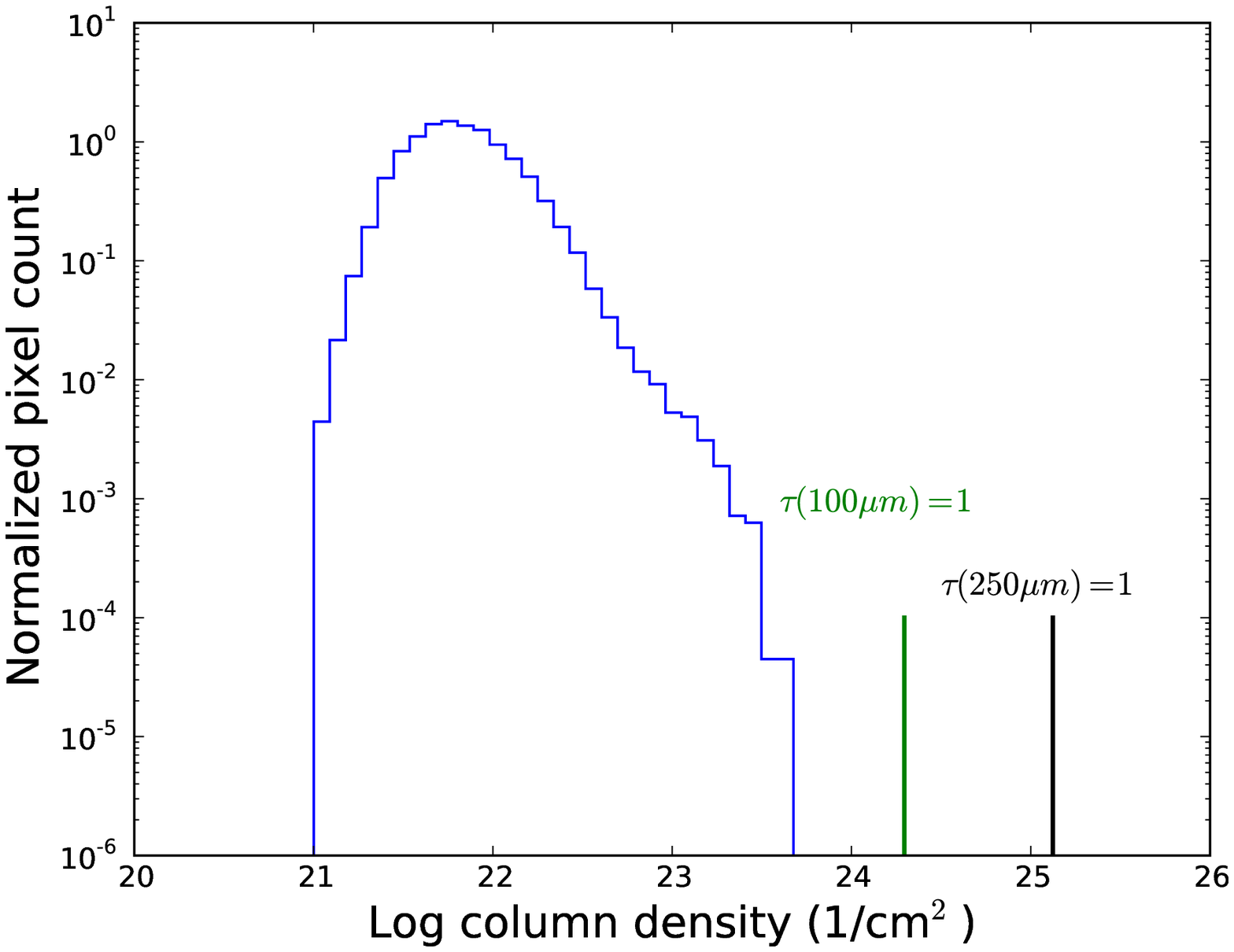}
\includegraphics[width=6cm]{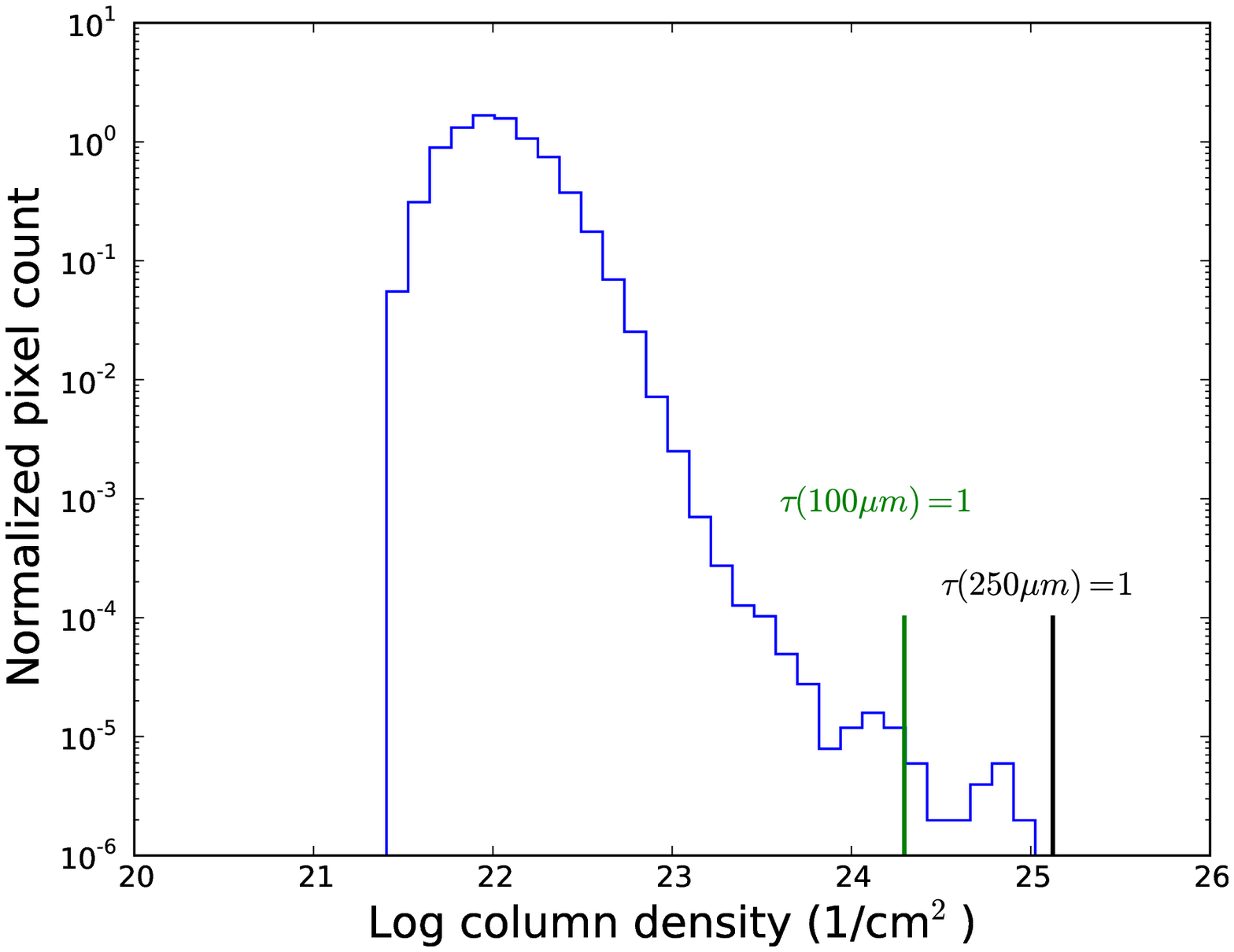}
\includegraphics[width=6cm]{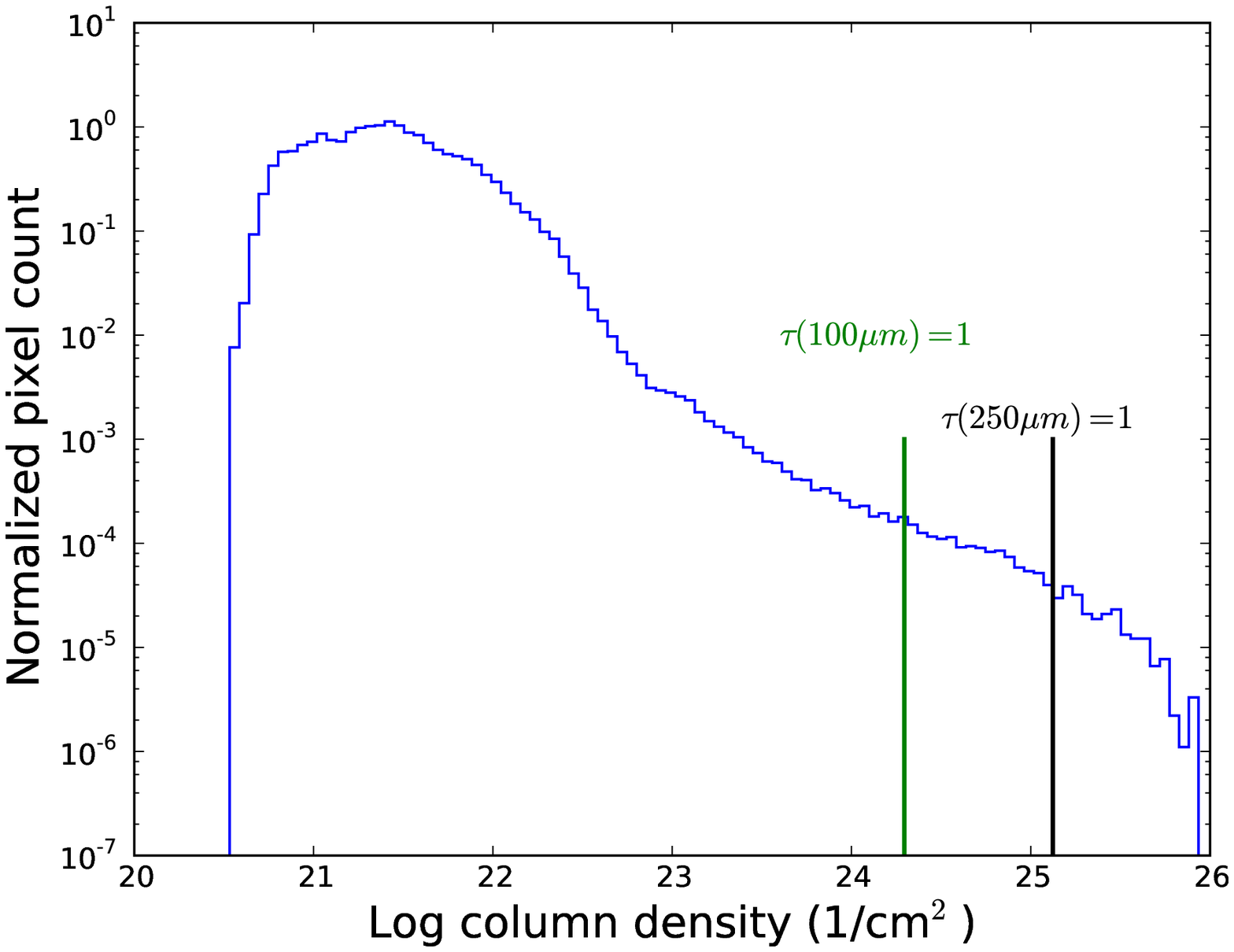}
\caption{
Column density maps (top row) and normalized column density
distributions (bottom row) of Model I (left), Model II (middle) and
Model III (right) in one viewing direction.
} 
\label{fig:models} 
\end{figure*}

\section{Methods}  \label{sect:methods}

\subsection{MHD simulations}

We used three different cloud models to investigate the effects of
different conditions in the dust clouds, mainly the opacity of the
formed cores.  The three models were all run in a similar fashion,
with only the density and resolution changed. An isothermal equation 
of state was assumed in all MHD runs. All three models begin
with the same non-gravitating turbulence simulation. A box of
$1000^3$ zones with initially uniform density and magnetic field along
the $\hat{z}$ axis was driven in a manner that maintained an rms sonic
mach number of $\approx 9$.  The initial plasma $\beta = 8 \pi P /
B^2$, the ratio of thermal to magnetic pressure, is 22.2.  The rms
Alfv\' enic Mach number, the ratio of rms velocity to the Alfv\' en
speed, is $\approx 2.8$.  Stirring continues for several shock
crossing times to statistically separate the turbulence from the
initial conditions, at which time the simulation is re-gridded to
three different resolutions, and gravity is switched on.  Introduction
of the gravity equation to the ideal MHD system imposes an additional
physical scale on the system, and this is chosen differently for each
of the three models.

Model I (Padoan \& Nordlund \cite{padoan2010}) was continued with the
Stagger code (Nordlund et al.~\cite{nordlund96}; Nordlund \& Galsgaard~\cite{nordlund97}) at a resolution of $500^3$.
Stagger is a fixed resolution (unigrid) high order finite difference
method. The code uses finite differences, with 6th order derivative operators, 5th order interpolation operators, and a 3rd order low memory
Runge-Kutta time integration scheme.  Because of the staggered mesh, $\nabla \cdot B$ is conserved to machine precision.
Box size and mean density are 6 pc and 450 $\rm{cm}^{-3}$, respectively.

Models II and III were continued with the MHD extension of the
adaptive mesh refinement (AMR) code Enzo (Collins et
al.~\cite{Collins10}). It is a higher order Godunov method, using the
patch solver of Li et al. (\cite{Li08}), the AMR technique of Balsara (\cite{Balsara01}), and constrains $\nabla \cdot B$ with the CT
method of Gardiner \& Stone (\cite{Gardiner05}).
Model II was re-gridded to $128^3$, with size and density of 10 pc and
$400 \rm{cm}^{-3}$, and used 4 levels of refinement for an effective
resolution of $2048^3$.  This model was studied in detail in Collins et al. 
(\cite{Collins10b}).
Model III used a root grid of $256^3$ and 4 levels of refinement, for an effective
resolution of $4096^3$.  It used the same box size as Model II, 10\,pc,
but a lower mean density of $144 \rm{cm}^{-3}$.  This model was
allowed to run for a full free-fall time ($t_{ff} = \sqrt{3 \pi / 32 G
\rho}$) longer than Model II, which allows the cores to reach higher
column densities, and thus opacities, while keeping a comparable number of
cores.

The column density maps of the models are presented in
Fig.~\ref{fig:models}. The figure shows also the distributions of the
column density values in the models. Going from Model I to Model III,
the peak column density increases more than two orders of magnitude,
from $\sim 10^{23.5}$ to almost $10^{26}$\,H\,cm$^{-2}$. In Model
II, when observed with the highest resolution (distance=100pc, 37"
beam), $\tau$(100\,$\mu$m)=1.0 is reached only in one core, within an
area smaller than the beam.
In Model III the peak value is $\tau$(100\,$\mu$m)$\sim$5 and 
$\tau$(100\,$\mu$m)=1 is found in several cores, still limited to the
size of a single beam (0.01\% of all pixels in the full map).
Therefore, even in Model III and even at 100\,$\mu$m, the mean optical
depth of the 'cores' is always $\sim$1 or less.

The model of super Alfv\' enic turbulence has been compared to
observations by a number of authors. Padoan et al. (\cite{Padoan99})
found that models with high Alfv\' en Mach number match column density
and extinction statistics in the Perseus and IC 5146 clouds quite
well, while models with stronger magnetic fields, thus lower Alfv\' en
Mach number, do not. Crutcher et al. (\cite{Crutcher09}) measured the
change in mass to flux from the envelope around a core to the core
itself, and found the ratio, $\mathcal{R}$, to be inconsistent with
the strong magnetic field prediction. Lunttila et al.
(\cite{Lunttila08}) found similar values for $\mathcal{R}$ from a
model similar to the turbulent initial conditions for all 3 models
presented here. Collins et al. (\cite{Collins10b}) find excellent agreement
between the data in Model II and Zeeman splitting measurements and
mass distributions of dense cores in a number of different clouds.

\begin{figure*} 
\centering
\includegraphics[width=8cm]{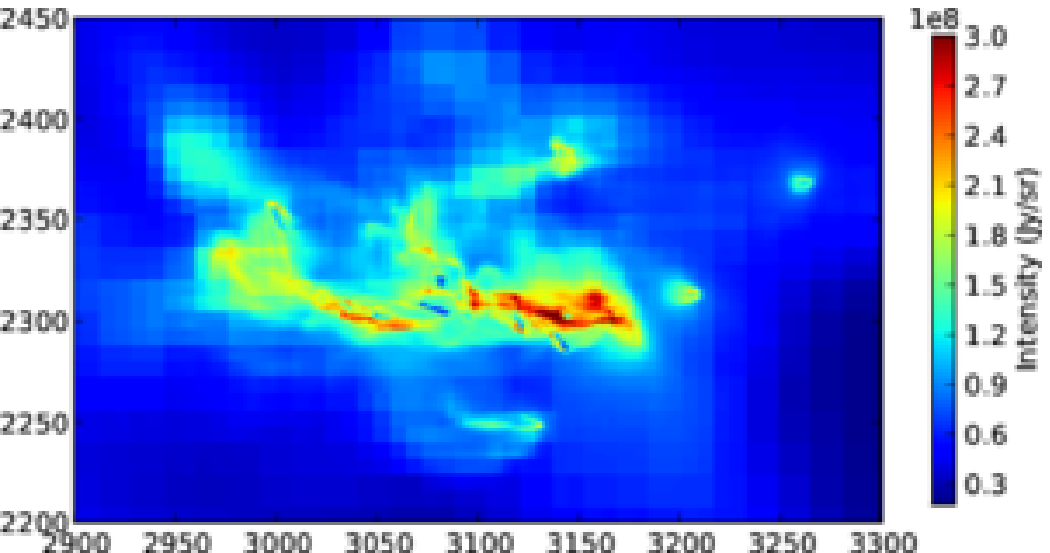}
\includegraphics[width=8cm]{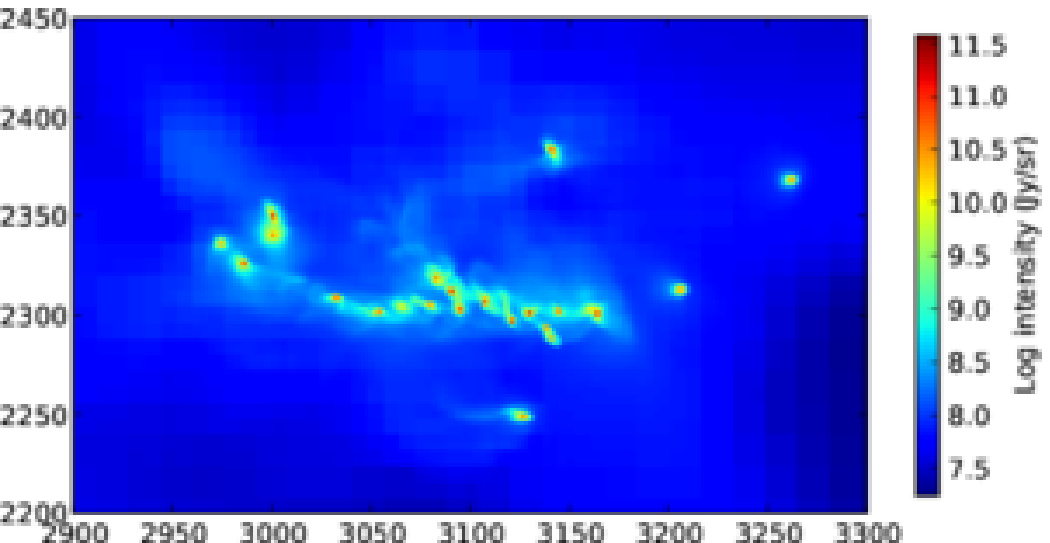}
\includegraphics[width=8cm]{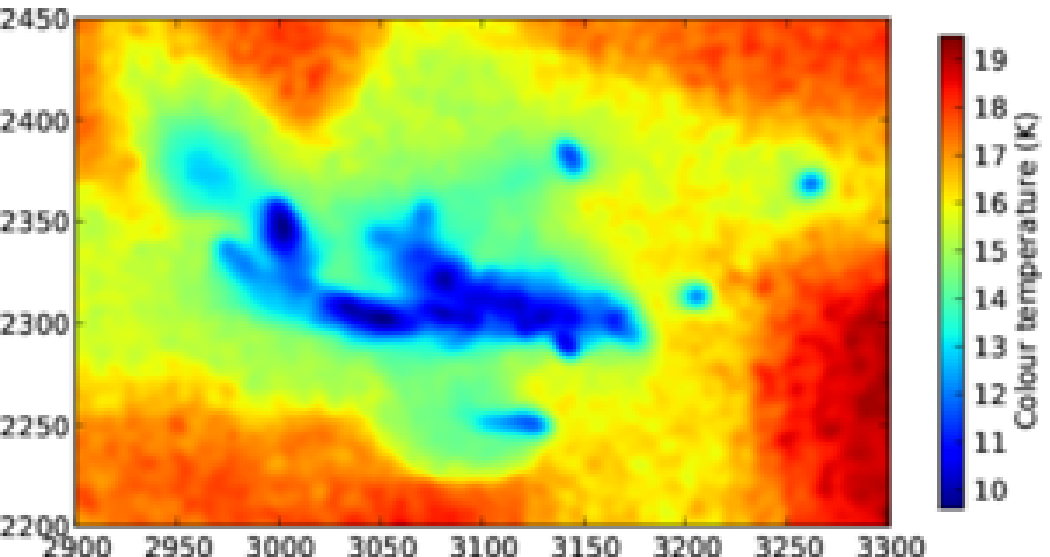}
\includegraphics[width=8cm]{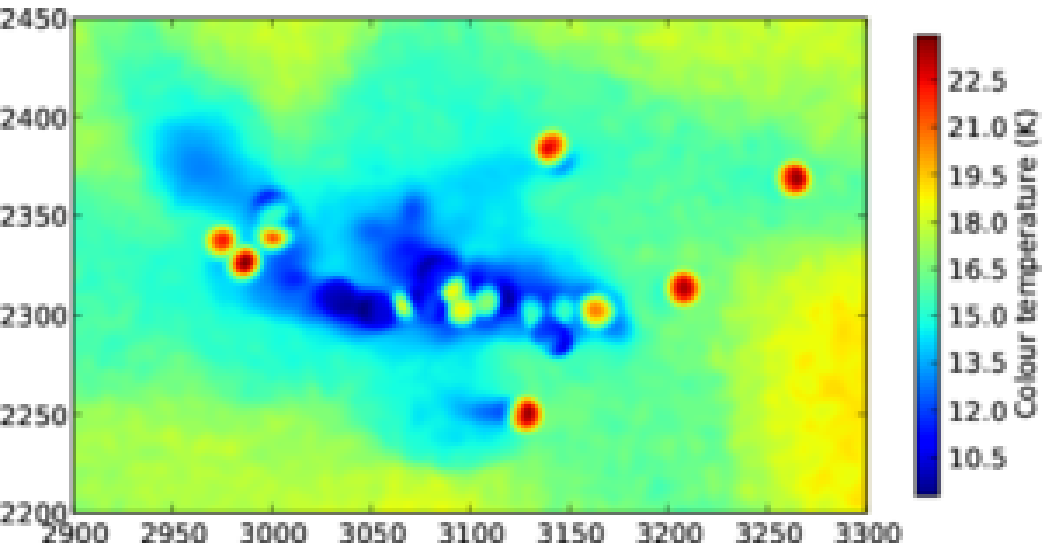}
\caption{
Closeups of the 250 $\mu$m intensity (top row) and colour temperature
(bottom row) maps of Model III. The maps are shown before the addition
of the internal radiation sources (left frames) and with the sources
(right frames). The colour temperature was calculated from the ratio
of 250\,$\mu$m and 500\,$\mu$m surface brightness.
} 
\label{fig:closeups} 
\end{figure*}

\subsection{Radiative transfer calculations}  \label{sect:RT}

The MHD calculations provide the density distributions for our
modelling. The clouds are illuminated externally by the interstellar
radiation field (ISRF) (Mathis et al. \cite{Mathis83}) and, for the
main part, the dust is assumed to have the average properties found in
the Milky Way (Draine \cite{Draine2003})
with a gas-to-dust ratio of 124 and $R_{\rm V}$=3.1.
\footnote{http://www.astro.princeton.edu/$\sim$/draine/dust/dustmix.html}
Radiative transfer (RT) calculations are used
to solve the dust temperature for each
cell in the model and are independent from the gas temperature assumed
in the MHD runs. The same RT program is used to calculate the emerging
intensity by integrating the radiative transfer equation and without
resorting to the assumption of optically thin emission.

The radiative transfer calculations of the unigrid model were carried
out with our Monte Carlo radiative transfer program (Juvela \& Padoan
\cite{juvela03}). A separate radiative transfer program was used for
the calculations on hierarchical grids. This second program is based
on the unigrid code which has been modified to work with AMR grids and
contains several associated improvements. The code will be described
in detail elsewhere (Lunttila et al. \cite{Lunttila2010}).

To investigate the effect of internal heating, black body radiation
sources in the range of 0.3-120 solar luminosities were later added
inside the gravitationally bound cores in Models II and III. The source temperatures were
selected randomly between 200\,K and 2000\,K so that the distribution
of log(T) was uniform. We assumed that a 0.1\,M$_{\sun}$ protostar
has a luminosity of 10$^{0.3} L_{\sun}$ and a 10\,M$_{\sun}$ protostar
has a luminosity of 10$^2 L_{\sun}$. We applied linear
interpolation on logarithmic scale for the other masses. The
prescription is roughly consistent with the theoretical predictions of
pre-main sequence evolution (see Wuchterl \& Tscharnuter
(\cite{wuchterl03}, Fig. 3). The number of sources was 34 in Model II
and 105 in Model III. In Model II the mass of the protostar was
assumed to be 20\% of the mass of the gravitationally bound core and
in Model III the corresponding percentage was 50\%. The model is not
entirely self-consistent but sufficiently realistic so that we can study
the basic observational effects caused by internal heating.

The cells in which the sources are located (`source cells') are not
optically thin and the re-radiated dust emission is important in
determining the dust temperature of the surrounding areas. Very close
to the source, the assumption of a constant temperature within the
cell also fails because of purely geometric effects. For these
reasons, in the case of Model III, we used subgrid models to describe
the temperature distribution of the source cells. We ran for each
source a spherically symmetric (1D) radiative transfer model that was
discretised to 100 shells and had an opacity equal to the opacity of
the cell in the AMR model. In addition to the central sources, we
assumed the 1D models to be illuminated externally by the ISRF. This was done
in spite of the fact that the central sources completely dominate the
radiation field even in the outer parts of the 1D models. The dust
temperature profiles of the 1D models were solved iteratively taking
into account the effects of re-radiated dust emission. In the full AMR
model, the dust temperatures outside the source cells are below
$30$\,K which justifies the omission of the coupling with the dust opacity
emission that would enter the 1D models from the outside. The 1D
calculations also ignore the effect of other nearby sources. However,
the sources are always tens of voxels apart so that contribution of
other sources is negligible compared to the source residing inside the
cell.

The 3D radiative transfer calculations were completed including the
background radiation (ISRF), the radiation of the internal sources,
and the dust emission from the medium. 
The average radiation field is dominated by the ISRF. 
When internal heating sources are included, they dominate the radiation field
but only in their immediate vicinity.
The re-radiated emission is
completely insignificant far from the sources but very close to the
sources can induce a small temperature raise (typically less than 1K).
For the source cells the
emission was already calculated with the 1D models but had to be still
solved for the other cells. Because of the high optical depths, the
calculations required iterations so that the effect of the sources
(and the hot dust near the sources) could be transported outwards. The
possibility of iterating arbitrary branches of the grid hierarchy
separately (e.g., doing first sub-iterations just for the grids
containing the sources) makes it possible to reach convergence much
faster than by iterating the full model (see Lunttila et al.
\cite{Lunttila2010}).

Once the dust temperatures were solved, we do line-of-sight
integration of the radiative transfer equation to calculate surface
brightness maps at the finest resolution of the AMR model. 
Thus the maps of Model II contain 2048$\times$2048 pixels that correspond to a
linear physical scale of 0.00488\,pc or 1007\,AU. The maps of Model
III have 4096$\times$4096 pixels and a resolution 0.00244\,pc or
503.6\,AU. Maps were calculated for three orthogonal directions of
observation ($X$, $Y$, and $Z$). As the final step in the simulation
of surface brightness maps, we added to the maps noise typical of
current Herschel observations (see Table~\ref{tab:noise}) and
convolved the maps to the resolution of the Herschel instruments at
the corresponding wavelengths. The procedure depends on the assumed
distance of the model cloud which was set to either 100, 400 or
1000\,pc.

\begin{table}
\caption{Noise levels used to simulate typical Herschel observations.}
\label{tab:noise}
\begin{tabular}{ll}
\hline
Wavelength ($\mu$m) &  $\sigma$ (MJy/sr/beam) \\
\hline
100 & 8.1 \\
160 & 3.7 \\
250 & 1.20 \\
350 & 0.85 \\
500 & 0.35 \\
\hline
\end{tabular}
\end{table}

Closeups of Model III 250 $\mu$m intensity maps (without noise and convolution) and colour temperature
maps before and after adding the sources are shown in
Fig.~\ref{fig:closeups}. In the original intensity maps of Model III
there are some dark 'worms' in the densest and coldest cores. These
are analogous to infrared dark clouds (e.g. Henning et al.~\cite{henning2010}) but, because of the high
opacity, are at high resolution visible up to sub-millimetre
wavelengths. However, convolution of the maps hides these features.
At longer wavelengths (e.g. 500\,$\mu$m) the relative
emission of these cold regions increases and such sharp absorption
lanes are no longer seen.

\subsection{Mass estimation}   \label{sect:mass}

We begin the analysis by presuming the observer has only two wavelenghts available (e.g., Schnee \& Goodman~\cite{schnee2005}, Kramer et al.~\cite{Kramer2003}, and Mitchell et al.~\cite{Mitchell2001}).
We use different wavelength pairs to compare how the results depend on the used
wavelengths. We also compare these results to the case when all five Herschel wavelenghts are available.

The masses were first estimated with just two wavelengths,
convolving the data to the resolution of the longer wavelength map.
The wavelengths used were either
100\,$\mu$m and 350\,$\mu$m or 250\,$\mu$m and 500\,$\mu$m.
The calculations
were mostly performed with the correct value of the spectral index
$\beta$.
The analysis assumes optically thin emission and a dust opacity law
of the form $\kappa \sim \nu^{\beta}$ so that the observed intensity
can be written
\begin{equation}
I_{\nu} = B_{\nu}(T_{\rm C})(1-e^{-\tau}) \approx B_{\nu}(T_{\rm C}) \tau = B_{\nu}(T_{\rm C}) \kappa N \propto B_{\nu}(T_{\rm C}) \nu^{\beta}.
\label{eq:beta}
\end{equation}

In our dust model the spectral index is 2.12 between 100\,$\mu$m and
350\,$\mu$m and 2.09 between 250\,$\mu$m and 500\,$\mu$m. 
In the case of two wavelengths and a fixed value of spectral
index the modified black body law gives the temperature without any ambiguity (e.g., from the
relative weight given for observations at different wavelengths).
The dust colour temperatures $T_{\rm C}$ 
were determined from the ratio of the surface brightness at two
wavelengths from the following equation
\begin{equation}
\frac{I(\nu_1)}{I_(\nu_2)} = \frac{B_{\nu_1}(T_{\rm C}) \nu_1^{\beta}}{B_{\nu_2}(T_{\rm C}) \nu_2^{\beta}}.
\label{eq:T_C}
\end{equation}
%We used two pairs of wavelengths to examine the effect the selection of
%the wavelengths has on the results. 
We also compared the results with an analysis where the colour
temperature is estimated by fitting the modified blackbody function
$B_{\nu}(T_{\rm C}) \nu^{\beta}$ to observations at all five Herschel wavelengths.
In the fit the free parameters were the scaling of the absolute level of the
SED curve and the colour temperature. The spectral index $\beta$ was set to a
fixed value or included as another free parameter of the fit.

Because of the added noise, the colour temperature estimates become very
uncertain in the regions of the lowest column density. As a technical
precaution the temperature was set to a constant value below a certain
surface brightness level. This does not affect any of the discussion
below because the regions containing the cores are well above this
threshold and for them the colour temperature was always calculated
pixel by pixel.

Once the colour temperatures were obtained, the column densities were
calculated from the formula
\begin{equation}
    N = \frac{I_{\nu}}{  B_{\nu}(T_{\rm C})  \kappa }
\label{eq:colden}    
\end{equation}
using the longer observed wavelength map. The selection of the wavelength
(shorter vs. longer) is important only so far as the noise levels of
the maps at the two wavelengths are different. 
When $T_{\rm C}$ was estimated by fitting the model SED to
observations at several wavelengths, the column density was calculated
from the same formula with the observed intensity and the correct opacity at
the wavelength 250\,$\mu$m. The column density is often calculated using the best-fit value of intensity.
However, for the cores there is not much difference between using the best-fit or
observed $I_{\nu}$ because of good S/N. The difference between the fitted and
observed value of $I_{\nu}$ is about 1\% at the wavelength 250\,$\mu$m.
With the dust opacity $\kappa$ expressed as area per hydrogen atom,
the column density $N$ is obtained in units $H/$cm$^2$. In the
calculations we use the correct value of the dust opacity that is
obtained directly from the dust model. Therefore, neither the employed
$\beta$ nor the $\kappa$ values should bias the subsequent mass
estimates.  The (hydrogen) mass per map pixel is calculated by
multiplying the column density with the physical size of the pixel and
the mass of the hydrogen atom. The masses are calculated either for
fixed regions around the known core positions or for the clumps found
with the Clumpfind algorithm (see Sect.~\ref{sect:clumpfind}). In both
cases, the masses are obtained by summing the pixels assigned to that
object. In spite of the convolution, the original pixel sizes were
used throughout the analysis.

We use the term 'observed' to describe also quantities derived
from observations (e.g. temperature, column density, mass, spectral
index or mass spectrum), in contrast with the corresponding 
quantities that are obtained directly from the model clouds. We
compare the observed mass spectra with the `true mass spectra'. In
both cases the clumps are identical and are extracted from the
observed column density maps. While the observed mass spectra are
calculated with the column densities derived from the simulated
surface brightness maps, the `true mass spectra' are derived using
column densities that are read directly from the cloud model.
Therefore, it is the `real' mass spectrum only in the sense of not
containing errors related to the derivation of column density 
from observations (`observational errors').

%%%
We compare the mass spectra with respect to their location on the mass
axis and their general shape. To assist the visual inspection we use the slope
of a least squares line fitted to a linear part as a numerical indicator of
the shape. 
In some cases the result from the fit is not very reliable because of
the absence of a clear linear part and the dependence on the mass
range fitted.

\subsection{Defining and finding cores}  \label{sect:clumpfind}

We use the CUPID implementation\footnote{http://starlink.jach.hawaii.edu/starlink/CUPID} of the automatic clump finding method Clumpfind (Williams et
al.~\cite{williams94}) to locate cores in 2D column density maps in a
similar manner as observers do. Because the number of pixels in our maps is larger than the resolution we have to scale the rms noise level of the background subtracted maps before using Clumpfind.
We adopt the terminology where \emph{clump} refers to larger
structures and \emph{core} to the densest condensations of the clumps
(e.g. Stamatellos et al.~\cite{stamatellos07}). 

There has been some debate about the usability of Clumpfind in defining the
shape of the CMF (see Smith et al.~\cite{smith08}; Goodman et
al.~\cite{goodman09} and Pineda et al.~\cite{pineda09}). 
However, our main goal is not to study the absolute shape of the
CMF but rather to examine the biases resulting from the way 
observations are usually analysed. 
Clumpfind also appears to find
the densest cores in our 2D position--position data rather reliably with the
right parameter values.  However, parameter values must be determined
independently for each case. Otherwise Clumpfind can easily start picking up
sparse filaments as clumps in our high resolution and confusion-limited data.
A problem with Clumpfind is that small changes in the parameter values can
change the size and shape of the clumps significantly, especially when clumps
become combined or separated.

The observed 2D clumps do not necessarily represent true 3D structures
in the clouds due to projection effects (see eg. Shetty et
al.~\cite{shetty10}). This could also change the shape of the
observed mass spectrum relative to the real spectrum that can be
determined from the 3D data only. However, we do not have to take 
this effect into account because we are mostly comparing the observed
and the true masses along the same full line-of-sight. 

For the purpose of adding radiation sources into the gravitationally
bound cores we have determined the positions by running the modified
3D Clumpfind over the 3D cloud model as in Padoan et al.~(\cite{padoan2007}) and then 
selecting the sources for which $GE/(TE+KE+BE) > 1$, that is the
clumps in which the gravitational energy dominates over the sum of
thermal, kinetic, and magnetic energy. In order to have a more
objective definition of the core regions, and to eliminate the effects of Clumpfind algorithm, we also calculate the core
masses within a fixed radius of these positions (projected to a map).
With fixed radii we can also study how the mass estimates vary with the size of the region.
This would of course not be possible in the analysis of normal
observations but, in the case of models, provides a useful test where
the effects of the clump selection are eliminated.

\section{Results}  \label{sect:results}

\subsection{Spherically symmetric reference models} \label{sect:results_1d}

To help the interpretation of the results of the 3D MHD models, we
first study a series of spherically symmetric model clouds and
examine the accuracy of the mass estimates in this simplified
setting.

\begin{figure}
\resizebox{\hsize}{!}{\includegraphics{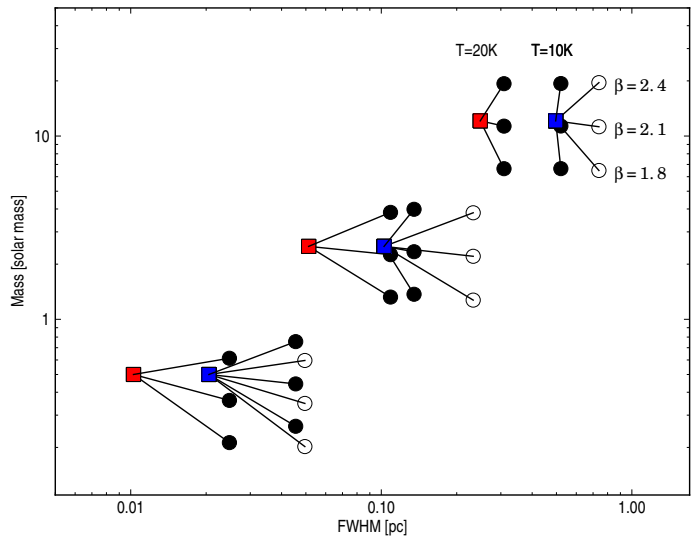}}
\caption{
Observed properties of Bonnor-Ebert spheres. The solid squares denote
the FWHM of the column density distribution and the true mass of the
cores at $T_{\rm gas}=$10\,K (blue symbols) or at 20\,K (red symbols
left of the 10\,K points). The circles show the FWHM of the
350\,$\mu$m surface brightness and the mass estimated from simulated
observations at 100\,$\mu$m and 350\,$\mu$m. The cores are heated
either by the full ISRF (open circles, shown for $T_{\rm gas}=$10\,K
only) or by the ISRF attenuated by A$_{\rm V}$=4$^{\rm m}$ (filled
circles). Masses are calculated with the correct spectral index,
$\beta \sim 2.1$, and with one lower and one higher value.
}
\label{fig:Bonnor}
\end{figure}

\subsubsection{Bonnor-Ebert cores}

The density profiles of the spherically symmetric models follow the
Bonnor-Ebert solution (\cite{Bonnor1956}) with a stability parameter
$\xi=6.5$ and a gas temperature of 10\,K or 20\,K. The cores are
heated only by external radiation. Surface brightness maps are
obtained from radiative transfer modelling and, as explained in
Sect.~\ref{sect:mass} the core masses are calculated using
observations at two wavelengths. The cores are assumed to be resolved
so that the mass estimates are based on the derived column density
profiles. We study three Bonnor-Ebert spheres that have masses of
0.5\,$M_{\sun}$, 2.5\,$M_{\sun}$, and 12.1\,$M_{\sun}$. The
corresponding average densities of the spheres are $2.2 \times 10^5$,
$8.8 \times 10^3$, and $3.8 \times 10^2$ H$_2$cm$^{-3}$.

Fig.~\ref{fig:Bonnor} shows the results for a series of Bonnor-Ebert
spheres comparing the real mass to the mass derived from 100\,$\mu$m
and 350\,$\mu$m observations. The FWHM of the column density
distribution is also compared to the FWHM of the 350\,$\mu$m surface
brightness. The open circles correspond to a case where the cores are
illuminated by the full ISRF, the filled circles to a case where the
ISRF is assumed to be attenuated by an external dust layer of $A_{\rm
V}=4^{\rm m}$. The emission of this external layer is not included as
would be the case if the analysis was carried out using background
subtracted maps. The first scenario applies to isolated cores, the
latter scenario would be more appropriate for cores inside molecular
clouds.

When analysis is done with the correct spectral index, $\beta=$2.1,
the mass estimates are mostly accurate to a few percent. Only for very
compact clouds, represented here by the $0.5\,M_{\sun}$ core, the mass
is clearly underestimated. For smaller Bonnor-Ebert spheres the column
density and the temperature gradients between the centre and the
surface are higher. The warm dust at the surface of the cores begins
to dominate the signal, the colour temperature is higher than the
average dust temperature, and the mass is underestimated. For the
isolated core with $M=$0.5\,$M_{\sun}$ and $T_{\rm gas}$=10\,K the
bias is $\sim$30\%. If the incoming radiation is attenuated by $A_{\rm
V}=4^{\rm m}$, the temperature gradients within the core decrease and
the bias is reduced to a few percent. However, for $T_{\rm gas}$=20\,K
model the maximum error remains at $\sim$30\%. At longer wavelengths
the dependence on colour temperature is weaker and, for example, in
typical Herschel observations these biases would not be a significant
source of error. The uncertainty resulting from the unknown value of
$\beta$ should be insensitive to the cloud properties. According to
Fig.~\ref{fig:Bonnor} the $\pm$0.3 uncertainty of the spectral index
translates to a $\sim$30\% error in the mass, almost irrespectively of
the mass of the BE core. The mass estimate depends, of course,
directly on the assumed dust opacity that in the
case of real observations has a similar or even larger uncertainty.

Because of the cold centre of the cores, the intensity profiles are
flat compared to the column density and the FWHM of the 350\,$\mu$m
intensity is higher than the FWHM of the column density. The
difference is largest for the most compact clouds where, in fact,
strong limb brightening is already seen at 100\,$\mu$m.

\subsubsection{High opacity cores}

The previous tests showed that mass estimates are quite reliable for
Bonnor-Ebert type cores. However, the errors should increase as the
optical depth increases. The Bonnor-Ebert spheres represent a special
case where the external pressure and gravity are balanced by the
thermal pressure only. In the case of strong turbulent motions,
rotation, or magnetic fields, stable cores may have higher opacity.
For unstable cores, the optical depths will strongly increase during
the collapse and, before the internal heating becomes important,
observations might miss a larger fraction of the dust mass.

\begin{figure}
\resizebox{\hsize}{!}{\includegraphics{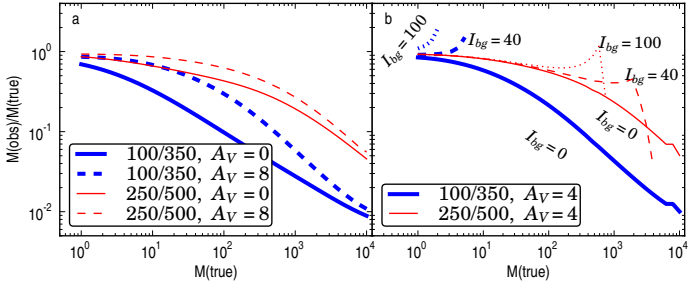}}
\caption{
The ratio of observed mass and the real core mass for a series of
models obtained by scaling the density of a one solar mass
Bonnor-Ebert sphere. In frame $a$, the masses are determined with
the 100\,$\mu$m and 350\,$\mu$m observations (thick lines) or
the 250\,$\mu$m and 500\,$\mu$m observations (thin lines). The core is
illuminated by the full ISRF (solid lines) or with a field that has
been attenuated by $A_{\rm V}=8^{\rm m}$. Frame $b$ shows results for
background subtracted observations after including a uniform
background that at 100\,$\mu$m has an intensity of 0, 40, or
100\,MJy\,sr$^{-1}$.
}
\label{fig:opacity}
\end{figure}

We investigated the possible effects purely from the standpoint
of radiative transfer. We started with a one solar mass Bonnor-Ebert
sphere ($T_{\rm gas}$=10\,K, $\xi$=6.5) and, by multiplying its
density with different constant factors, produced a series of models
of increasing optical depth. Synthetic observations at two wavelengths
were again analysed. This time the cores were assumed to be unresolved
and the mass estimates were derived using the fluxes in an aperture
with the size equal to the diameter of the core. The results are shown
in Fig.~\ref{fig:opacity}. The relevant parameter is the cloud opacity
which, of course, is here directly proportional to the mass of the
core.

In Fig.~\ref{fig:opacity}a, in accordance with Fig.~\ref{fig:Bonnor},
the bias for the original Bonnor-Ebert sphere is about one third when
the 100\,$\mu$m and 350\,$\mu$m observations are used. The bias is
almost eliminated by either resorting to longer wavelengths
(250\,$\mu$m and 500\,$\mu$m) or by reducing the temperature gradients
by attenuating the external radiation field. In
Fig.~\ref{fig:opacity}, the attenuation corresponded to $A_{\rm
V}=8^{\rm m}$. When the optical depth is increased by a factor of ten,
the error is a factor of five for an isolated core ($A_{\rm V}=0^{\rm
m}$) observed at 100\,$\mu$m and 350\,$\mu$m. For the longer
wavelength pair, similar bias is found only for opacities three orders
of magnitude higher.
The longer wavelengths are less sensitive to
temperature errors and, unless observations contain significantly more
noise, tend to give more accurate results.

In Fig.~\ref{fig:opacity}b, we consider the effect of a constant
background that follows a spectrum $B_{\nu}($T=17\,K$) \times \nu^2$
and the level of which is defined by its intensity at 100\,$\mu$m,
$I_{\rm bg}$. The core masses are now estimated from background
subtracted observations. When 100\,$\mu$m is included
the core becomes colder and the 100\,$\mu$m surface brightness becomes lower than the surface brightness of the background, and the core disappears.
For the combination of 250\,$\mu$m and 500\,$\mu$m, the background has no
effect before the opacities are two orders of magnitude higher than
those of the original Bonnor-Ebert sphere. After that point, the mass
estimates again briefly increase (relative to the real mass) before
the absorption of the background again exceeds the emission.

In the analysis of 3D MHD runs, background subtraction will not be
used. In that case, according to Fig.~\ref{fig:opacity}a, a factor of
two errors in the mass estimates are not expected before the optical
depths exceed the optical depth of the one solar mass Bonnor-Ebert
sphere ($A_{\rm V} \sim 26^{\rm m}$ for $N(H) \sim 5\times
10^{22}$\,cm$^{-2}$) by at least one order of magnitude.

\subsection{MHD model clouds} \label{sect:results_3d}

\subsubsection{Model I: Unigrid calculations and modified dust}

\begin{figure} 
\centering
\includegraphics[width=8cm]{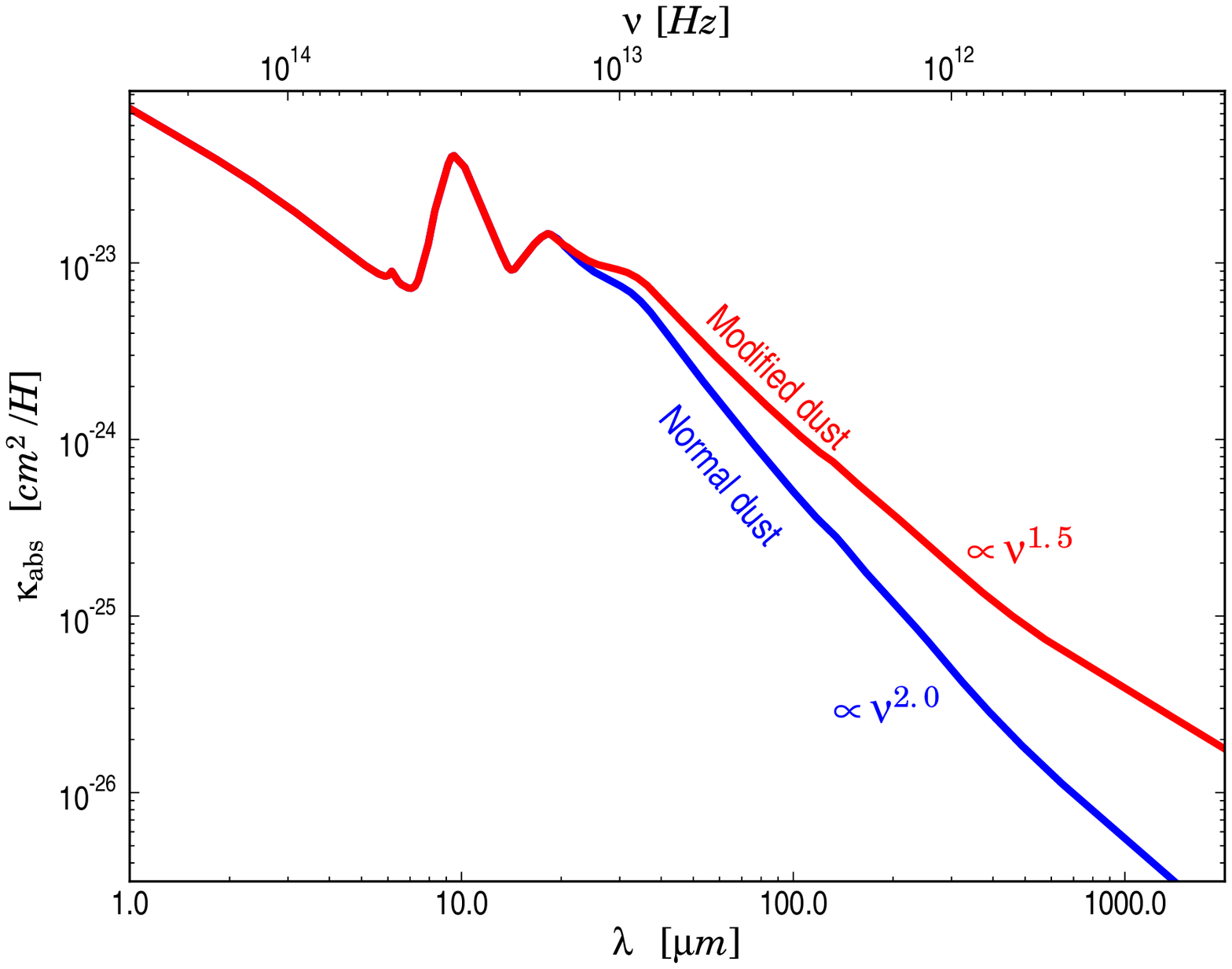}
\includegraphics[width=8cm]{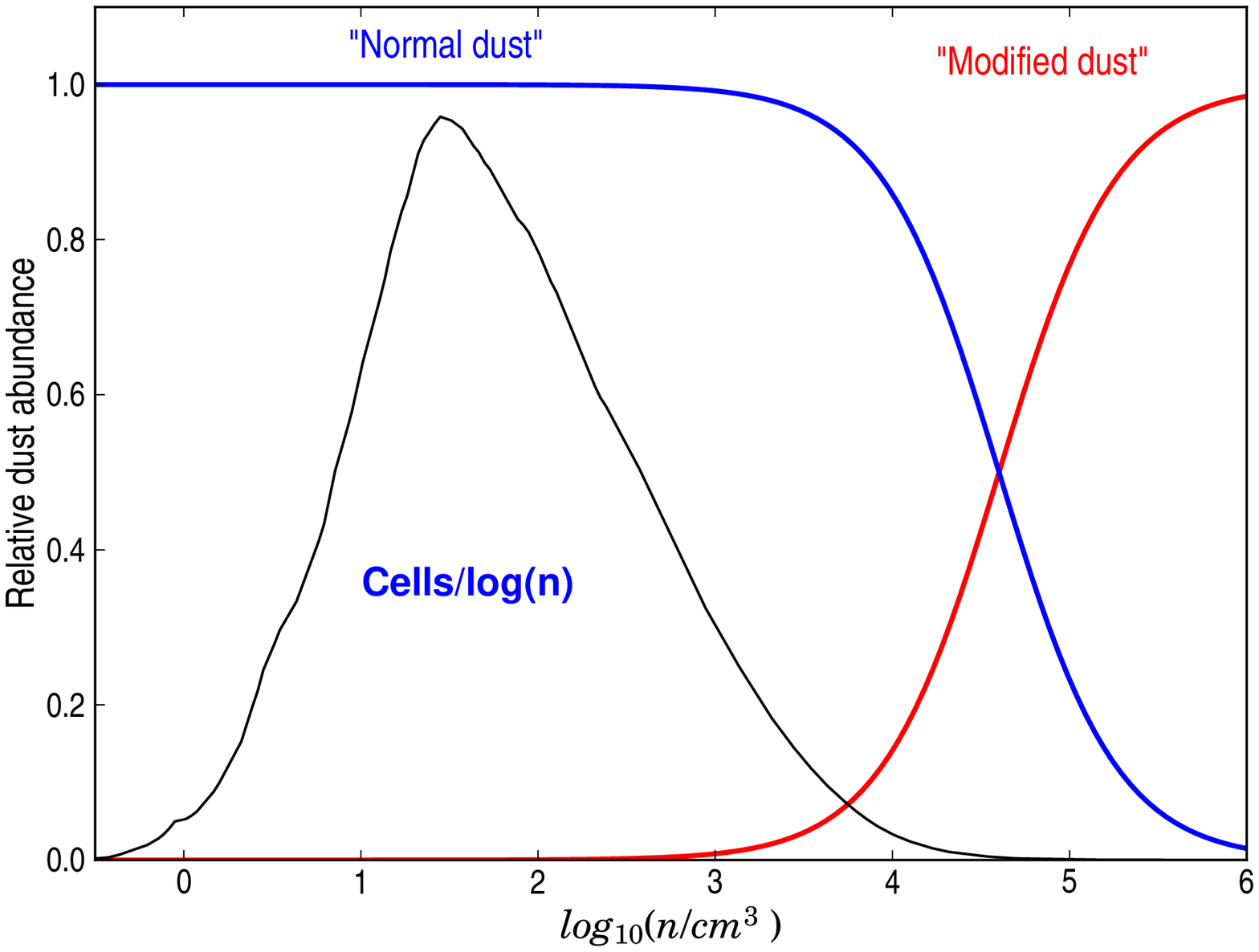}
\caption{
Dust opacities for the two dust models used (top) and
their relative abundances as a function of density (bottom). The
histogram shows the distribution of volume densities in Model I. The y-scale of the histogram is arbitrary.
} 
\label{fig:moddust} 
\end{figure}

As mentioned in Sect.~\ref{sect:intro}, there are indications that in
the dense and cold regions the dust sub-mm emissivity increases and
emissivity index $\beta$ changes. Following the results of Ossenkopf
\& Henning~(\cite{ossenkopf94}) we created a modified version of the
employed dust model by changing the emissivity as shown in
Fig.~\ref{fig:moddust}. The relative abundances of the normal and
modified dust were set according to the density so that the abundance
of the modified dust becomes significant only in the densest regions. 

\begin{figure}
\centering
\includegraphics[width=8cm]{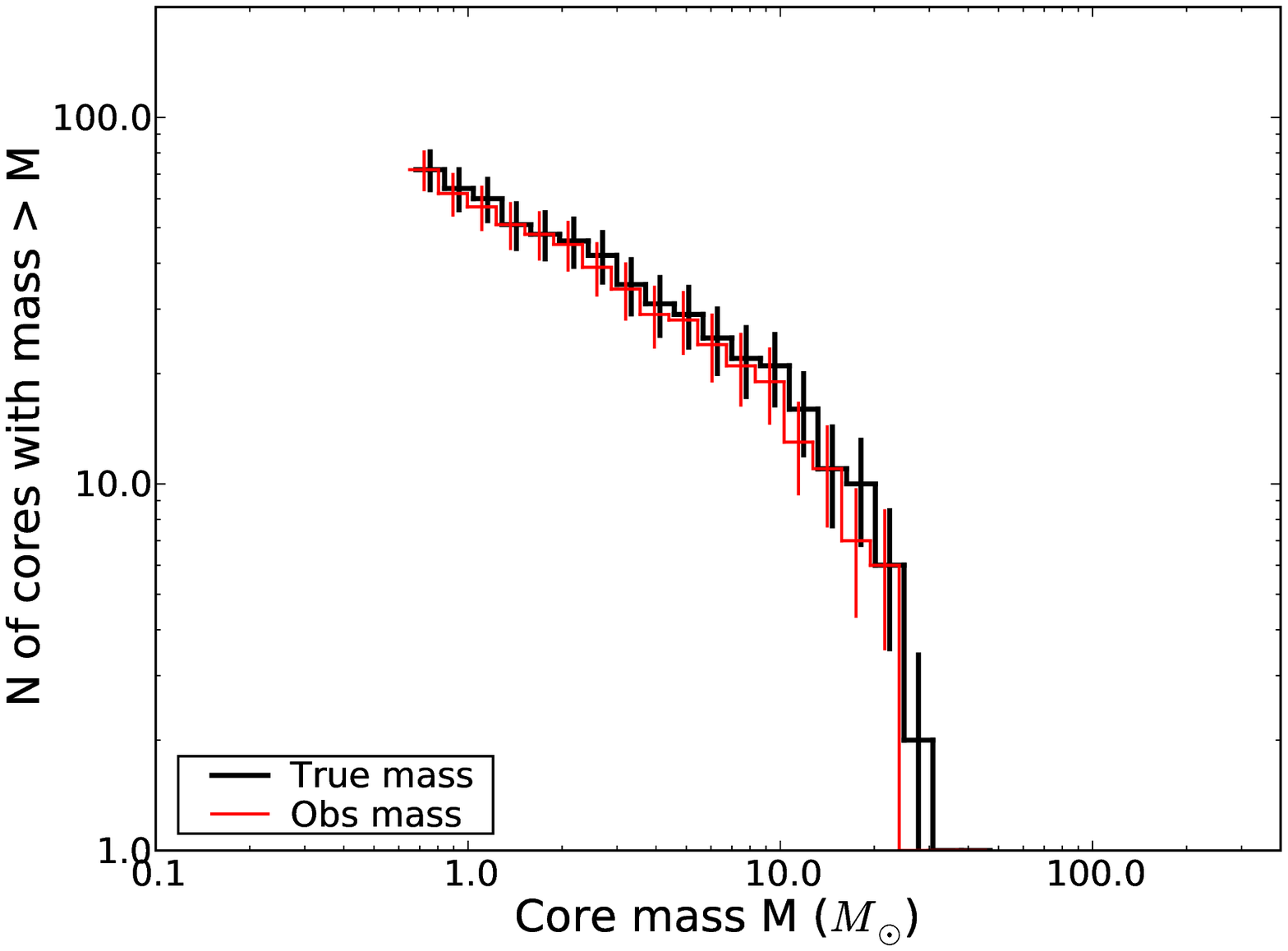}
\includegraphics[width=8cm]{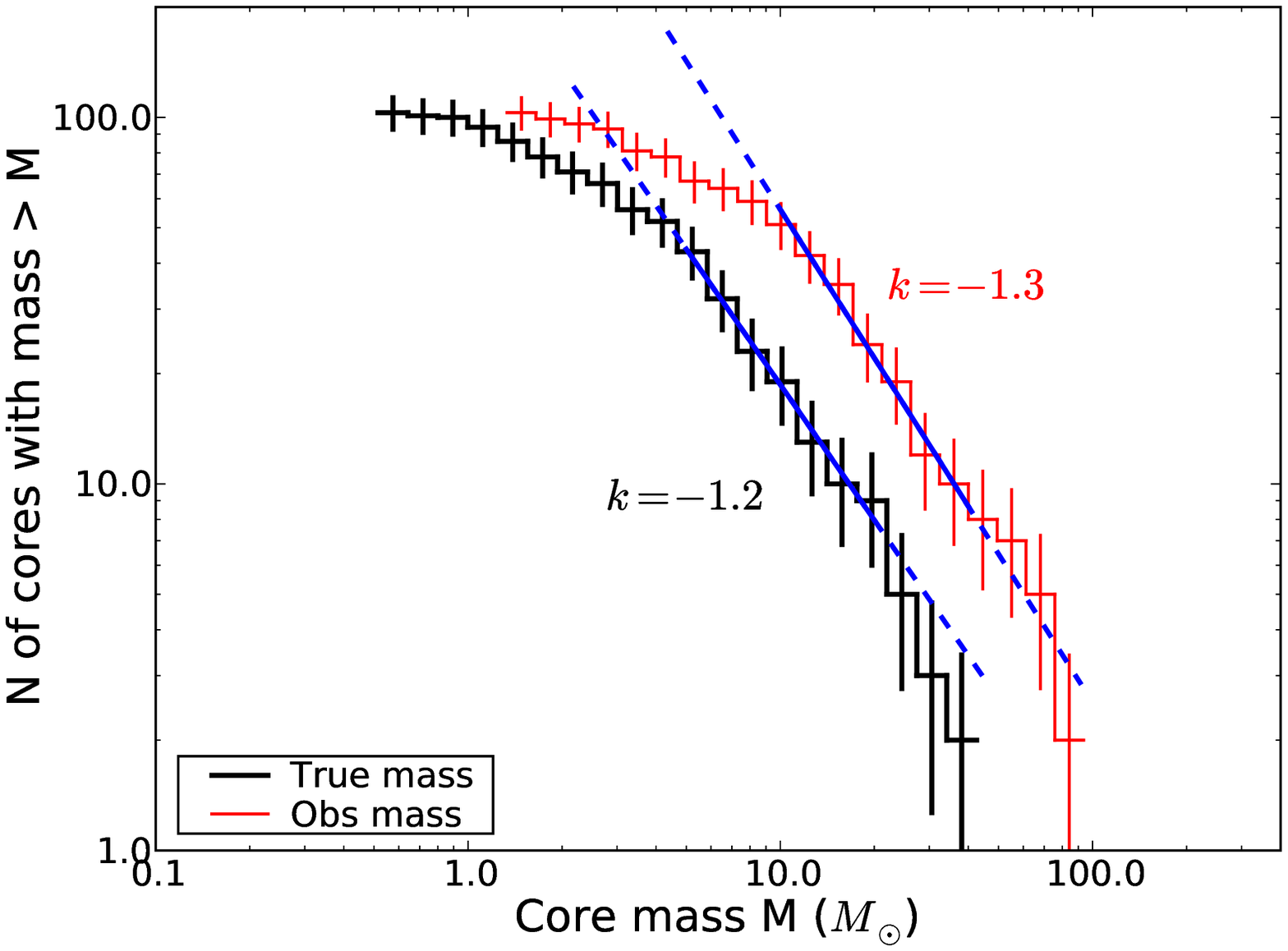}
\caption{
Model I: Cumulative mass spectra (with Clumpfind clumps) obtained with
normal dust (top frame) and a modified dust model (bottom frame; see
text). The analysis was carried out with observations at 250\,$\mu$m
and 500\,$\mu$m and an assumed cloud distance of 400 pc. The
slope (k) of the mass spectra is obtained by making a least squares
fit to the linear part (marked with continuous line).
} 
\label{fig:msp_mwd_mod} 
\end{figure}

The mass spectra obtained with the normal dust and with the mixture of
normal dust and modified dust are shown in Fig.~\ref{fig:msp_mwd_mod}.
The mass estimates were calculated with the 250\,$\mu$m and
500\,$\mu$m surface brightness maps and a distance of 400 pc. In the
analysis, the emissivity index $\beta$ and the dust opacity $\kappa$ both corresponded to the actual values of the
original dust model. Without the modified dust, the mass estimates
were found to be almost completely unbiased. However, in the model
containing also modified dust, the observed masses were strongly {\em
overestimated} because in that case most of the dust found in the
cores has a $\kappa$ value that is higher than what was assumed in the
analysis of the observations. 
However, the difference is not as large
as expected by the change in $\kappa$ alone which at 500\,$\mu$m would
be a factor of five. 
The mass errors increase when analysis is performed with $\beta=2$ that is larger than the actual value $\beta=1.5$
of the modified dust. The fact that masses are overestimated only by a factor of $\sim$3 (see Fig.~\ref{fig:msp_mwd_mod}) suggests that a
large fraction of the observed intensity still comes from normal dust in regions surrounding the densest parts of the cores.
The slope for the true mass spectrum 
(meaning one estimated with the true column densities) 
is $k = -1.2$ and for the observed mass spectrum $k = -1.3$,
suggesting that the shape of the spectra does not change notably.
The values correspond to the power law exponent -2.3 for $dN/dM$
similar to the values observed in real clouds (\cite{enoch08}, Motte et
al.~\cite{motte98}, K\"onyves et al.~\cite{Konyves2010}).

\subsubsection{Model II: AMR model with cores of moderate opacity}

\begin{figure*}
\centering
\includegraphics[width=0.285\linewidth]{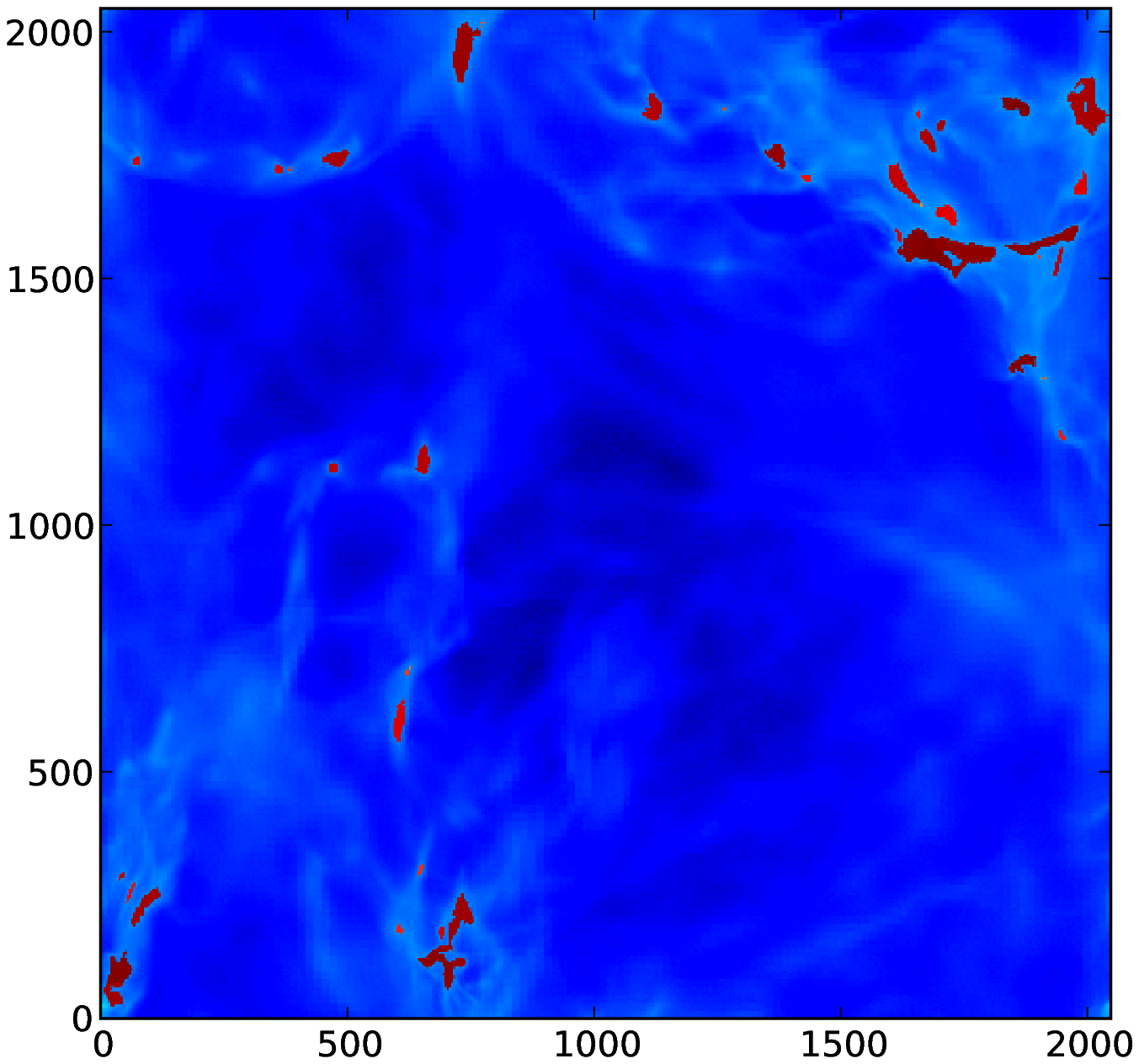}
\includegraphics[width=0.285\linewidth]{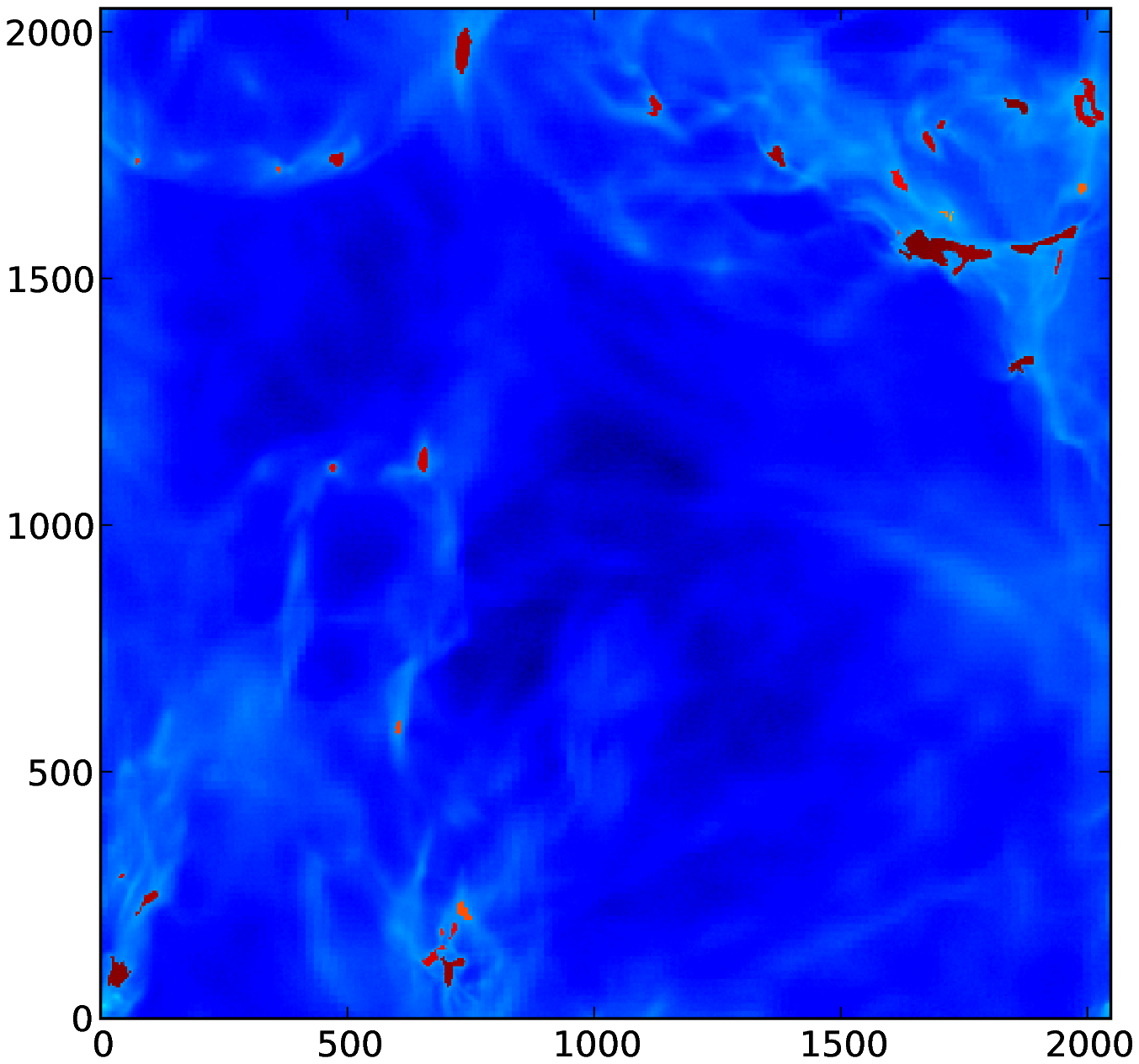}
\includegraphics[width=0.285\linewidth]{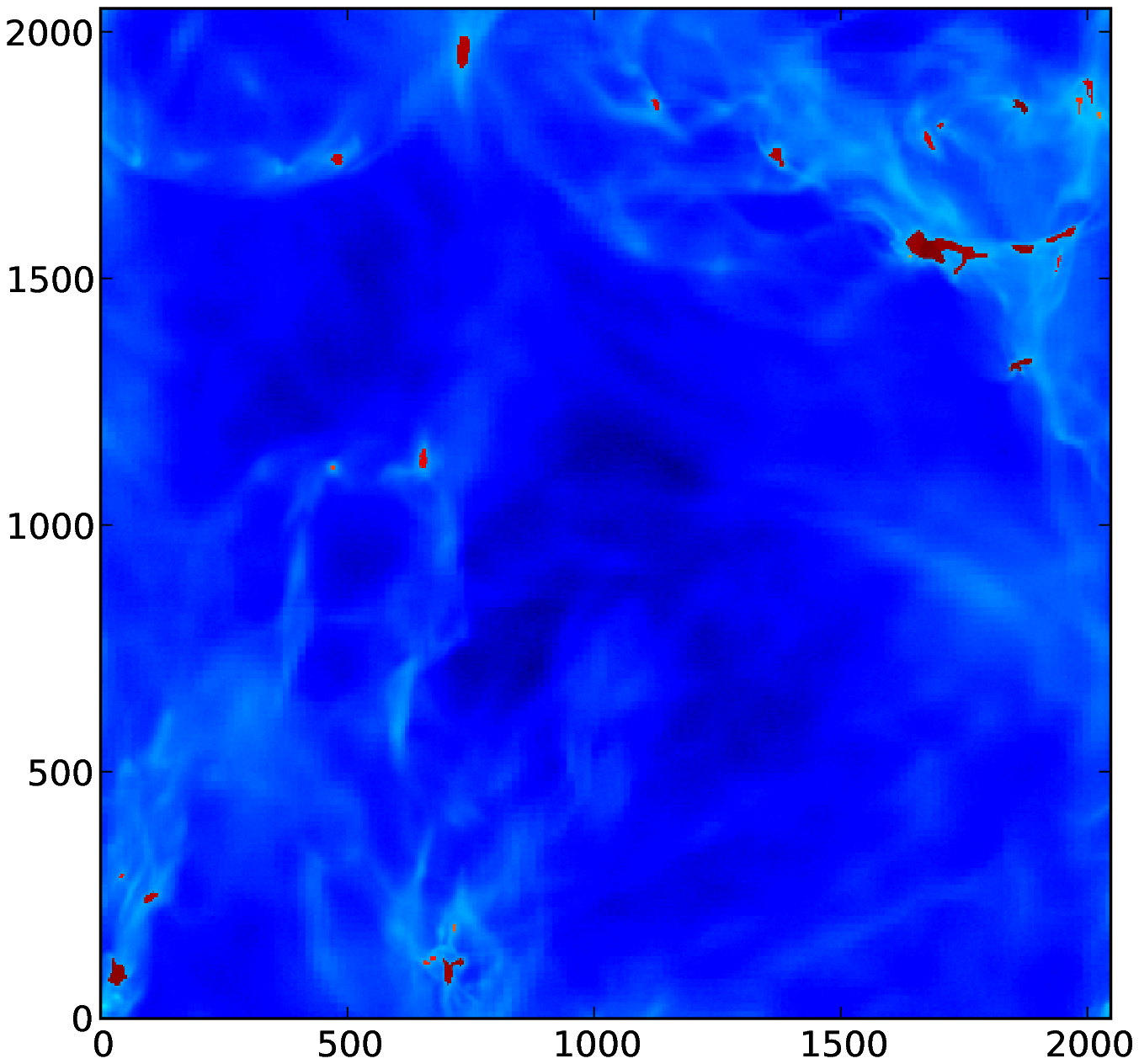} \\
\includegraphics[width=0.285\linewidth]{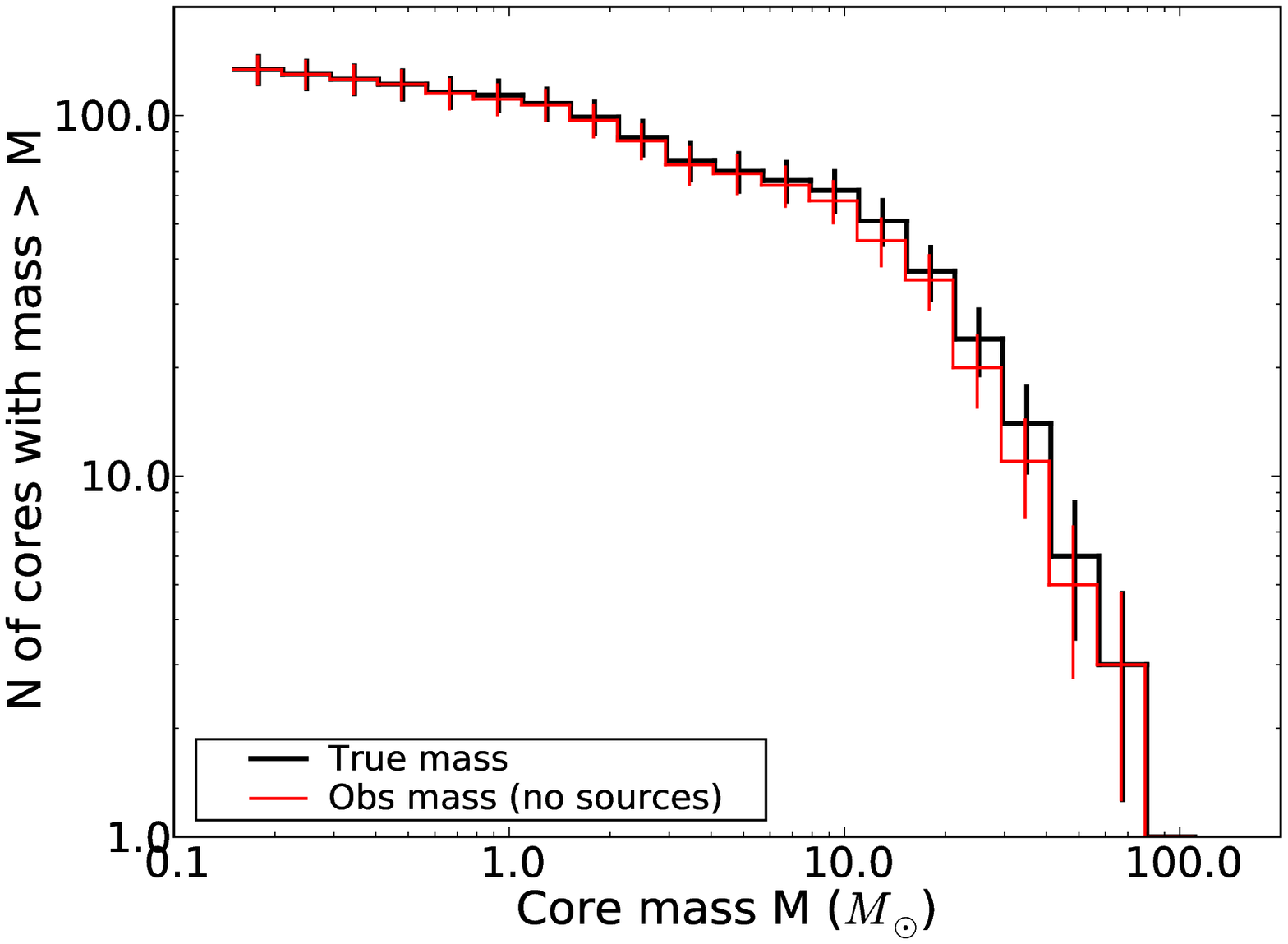}
\includegraphics[width=0.285\linewidth]{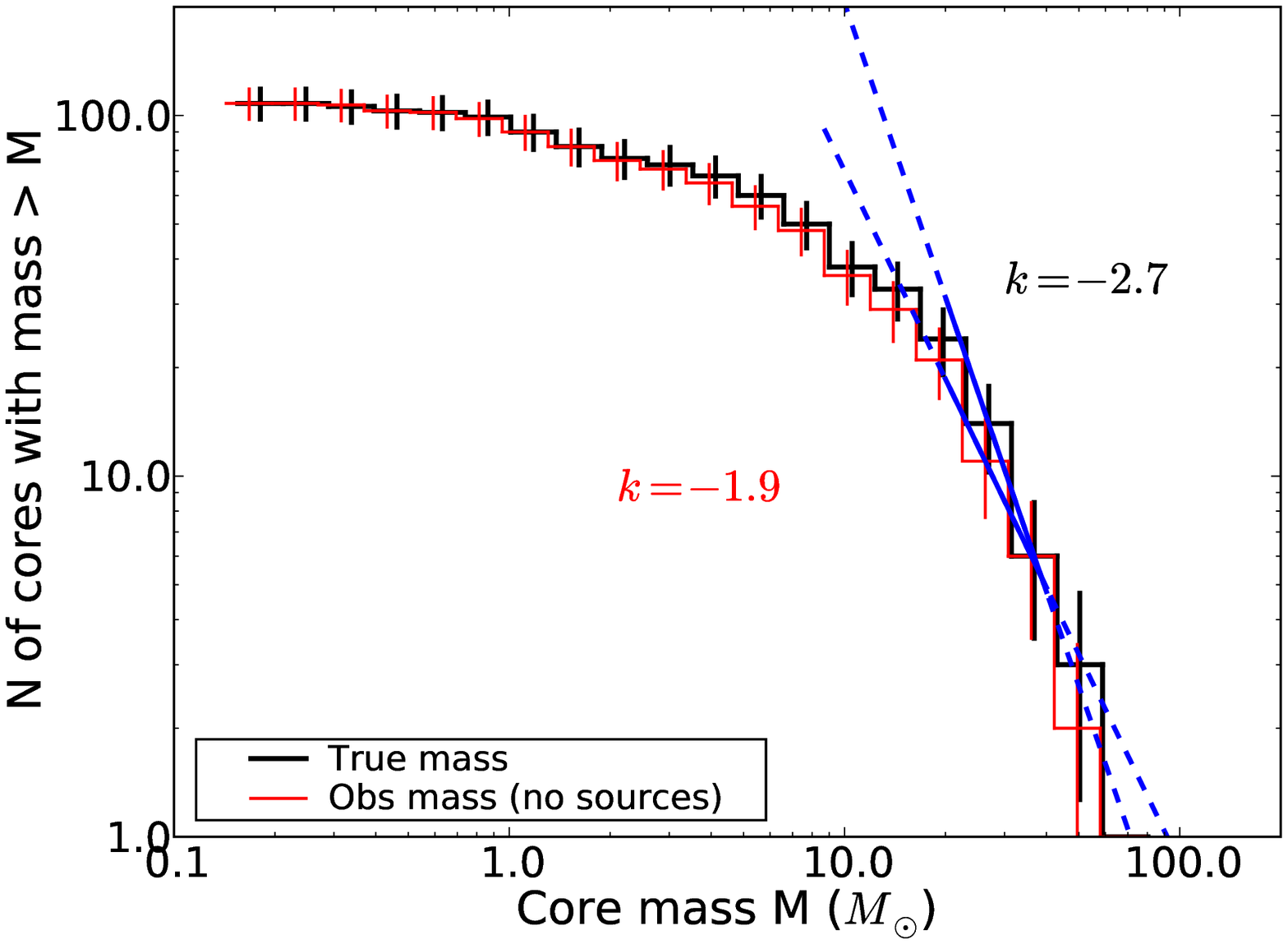}
\includegraphics[width=0.285\linewidth]{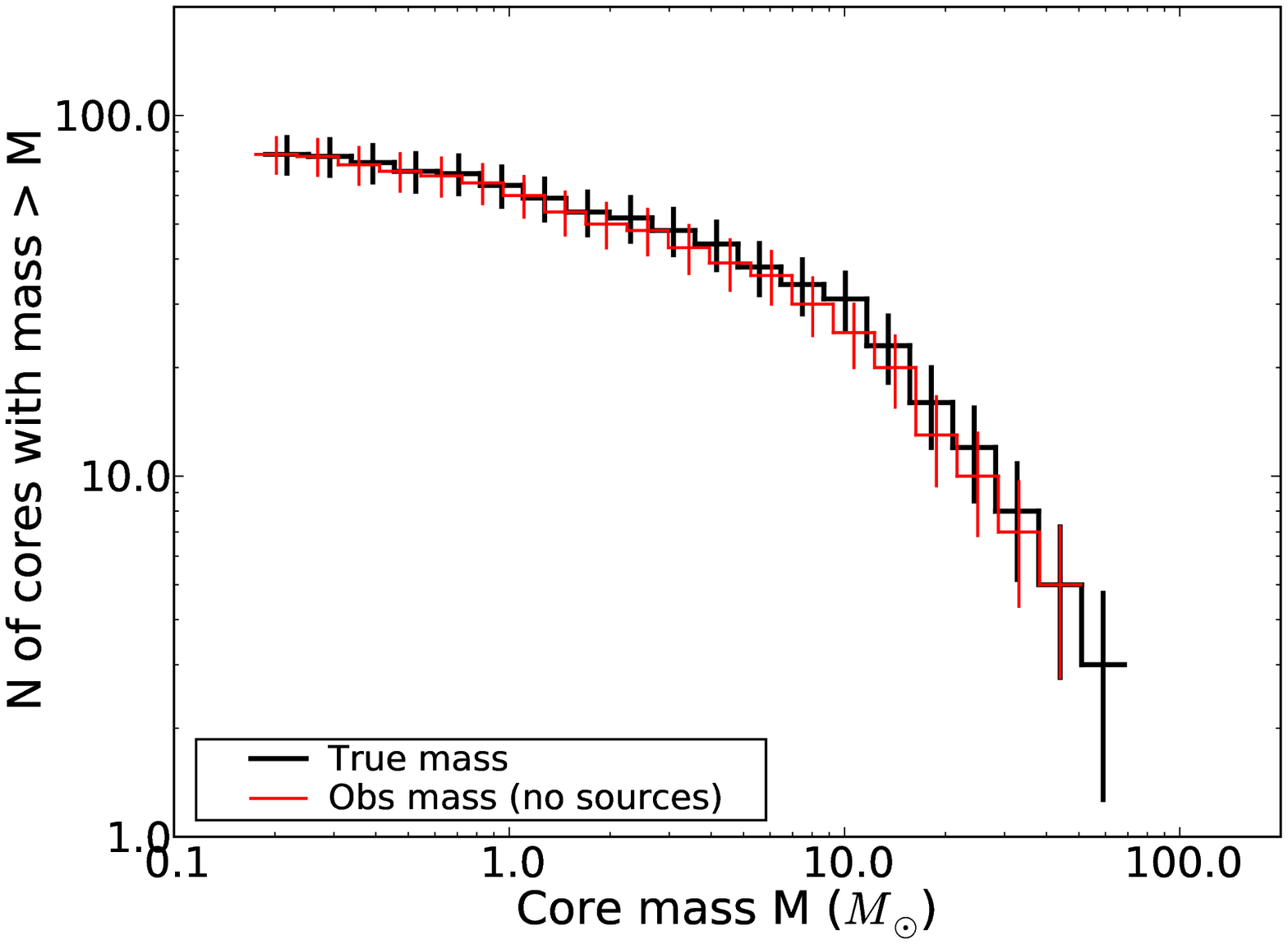} \\
\includegraphics[width=0.285\linewidth]{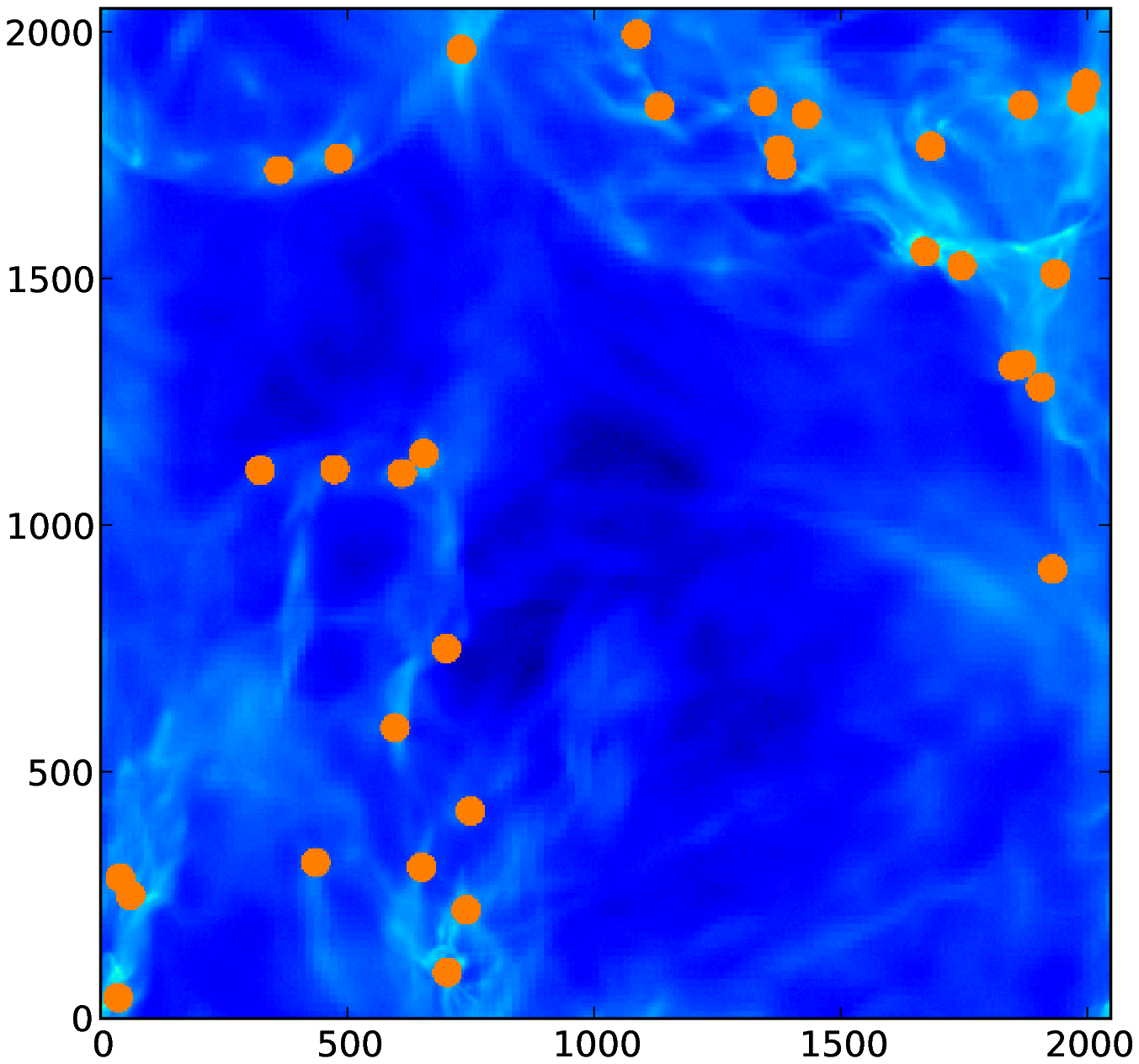}
\includegraphics[width=0.285\linewidth]{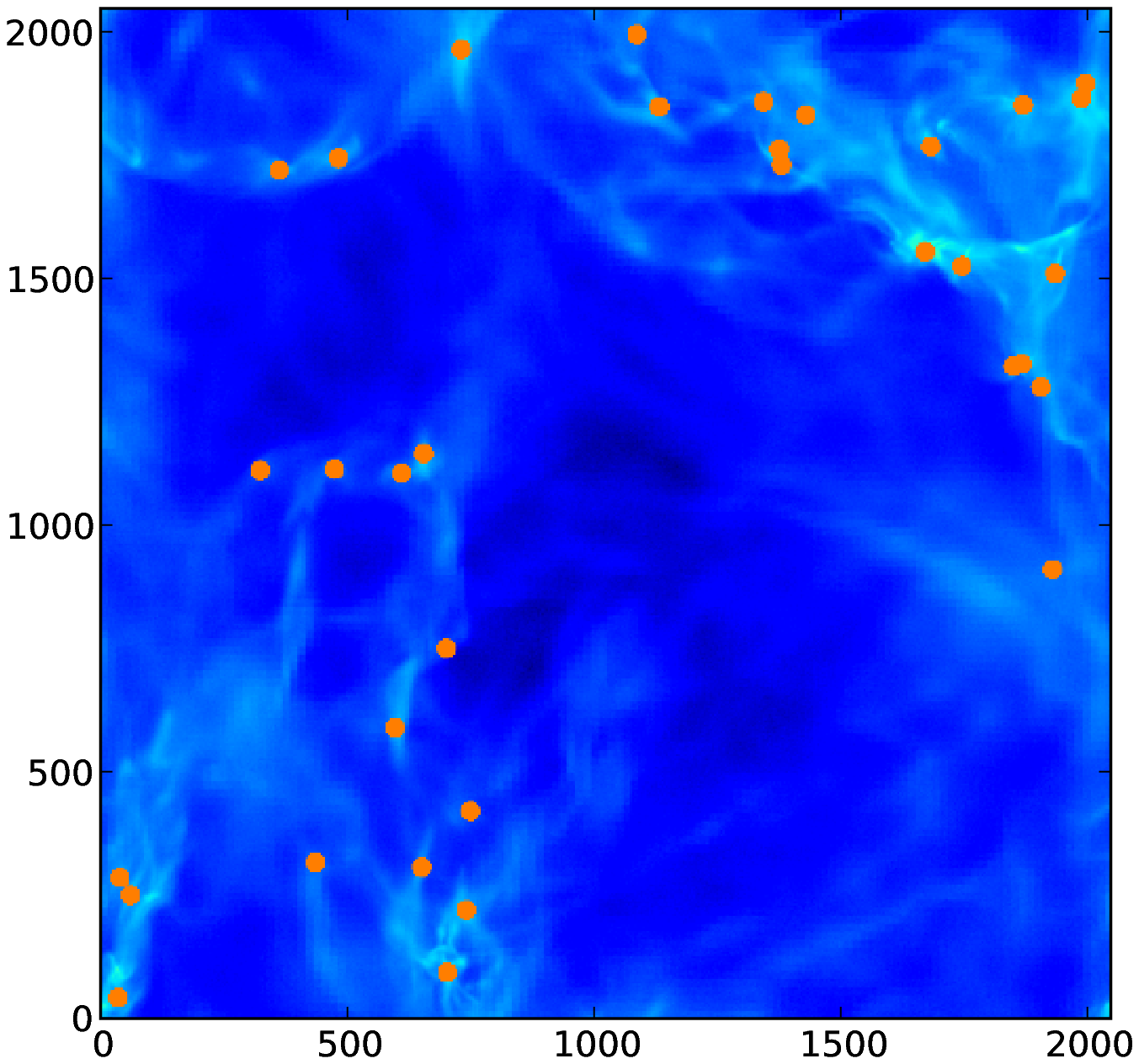}
\includegraphics[width=0.285\linewidth]{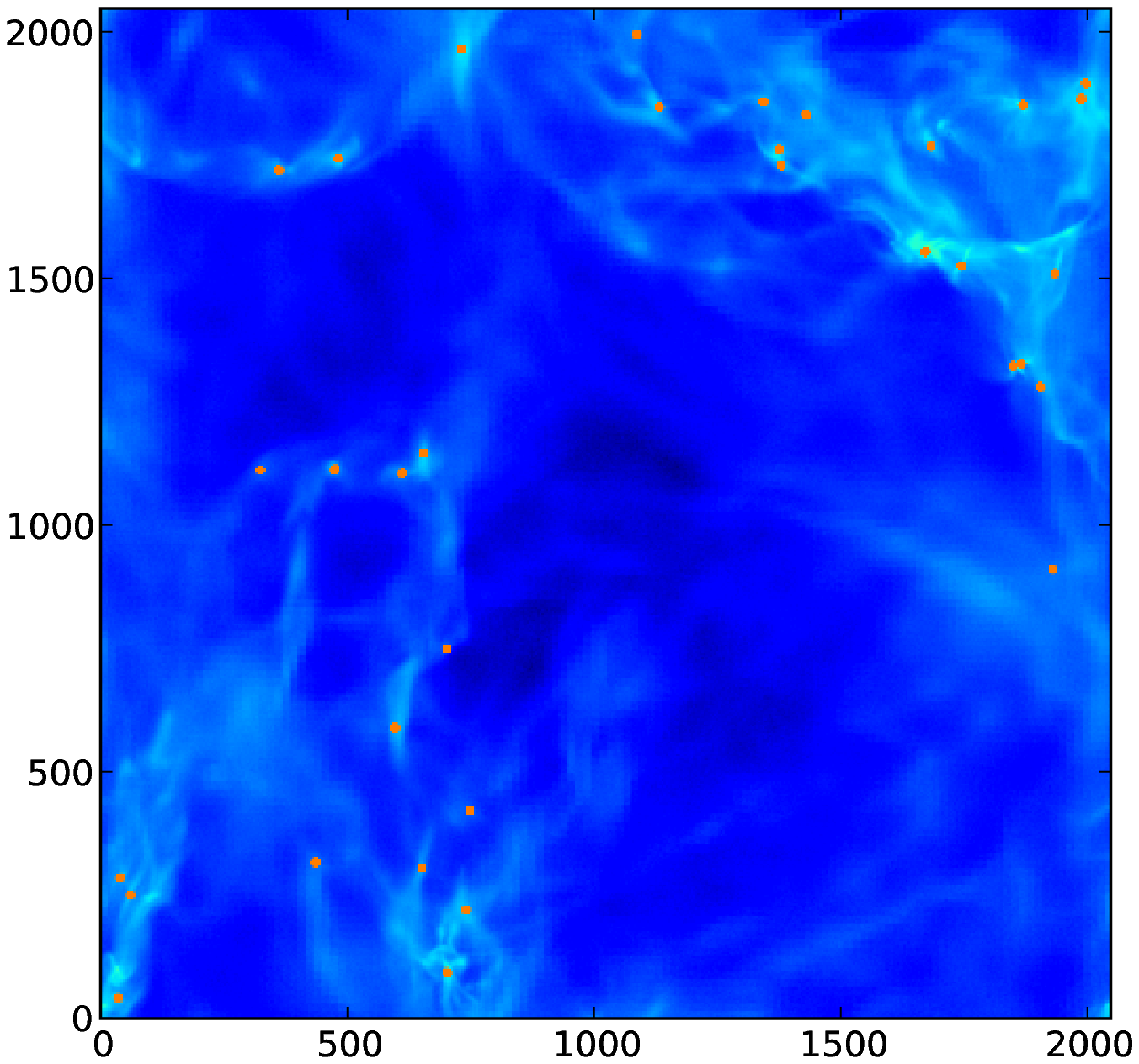} \\
\includegraphics[width=0.285\linewidth]{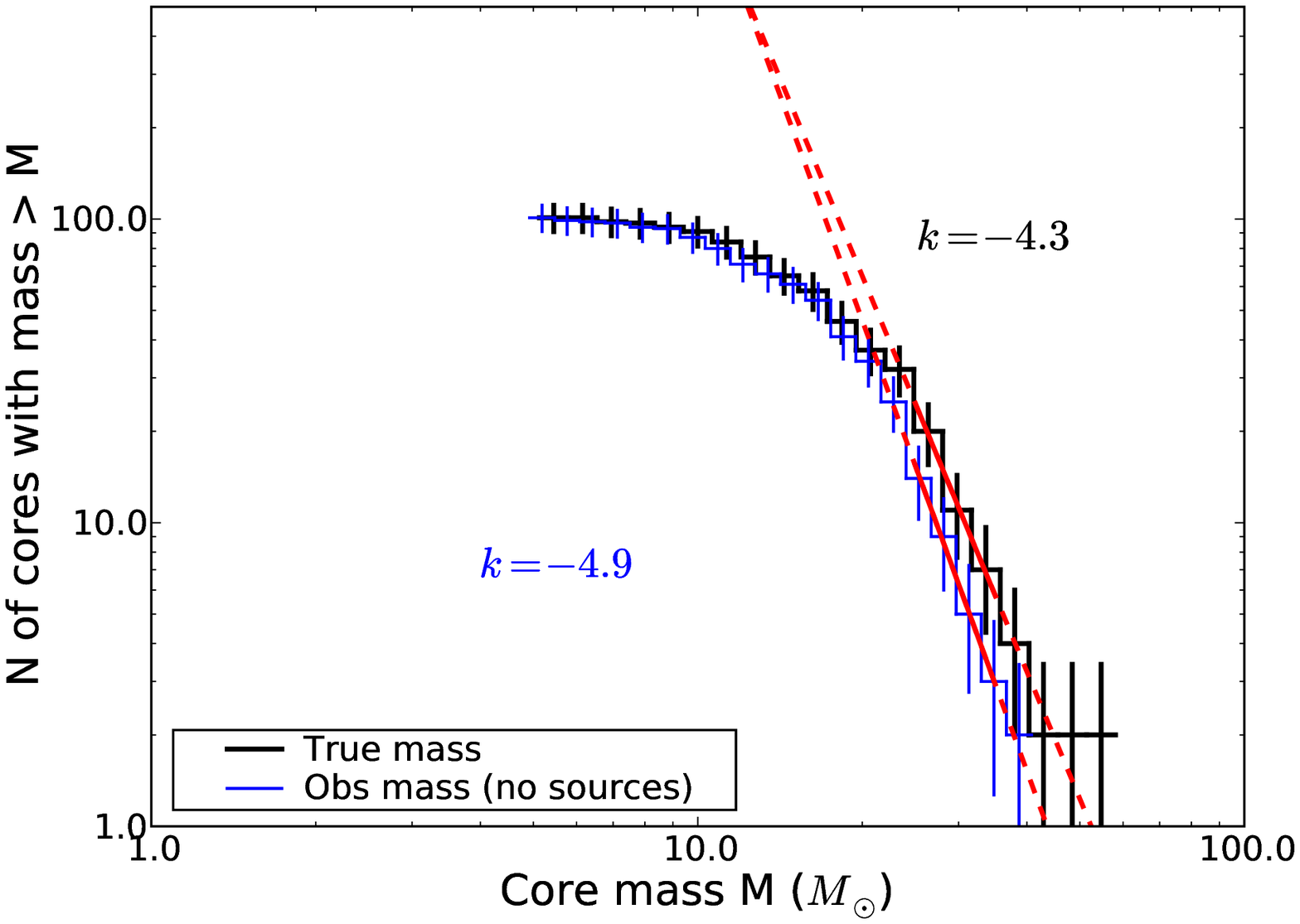}
\includegraphics[width=0.285\linewidth]{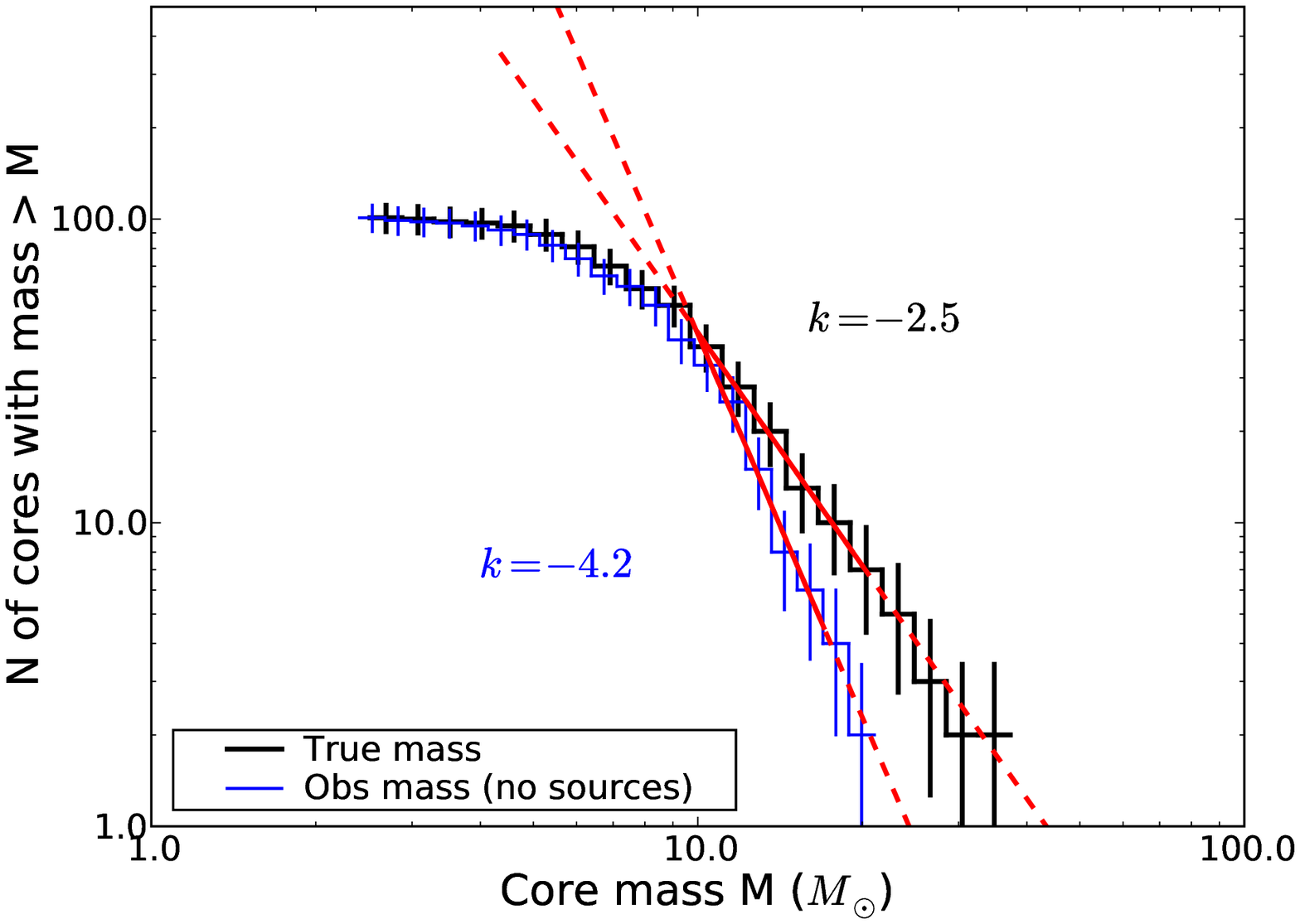}
\includegraphics[width=0.285\linewidth]{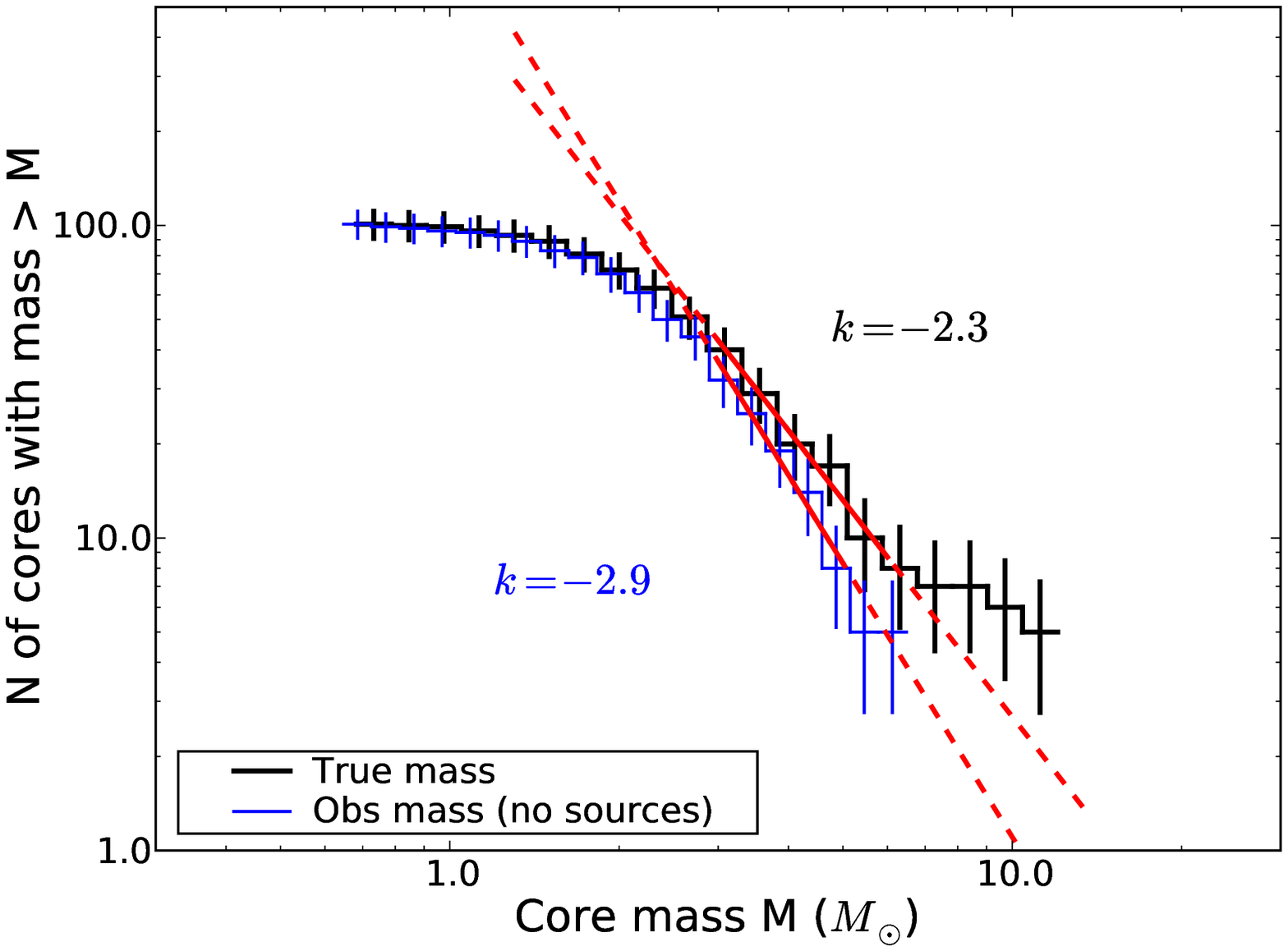} \\
\includegraphics[width=0.285\linewidth]{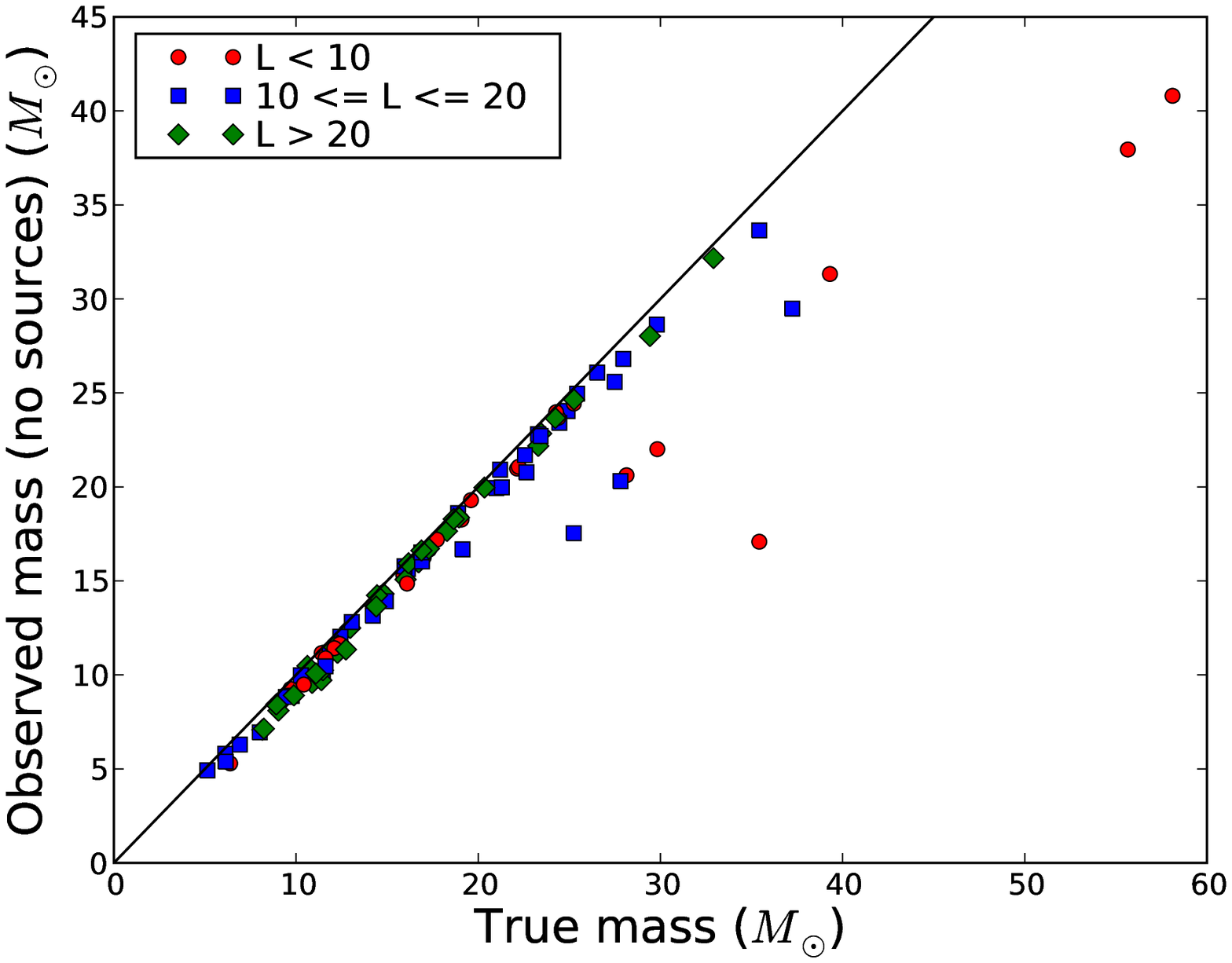}
\includegraphics[width=0.285\linewidth]{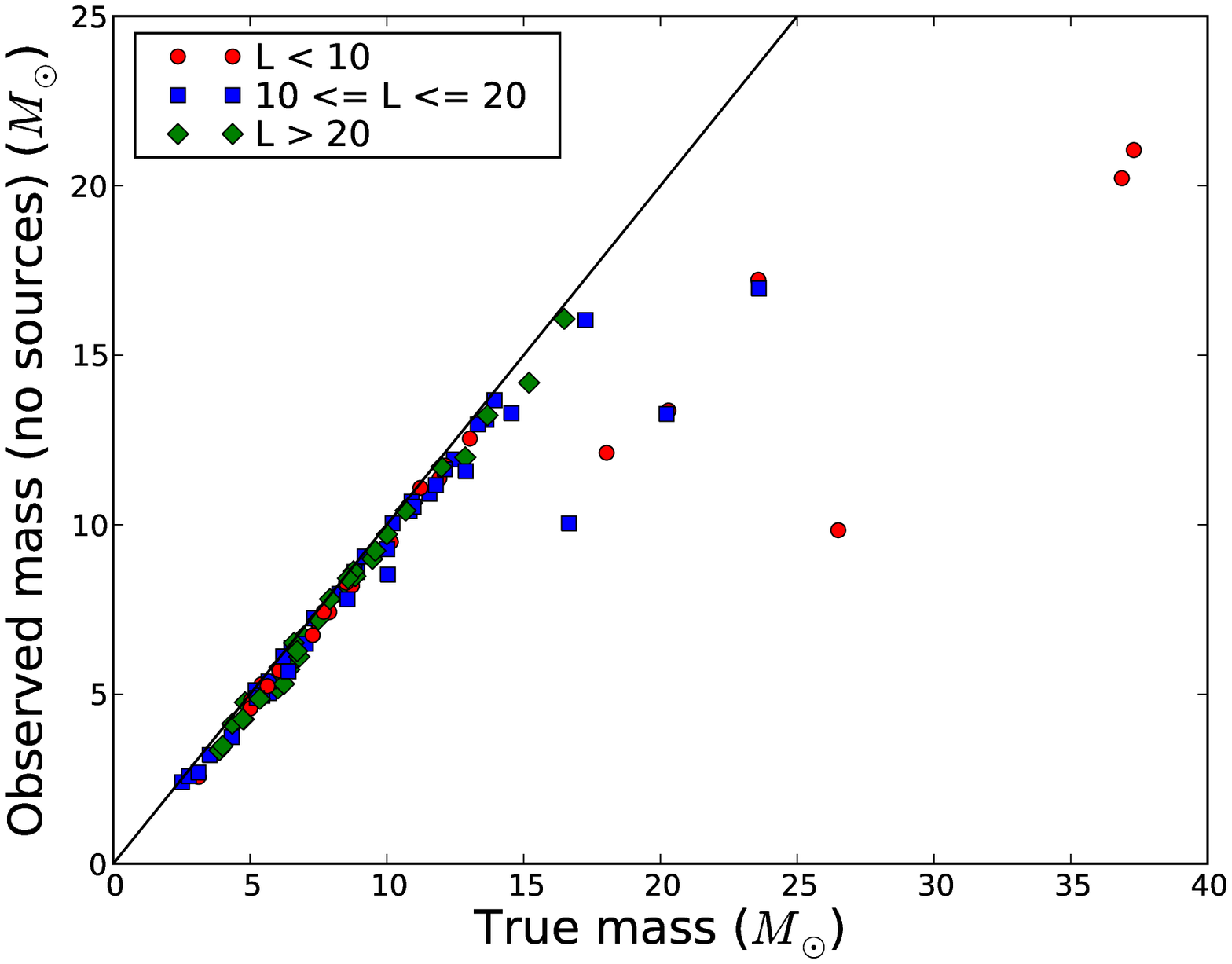}
\includegraphics[width=0.285\linewidth]{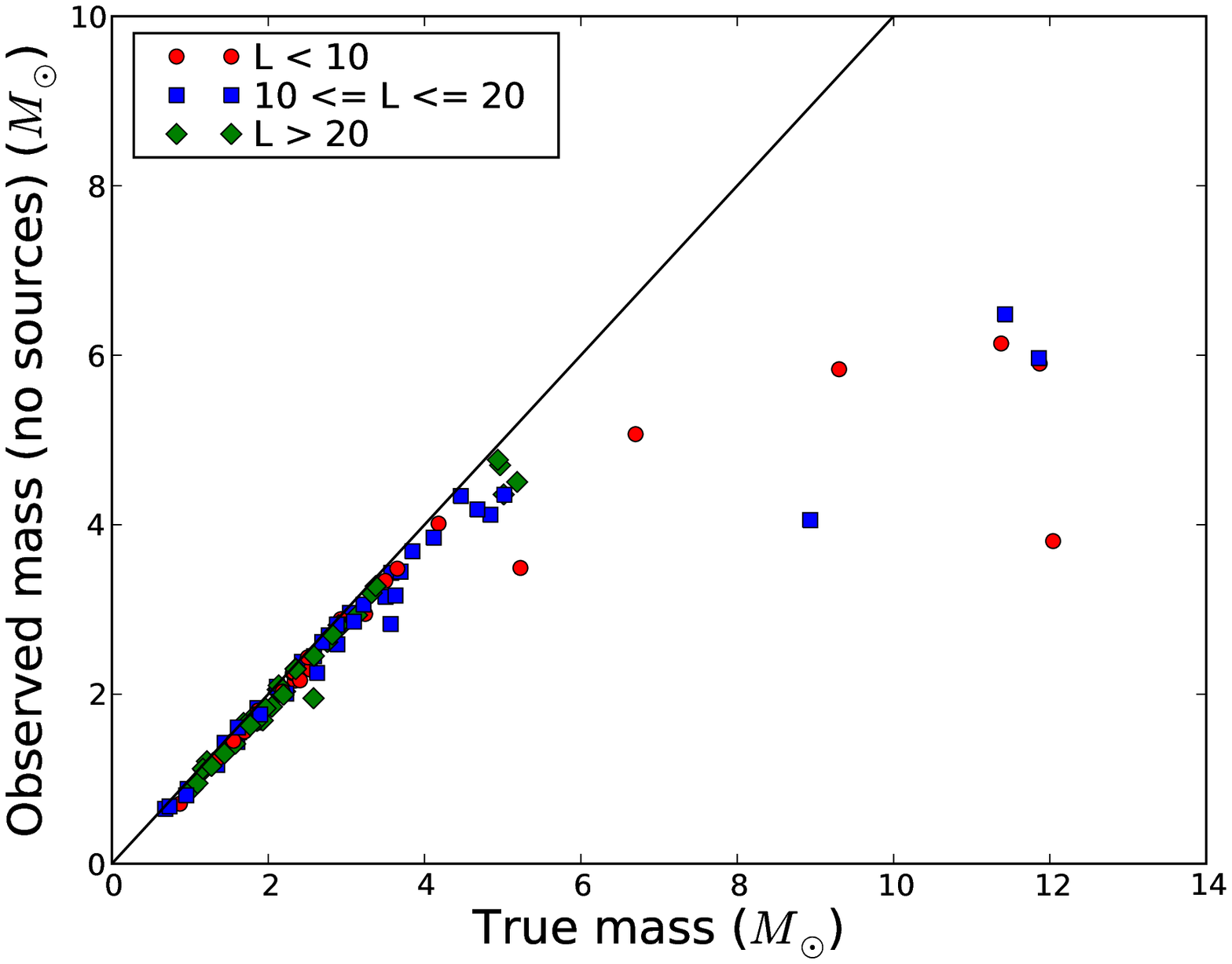}
\caption{
The comparison of Model II mass estimates obtained with different
methods. The analysis was performed with surface brightness maps at
250\,$\mu$m and 500\,$\mu$m and an assumed cloud distance of 100\,pc.
{\em Rows 1--2:} The clump maps and mass spectra obtained with three different 
scaling factors for the rms noise.
In the map the discovered clumps are
drawn in red. The mass spectra are plotted using both the observed
masses derived from the simulated observations (red line) and the true masses
(but identical clumps) obtained from the cloud model.
{\em Rows 3--4:} The results of similar analysis using areas within a
fixed radius around the positions of gravitationally bound cores. The
radius was 30 (leftmost frames), 20, or 10 pixels. The observed mass
spectra is drawn with blue line in the electronic article in order to
make a difference between the mass spectra obtained with Clumpfind
clumps.
{\em Row 5:} Relations of observed mass vs. true mass for the
fixed radius environments. $L$ denotes the luminosity of the sources
that are later added in the cores and is related to the mass of the original 3D clump.
Here it acts only as an identifier for the masses of the cores.
The clump areas are shown in only one direction, but in the mass
spectra and observed mass vs. true mass plots clumps in all three
directions are included. The slope (k) of some of the mass spectra is
obtained by making a least squares fit to the linear part (marked with
continuous line). 
Note that in middle frame on row 2 there is not a proper linear part. The slope values are very sensitive to the fitting interval.
}
\label{fig:clumps} 
\end{figure*}

Compared to Model I, Model II
contains denser and more evolved cores and the resolution of the model
is correspondingly much higher.
The derived mass spectra of Model II are shown in
Fig.~\ref{fig:clumps}. The analysis was done using the wavelengths
250\,$\mu$m and 500\,$\mu$m, the correct values of the $\kappa$ and
$\beta$ parameters, and an assumed cloud distance of 100\,pc. The
clumps were searched with Clumpfind using three different 
parameter values as scaling factors for the rms noise
(the uppermost frames in Fig.~\ref{fig:clumps}). For
comparison, the masses were also calculated inside fixed radius
regions around each projected centre position of the gravitationally
bound cores (see Sect.~\ref{sect:clumpfind}). Three different radii
were used: 30, 20, or 10 pixels (the third row of frames in
Fig.~\ref{fig:clumps}).

The cumulative histogram plots of Fig.~\ref{fig:clumps} compare the
observed mass spectra with the `true mass spectra'.  In both cases the
clumps are identical and are extracted from the observed column density maps.
With the Clumpfind clumps, the true and the observed mass spectra are
almost indistinguishable, independently of the Clumpfind parameters
and thus the number and spatial extent of the cores. Also the mass
spectra obtained with regions of a 30 pixel radius show a close
similarity and only at the highest densities the observed clump masses
are somewhat underestimated. However, if we look at masses within 20 or
10 pixel radii, the masses are clearly underestimated in the high mass
end. With 10 pixel radius the high mass end of true mass spectrum
develops a hump that also deviates from the typical power law shape of
mass spectra.

The mass spectra obtained with Clumpfind clumps
(Fig.~\ref{fig:clumps}, row 2, middle frame) do not have a proper
linear part. 
By fitting a slightly different mass interval
the slopes can vary at least between values 
$k$ = -1.9 and -2.7. The slopes for the observed and true mass spectra
obtained using clumps with a fixed 20 pixel radius
(Fig.~\ref{fig:clumps}, row 4, middle frame) are $k$ = -4.2 and -2.5,
respectively, indicating a clear change in the shape of the mass
spectra. The mass spectra obtained with constant radius clumps do
not adjust to the size of the clumps and because of this they depict
more the column density than mass. Therefore these slopes cannot be
directly compared to the CMF results reported from observations. The slopes for mass spectra with radius = 10 and 30 are given in the figure, but the values are highly dependent on the used mass interval.

In Fig.~\ref{fig:clumps} the last row of figures compares directly the
true and the observed masses within the circular regions around the 
positions of gravitationally bound cores. The effect that was seen in
the mass spectra is even more apparent in these correlations. 
For a majority of the cores the masses are almost unbiased within all the three radii
but in all cases there are several high density cores whose masses are
severely underestimated. When the radius is decreased, the errors of
the high density cores increase and the masses are systematically
underestimated, sometimes by up to a factor of three.

\begin{figure} 
\centering
\includegraphics[width=8cm]{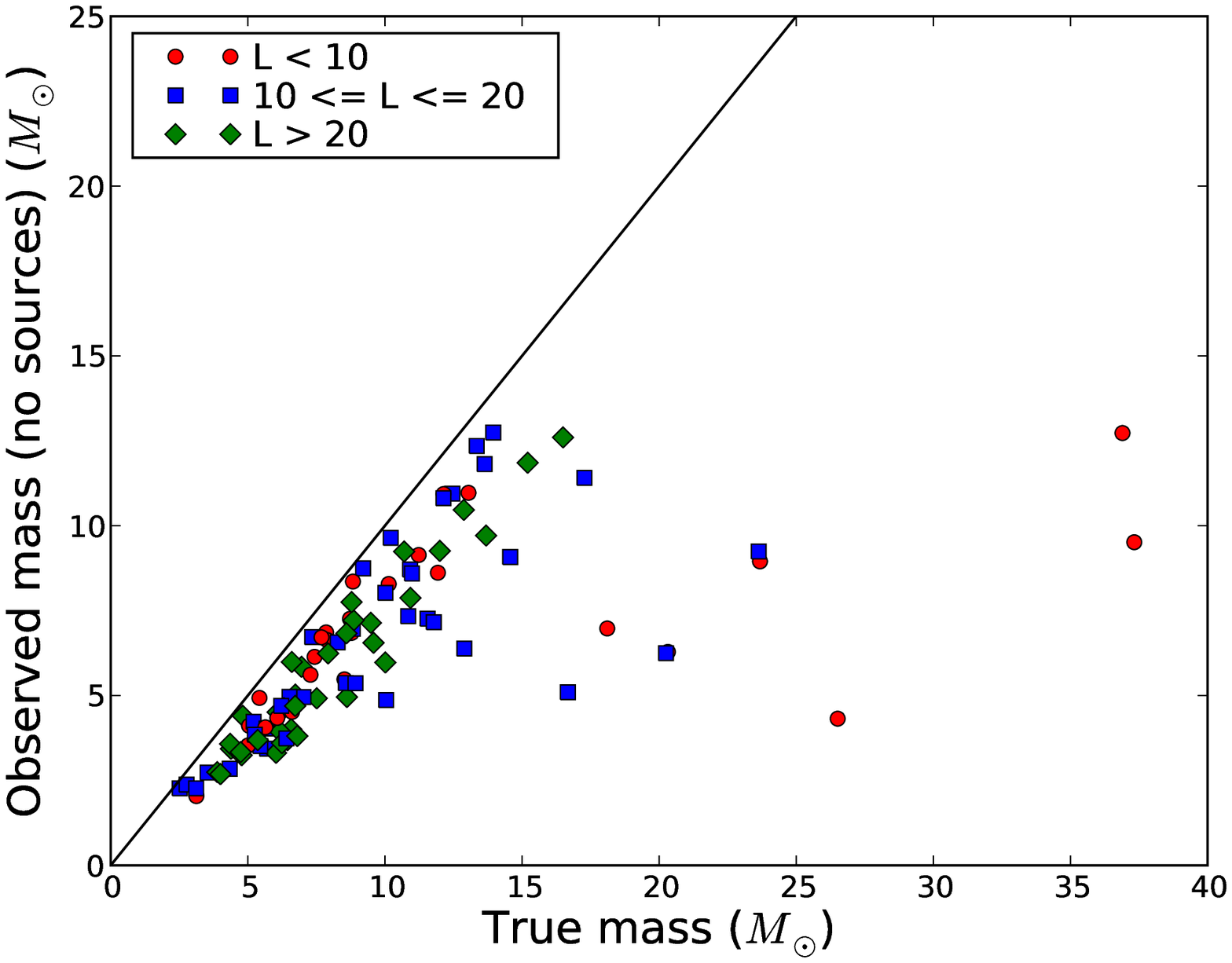}
\includegraphics[width=8cm]{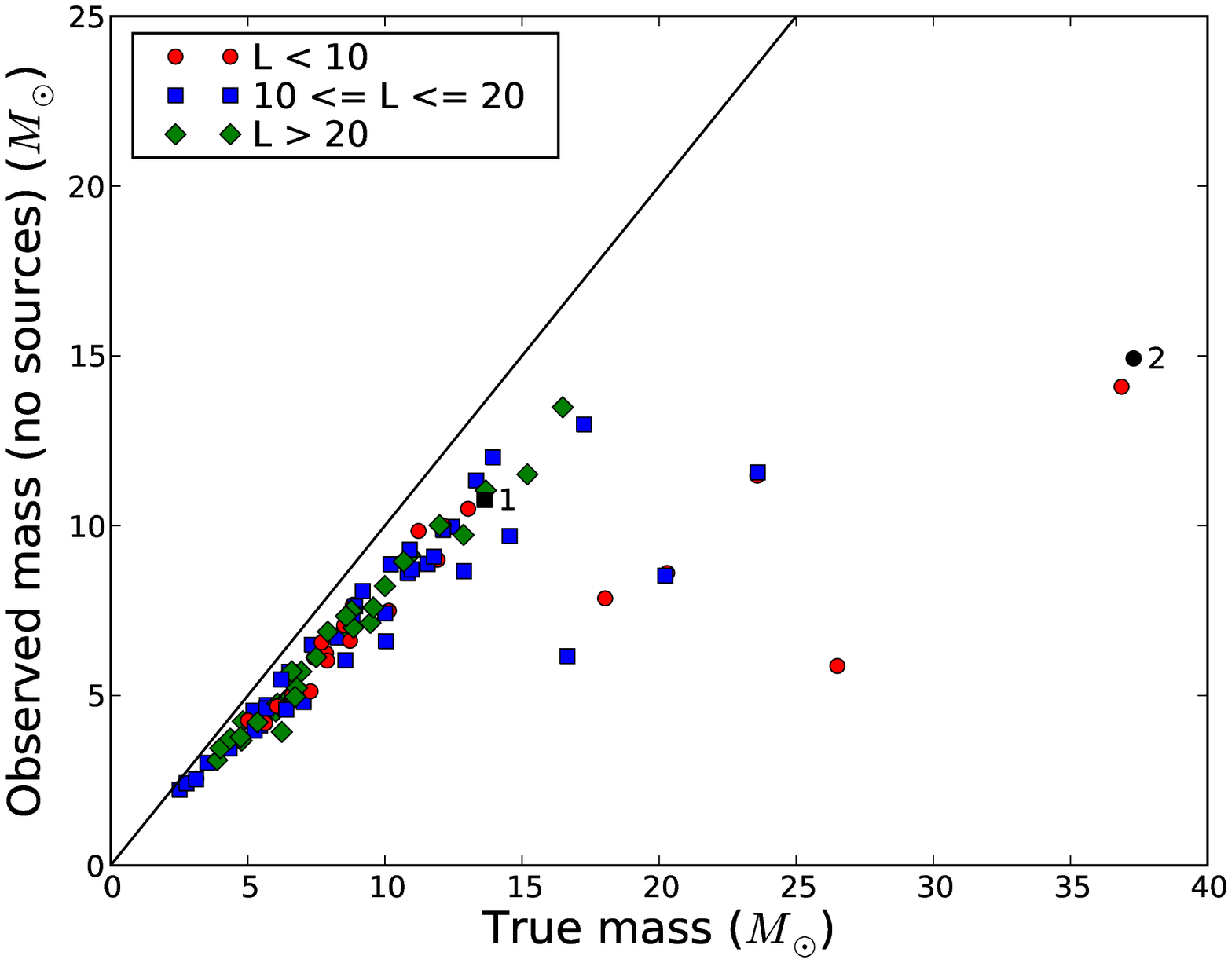}
\caption{
Model II: Relations of observed mass vs. true mass for the fixed
radius environments with radius 20, distance of 100\,pc and correct
$\beta$. Colour temperature is calculated using wavelength pair 100
and 350 $\mu$m (top frame) and five wavelengths (100, 160, 250, 350,
500 $\mu$m) (bottom frame). The SED for the cores in bottom frame
marked with black colour and numbers 1 and 2 are shown in
Fig.~\ref{fig:CIImultipleT_SED}.
}
\label{fig:CIIscatterplots} 
\end{figure}

We also calculated the colour temperatures using the wavelength
pair 100 and 350\,$\mu$m and the five wavelengths (100, 160, 250, 350,
500\,$\mu$m). The relations of observed mass vs. true mass are shown in
Fig.~\ref{fig:CIIscatterplots}. Compared to the case with the
wavelength pair 250 and 500\,$\mu$m (Fig.~\ref{fig:clumps}, bottom row,
middle frame), the observational bias is larger in both of these cases
where shorter wavelengths were used. The case with wavelength pair 100
and 350\,$\mu$m appears to have approximately the same mean value for
the $M_{\rm true}/M_{\rm obs}$ relation as the case with five
wavelengths, although with more scatter. Therefore smaller scatter
does not imply smaller bias that is caused by the use of shorter
wavelengths (100\,$\mu$m and possibly 160\,$\mu$m).

\begin{figure} 
\centering
\includegraphics[width=8cm]{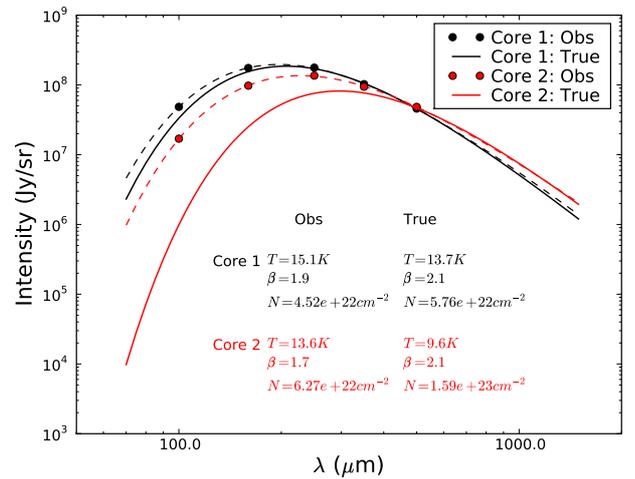}
\caption{
Model II: SEDs for the two cores (with radius = 20 pixels) that were
marked with numbers 1 and 2 in Fig.~\ref{fig:CIIscatterplots}.
The ``true'' SEDs (see text) are drawn with continuous lines.
The observed SEDs (dashed lines) are fits to five wavelengths
100, 160, 250, 350, and 500\,$\mu$m and the corresponding mean
intensities are marked with circles. Core 1 (with small
observational mass bias) is marked with black and core 2 (with large
observational mass bias) with red colour.
For true SEDs:
$T$ = true mean temperature of dust grains, $\beta$ = true spectral
index = 2.1, $N$ = true mean column density for the core area. For
observed SEDs: $T$ = temperature (given by fit), $\beta$ = spectral
index (given by fit), $N$ = observed mean column density for the core
area.
}
\label{fig:CIImultipleT_SED} 
\end{figure}

The observed and 'true' spectral energy distributions (SEDs) of two cores (one
with a small and one with a large mass error; see lower frame in Fig.~\ref{fig:CIIscatterplots}) are
shown in Fig.~\ref{fig:CIImultipleT_SED}. With the 'true' SED we mean the spectrum that corresponds to the true mean
temperature of the dust grains, the true column density of the core, and the true
$\beta$ of the dust. If a spectrum similar to the 'true' SED was observed, we would get the correct mass estimate. The
difference between the 'true' and the observed SED is a measure of the temperature variations along the line-of-sight.
At short wavelengths (less than $\sim$500$\mu$m) the intensity depends on the temperature nonlinearly which results in
a difference between the colour temperature and the average grain temperature. As the colder dust emits less
efficiently, the observed colour temperature is weighted towards the temperature in the warmest regions. This leads to
overestimation of temperature and therefore to the
underestimation of mass. As can be seen in
Fig.~\ref{fig:CIImultipleT_SED}, for the core with a small error in the mass the observed and 'true' SEDs are very
similar. In the case of the core with large observational mass bias (the high density core),
the line-of-sight temperature variations are apparently stronger leading to a larger difference between the 'true'
spectrum and the warmer observed spectrum.

\begin{figure} 
\centering
\includegraphics[width=8cm]{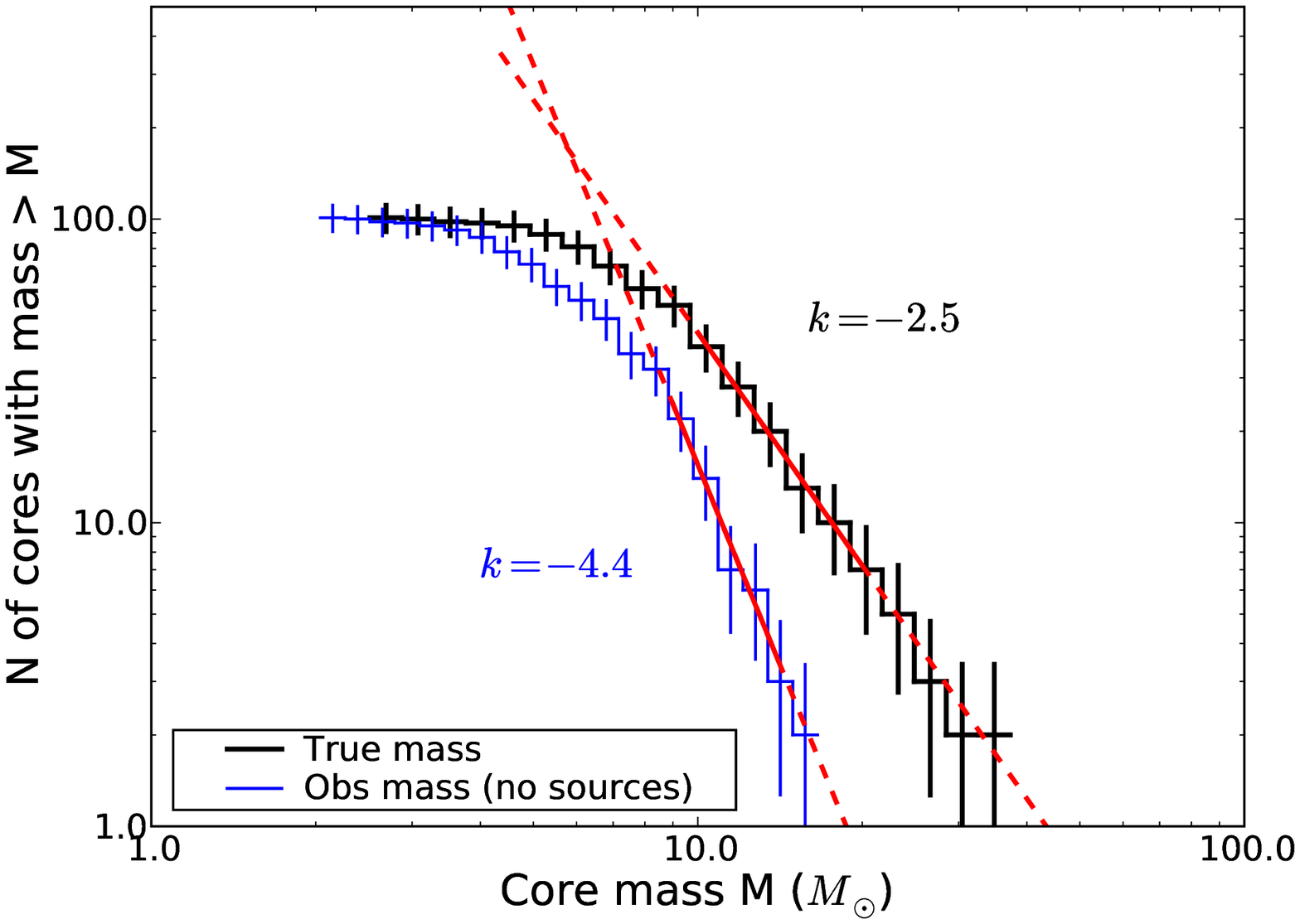}
\includegraphics[width=8cm]{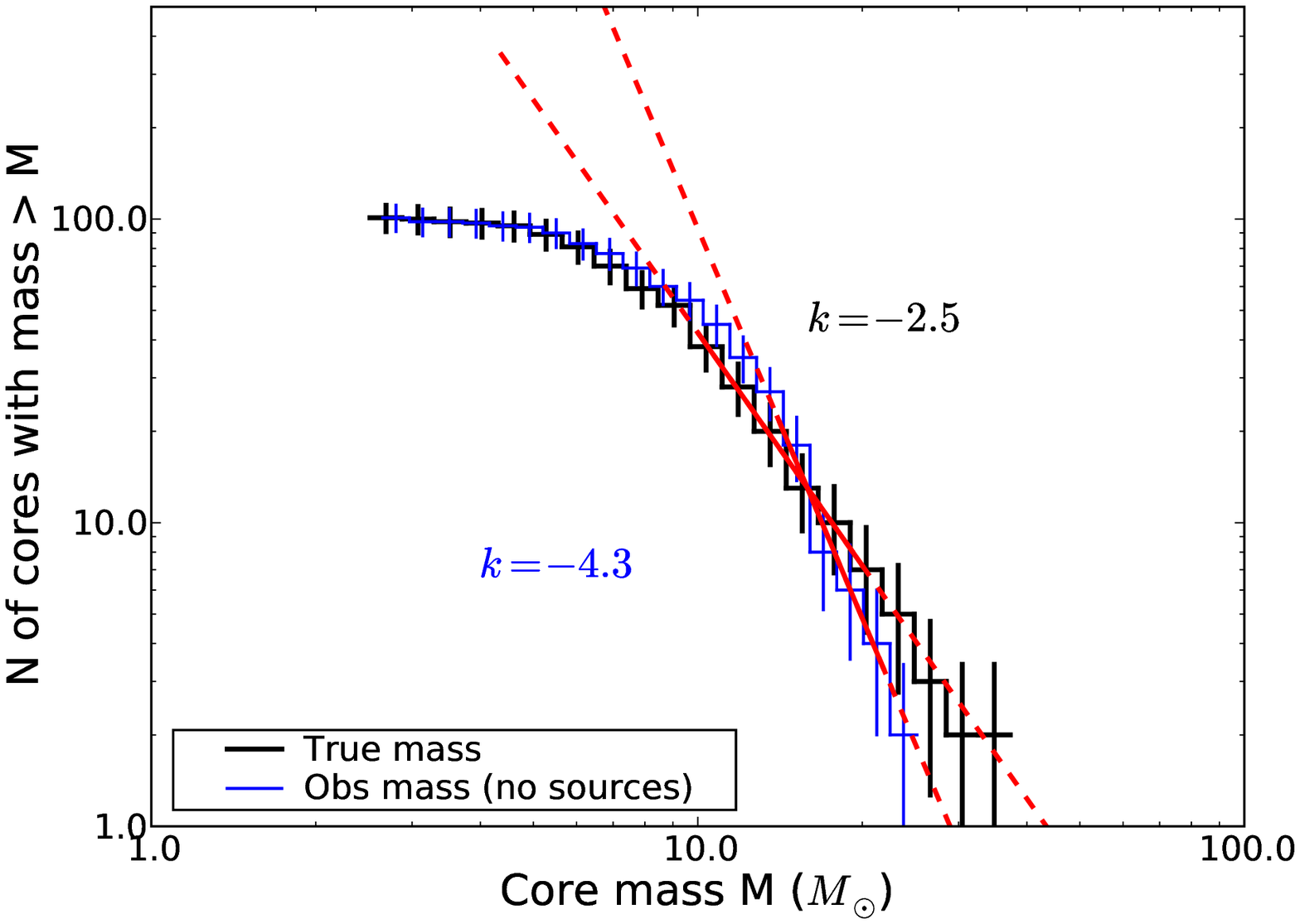}
\caption{
Model II: The effect of changing $\beta$ on mass spectra obtained with
the wavelength pair 250\,$\mu$m and 500\,$\mu$m, assumed distance of
100\,pc and environments of radius 20, $\beta =$1.8 (top) and 2.3
(bottom). The mass spectra obtained with correct $\beta \sim 2.1$ is
shown in Fig.~\ref{fig:clumps}.
}
\label{fig:CIIbeta} 
\end{figure}

The effect of changing the value of $\beta$ that is used in the 
estimation of the colour temperatures is shown in
Fig.~\ref{fig:CIIbeta}. These mass spectra of Model II are obtained
using wavelength pair 250\,$\mu$m and 500\,$\mu$m, assumed distance of 100\,pc and
regions of a fixed radius of 20 pixels. With a smaller value of
$\beta$ the masses are underestimated more severely while the increase
of $\beta$ by 0.2 units is enough to shift the observed mass spectrum
roughly on top of the spectrum obtained with the correct column
densities. However, the $\beta$ parameter does not affect the shape of
the observed mass spectrum which remains steep (slope $k$ varies between -4.2 and -4.4 with all the used $\beta$ values) as the masses of the densest cores are still underestimated.

\begin{figure*}
\centering
\includegraphics[width=0.3\linewidth]{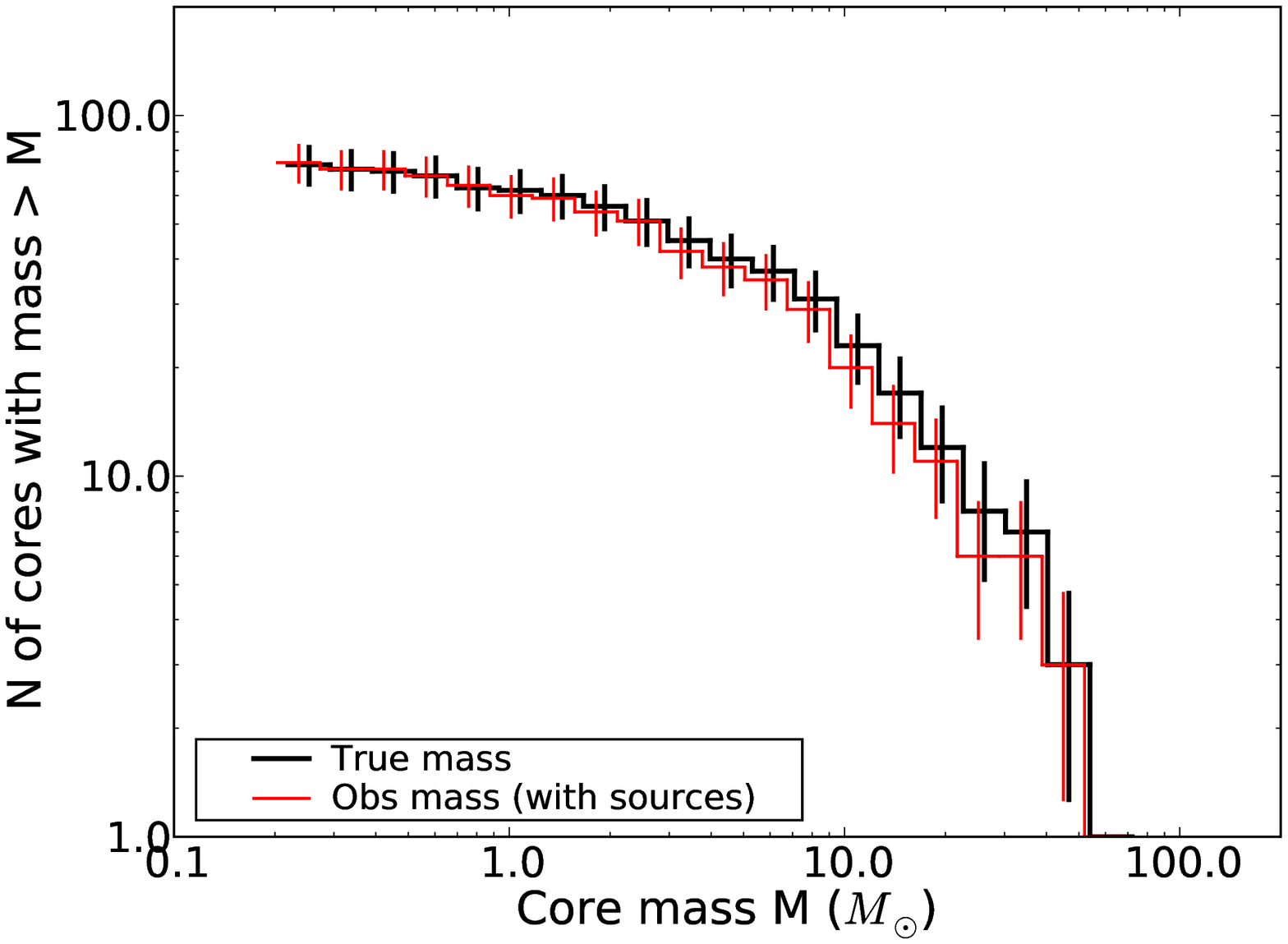}
\includegraphics[width=0.3\linewidth]{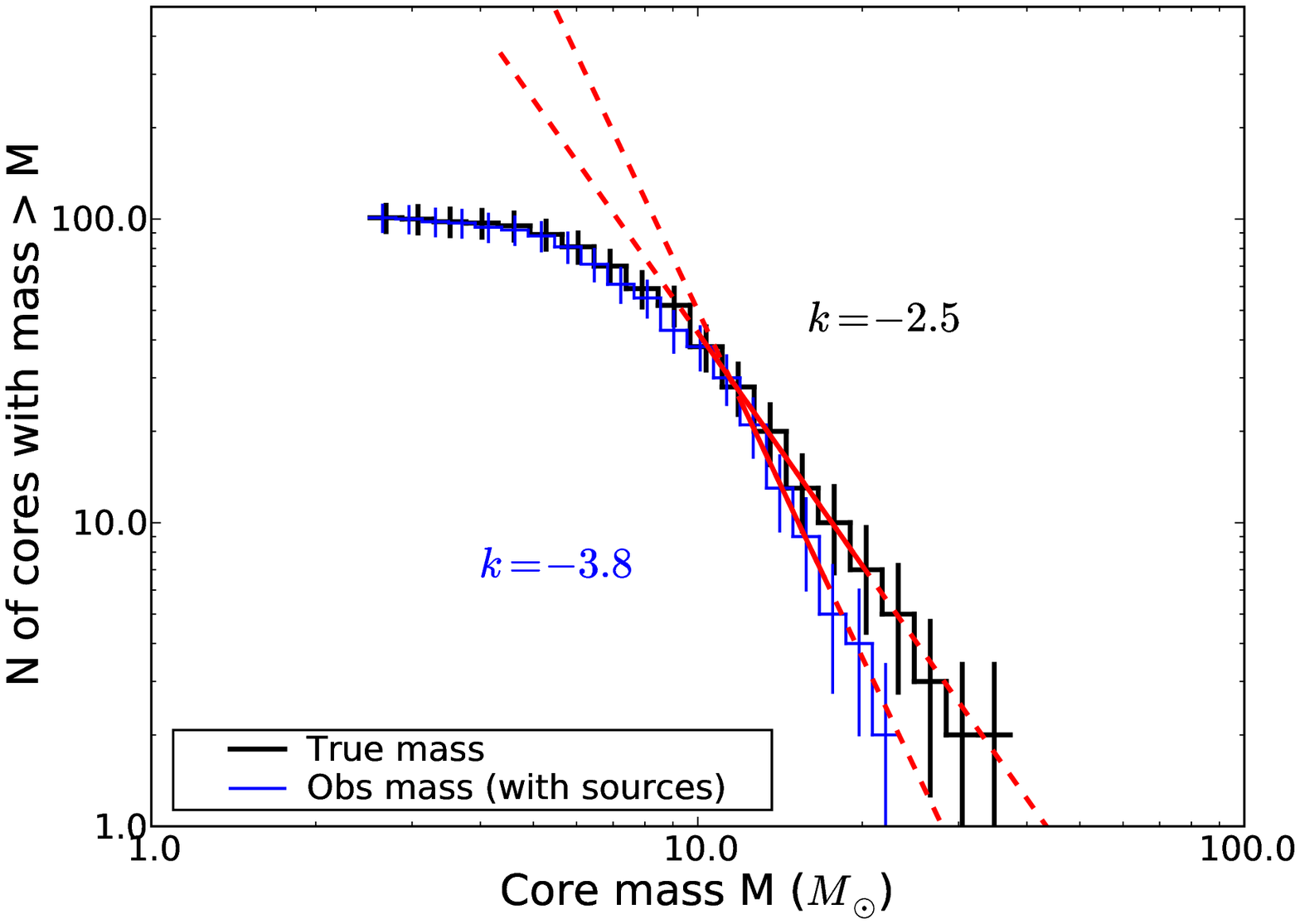}
\includegraphics[width=0.3\linewidth]{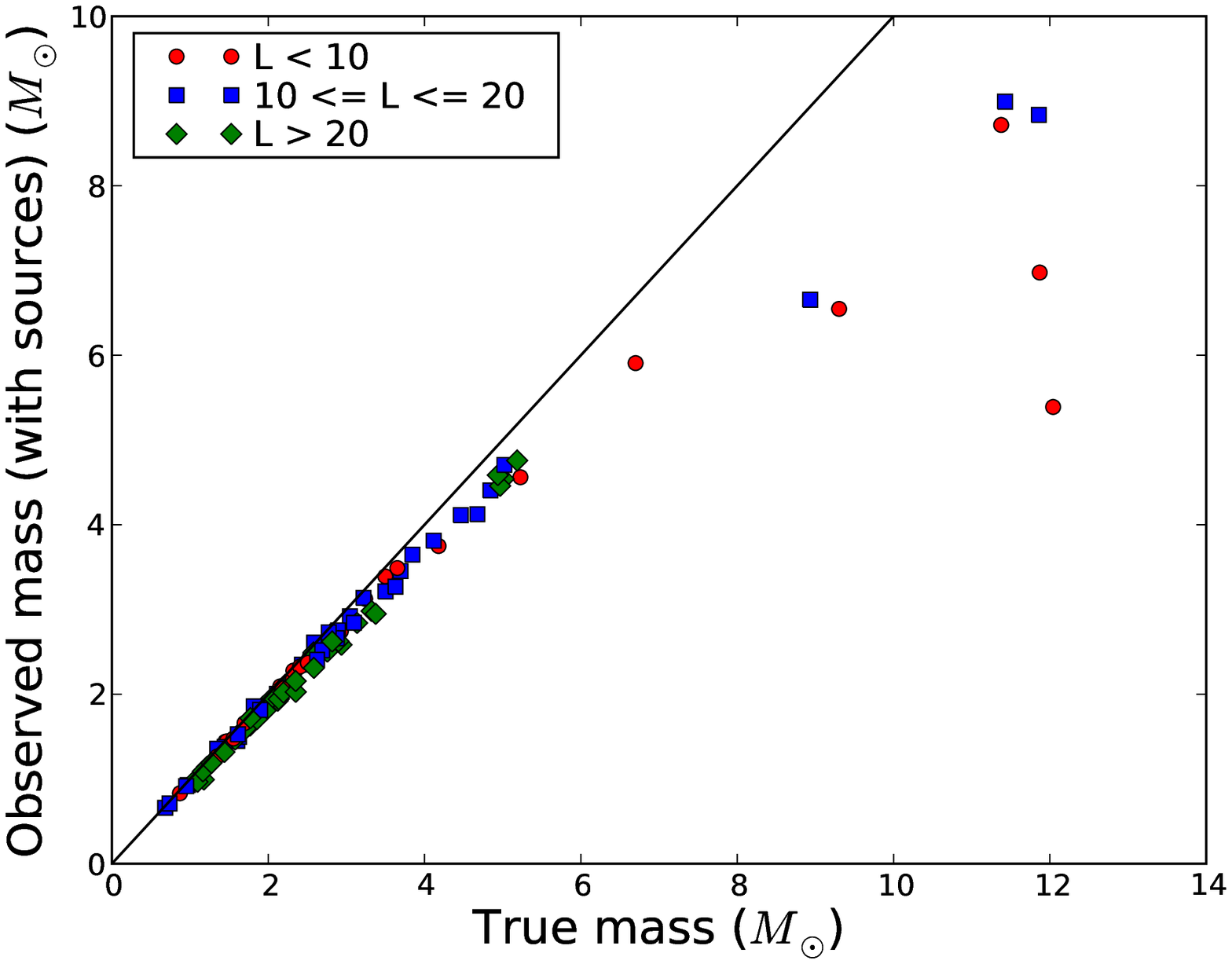}
\caption{
Model II with added sources using wavelength pair 250/500 $\mu$m, the
correct value of $\beta$, and a distance of 100 pc. {\em
(Left)} Mass spectra obtained with Clumpfind clumps. {\em (Middle)}
The mass spectra obtained with environments of 20 pixel radius. {\em
(Right)} The true vs. the observed masses within 10 pixel radii.
}
\label{fig:CIIallsum} 
\end{figure*}

The mass hidden in the cold dust should become visible if the cores
form protostars that start to heat up the cores from inside. To test
this hypothesis, we added a radiation source to each gravitationally
bound core as described in Sect.~\ref{sect:RT} and the analysis was
repeated using the newly calculated surface brightness maps. The
results are shown in Fig.~\ref{fig:CIIallsum} for one set of Clumpfind
clumps and the 3D clumpfind cores using regions with the 20 pixel
radius areas (giving slopes $k$ = -3.8 and -2.5 for observed and true mass spectrum, respectively). The Clumpfind parameters were the same as in the middle column frame of row 2 in Fig.~\ref{fig:clumps}. The addition of
heating sources has not changed the situation for the Clumpfind
spectra and, as before, the observed spectrum is consistent with the
spectrum drawn with the true masses. In
the case of fixed radius environments, the difference between observed
and true masses is decreased. This is visible in the mass spectra but
more clear when comparing directly the true and the observed masses of
individual cores.

\subsubsection{Model III: AMR model with cores of high opacity}

The effective resolution of Model III is 4096$^3$ cells. In our study,
it represents the most evolved case and contains some cores with very
high column density (Fig.~\ref{fig:models}).

\begin{figure*}
\centering
\includegraphics[width=8cm]{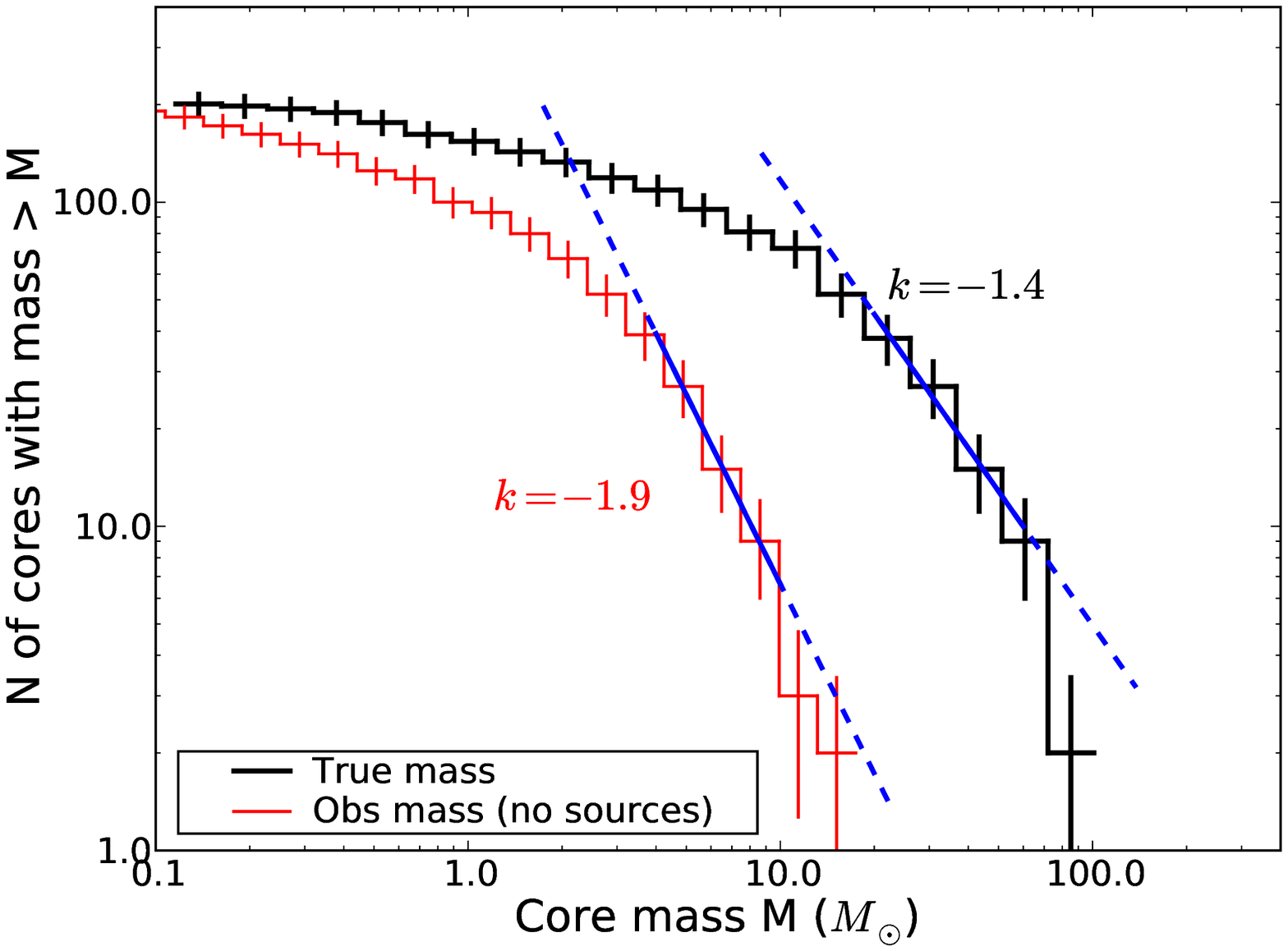}
\includegraphics[width=8cm]{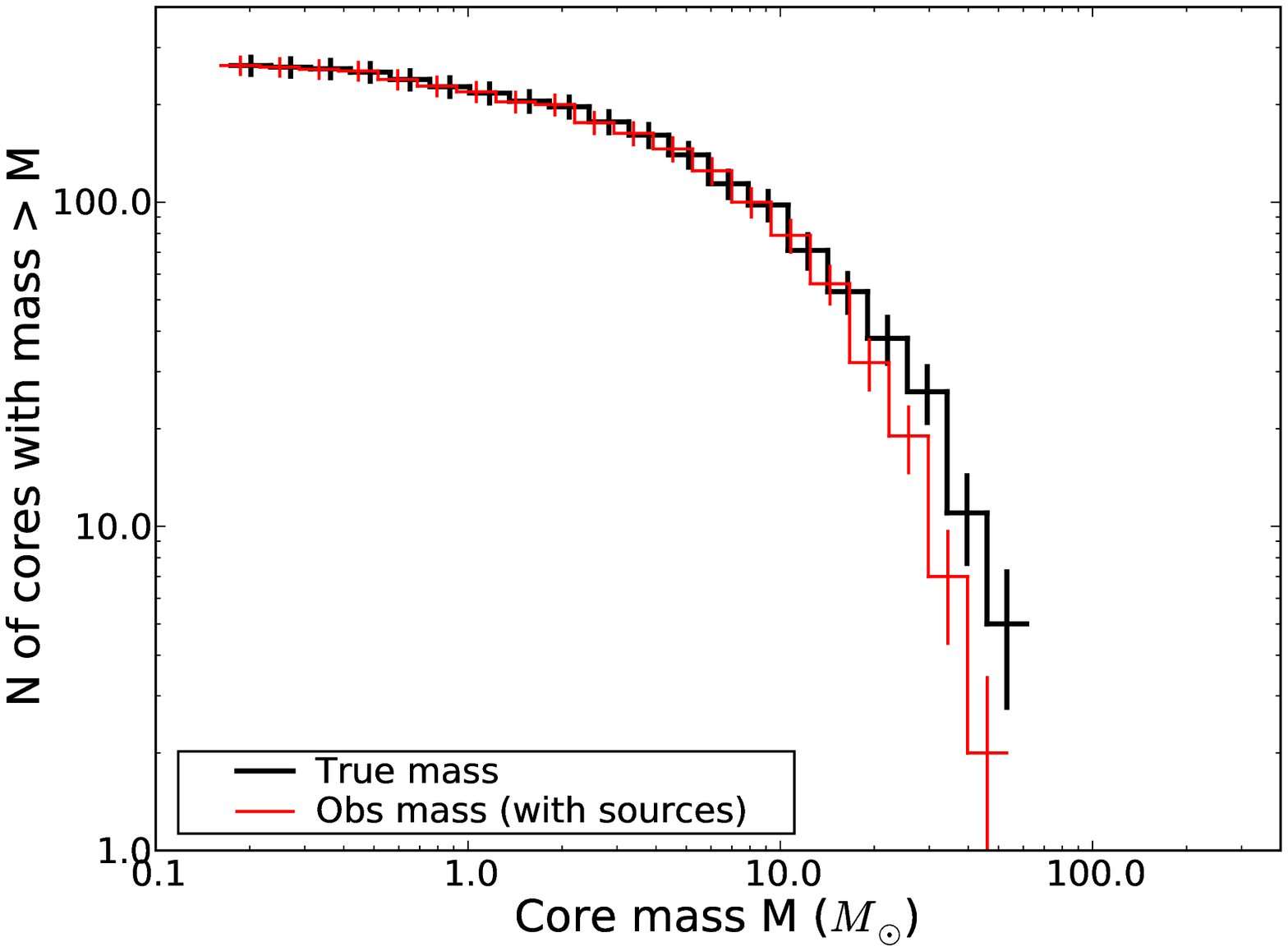} \\
\includegraphics[width=8cm]{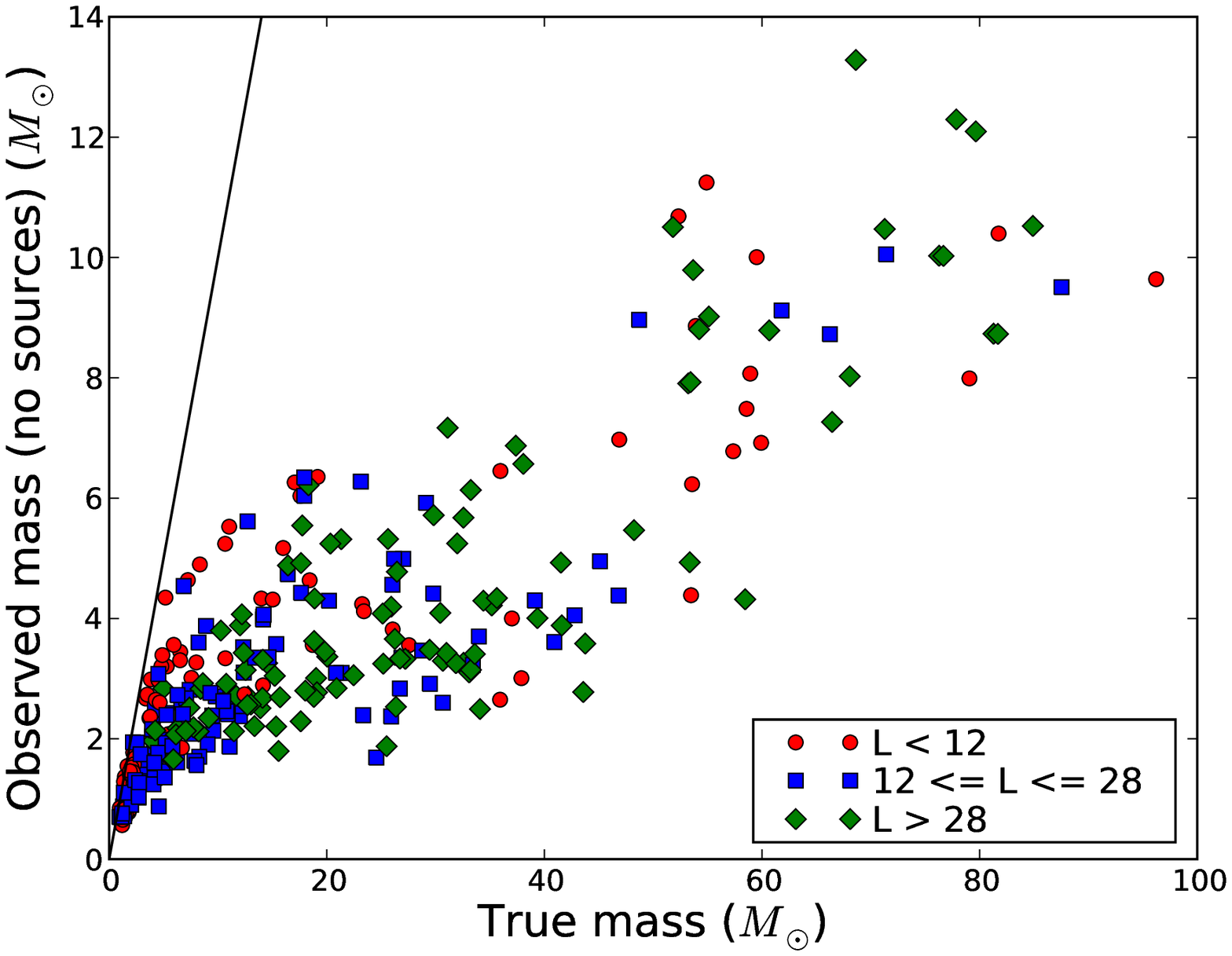}
\includegraphics[width=8cm]{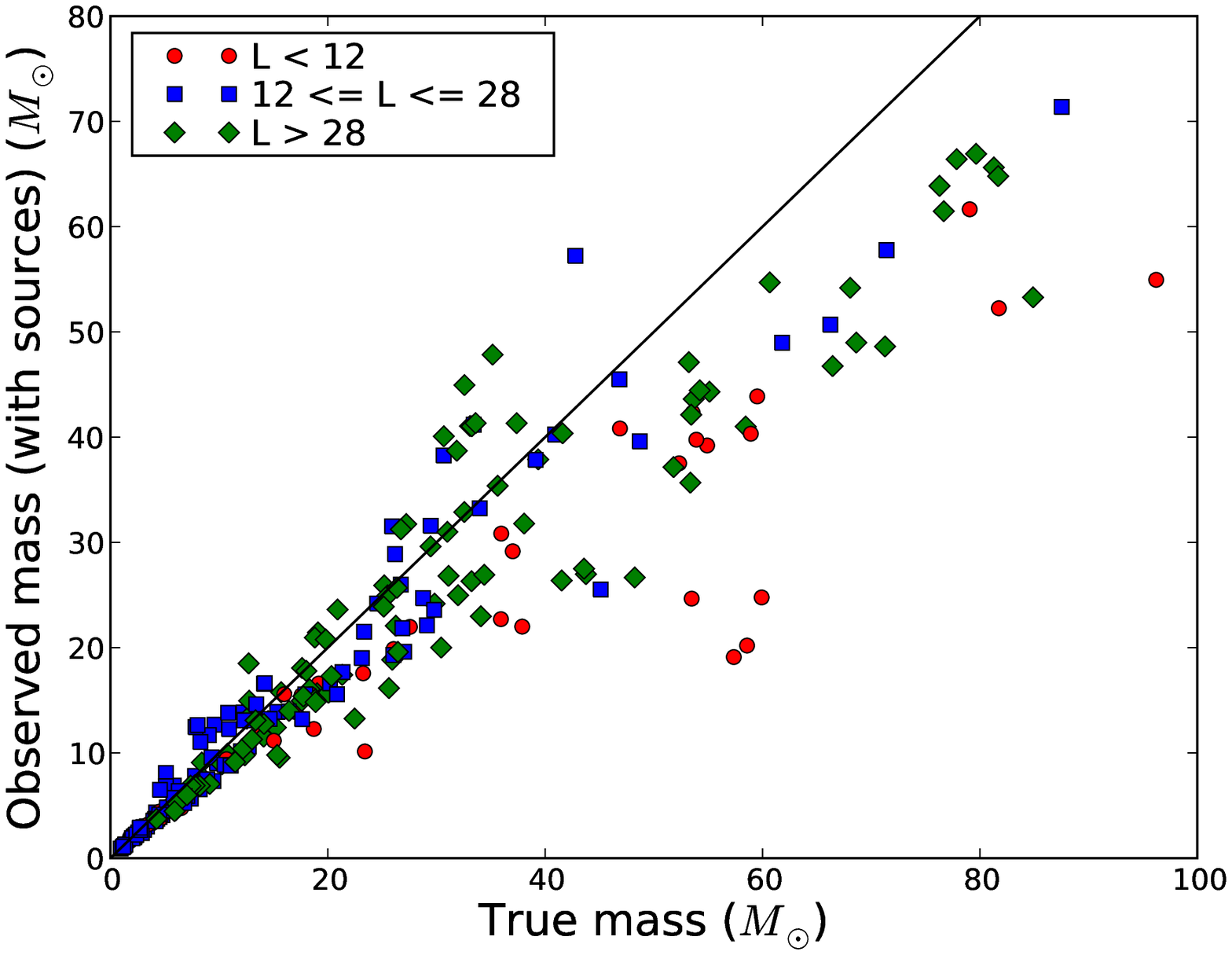} \\
\includegraphics[width=8cm]{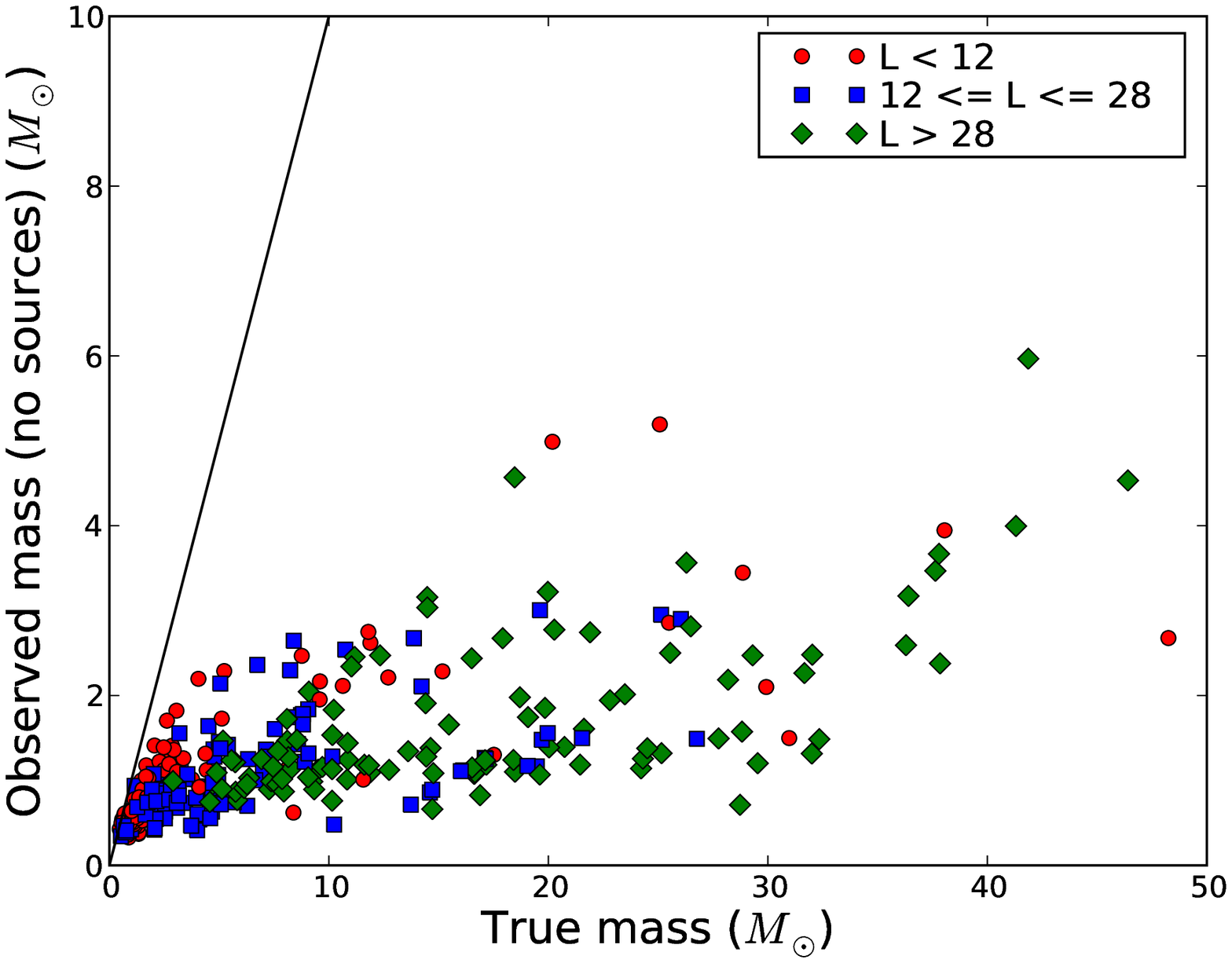}
\includegraphics[width=8cm]{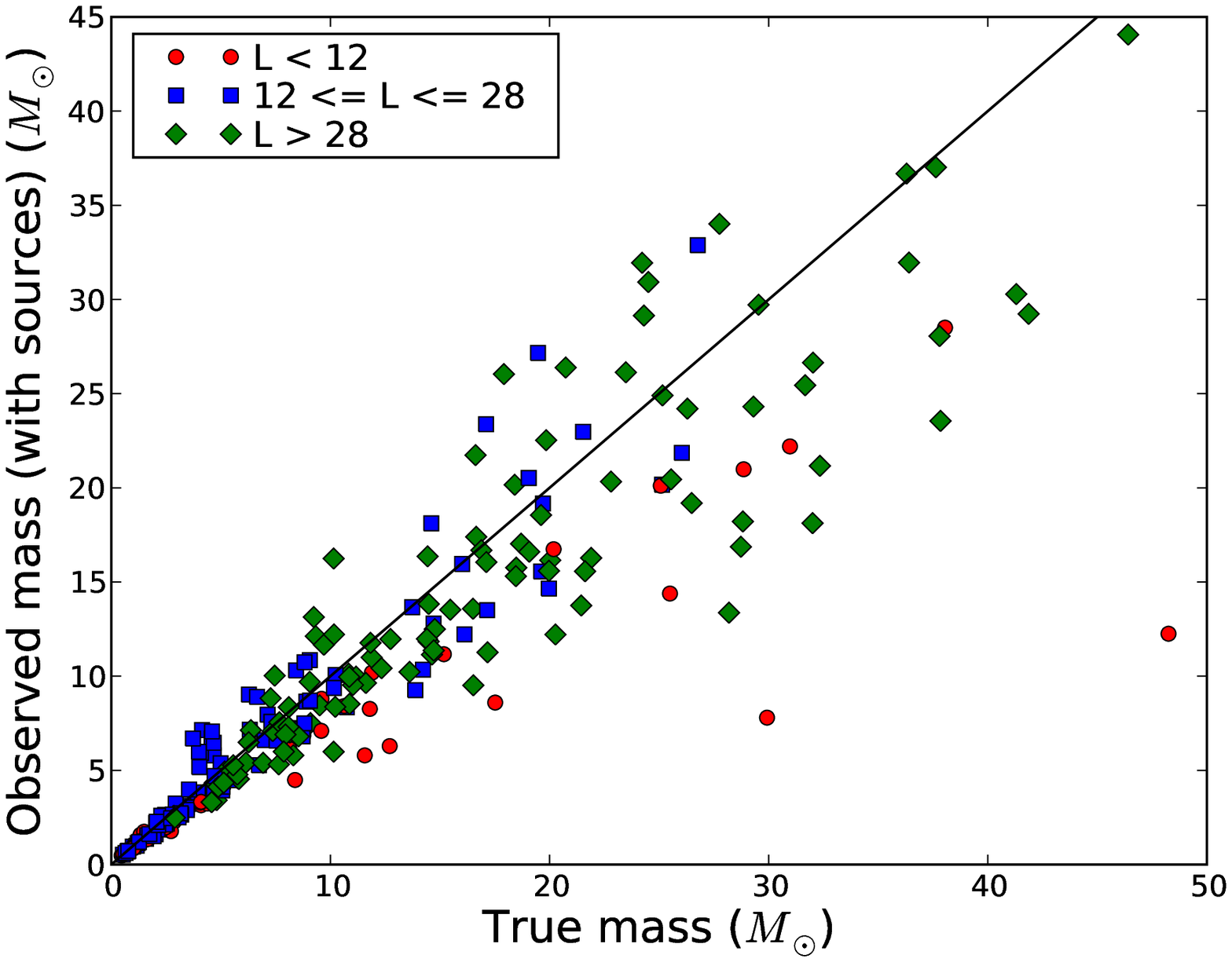}
\caption{
Model III: Mass estimates using wavelength pair 250/500 $\mu$m, the
correct $\beta \sim$ 2.1, and a distance of 100\,pc without (left
column) and with (right column) internal heating sources. Rows from
top: (1) Mass spectra obtained with Clumpfind clumps, (2) True vs.
observed mass within regions of 20 pixel radius, (3) True vs. observed
mass within regions of 10 pixel radius.
}
\label{fig:CIbo_as} 
\end{figure*}

For Model III the mass spectra with and without internal radiation
sources are shown in Fig.~\ref{fig:CIbo_as}. The spectra correspond to
the objects found with Clumpfind. Unlike in Model II, without the
internal radiation sources the core masses are significantly
underestimated. The observed mass spectrum is much steeper than the
spectrum obtained using the true masses. At the high mass end the
shift in the spectrum corresponds to one order of magnitude in mass.
The slopes for the observed and true mass spectra are $k$ = -1.9 and -1.4, respectively, indicating that there is a small change in the shape of the spectra.
Once the radiation sources are turned on, the observed mass spectrum
is almost completely rectified and a small difference remains only at
the very highest masses.

The second row of frames in Fig.~\ref{fig:CIbo_as} compares the
observed masses $m_{\rm obs}$ and the true masses $m_{\rm true}$ core
by core using regions of 20 pixel radius. Without internal heating,
the correct mass is recovered only for the smallest cores. When
$m_{\rm true}$ reaches 40\,$M_{\sun}$, the true mass is underestimated
by a factor of ten although there is also significant scatter. When
the internal sources are turned on, the bias disappears almost
completely and is only $\sim$20\% for the most massive cores. 
The cores are plotted with different symbols according to the
luminosity that was assigned to the added sources. The luminosities
depend monotonically on the mass of the gravitationally bound cores
(see Sect.~\ref{sect:RT}). However, the values plotted in
Fig.~\ref{fig:CIbo_as} correspond to areas of the projected cloud and
$m_{\rm true}$ often contains contribution from several 3D clumps that
are close in the 2D projection. 
This explains 
why there is not a tight correlation between the source luminosity and $m_{\rm true}$.
However, the luminosity assigned to the sources should still
act as an identifier for the mass of the object at the centre of the areas
examined. With this assumption one can say that, before the heating
sources are added, the bias in the mass estimates increases with the
object mass. This is clear below $\sim$10\,$M_{\sun}$ but disappears at
higher $m_{\rm true}$, presumably because these regions correspond to
tight clusters of sources. When the radiation sources are included, the
mass is underestimated mostly in the cores with the faintest sources.
This may suggest that the heating power of our sources is generally
only just enough to correct the mass estimates. The effects become
clearer when we select a smaller radius (bottom frames in
Fig.~\ref{fig:CIbo_as}).

\begin{figure} 
\centering
\includegraphics[width=8cm]{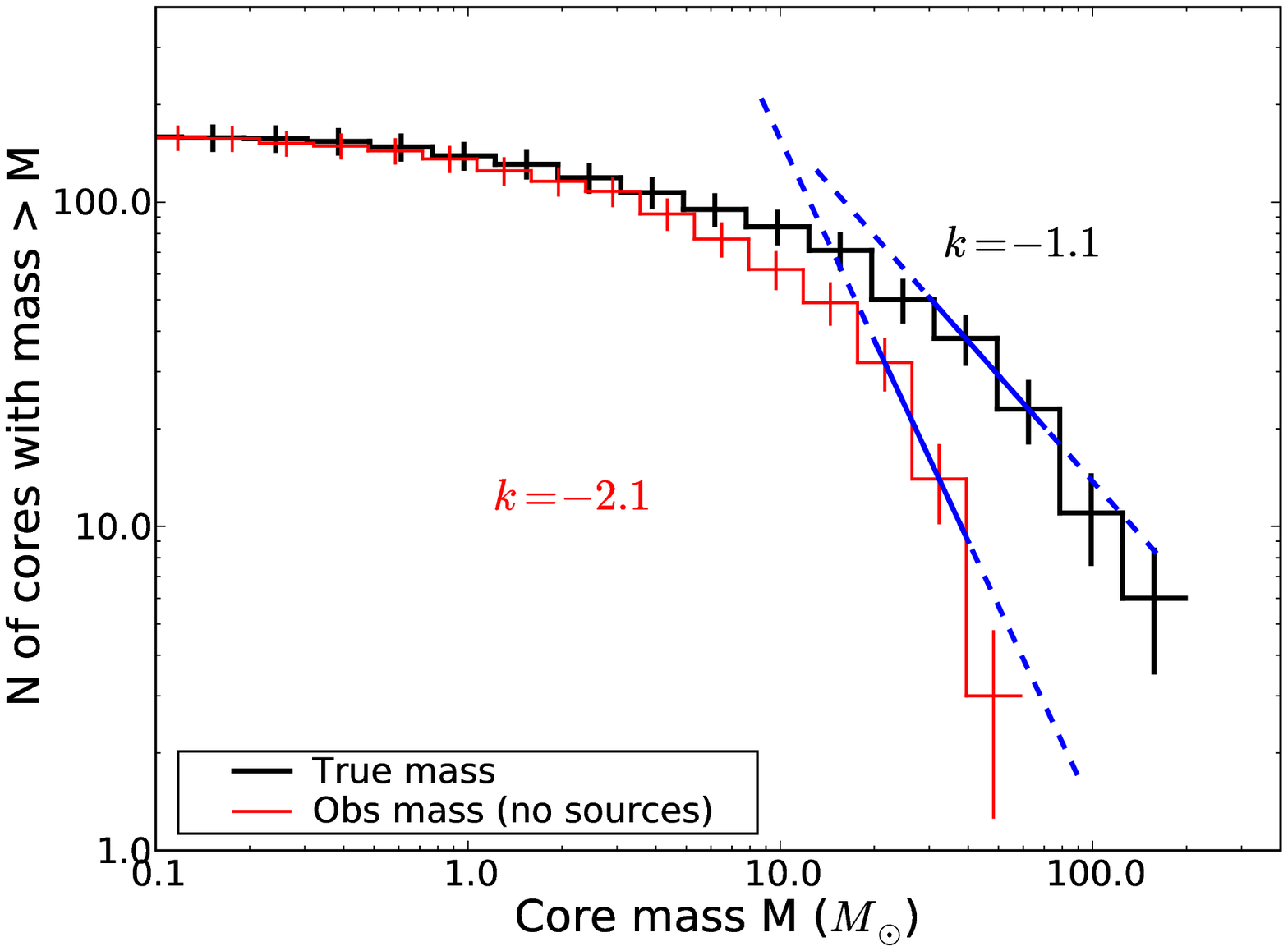}
\includegraphics[width=8cm]{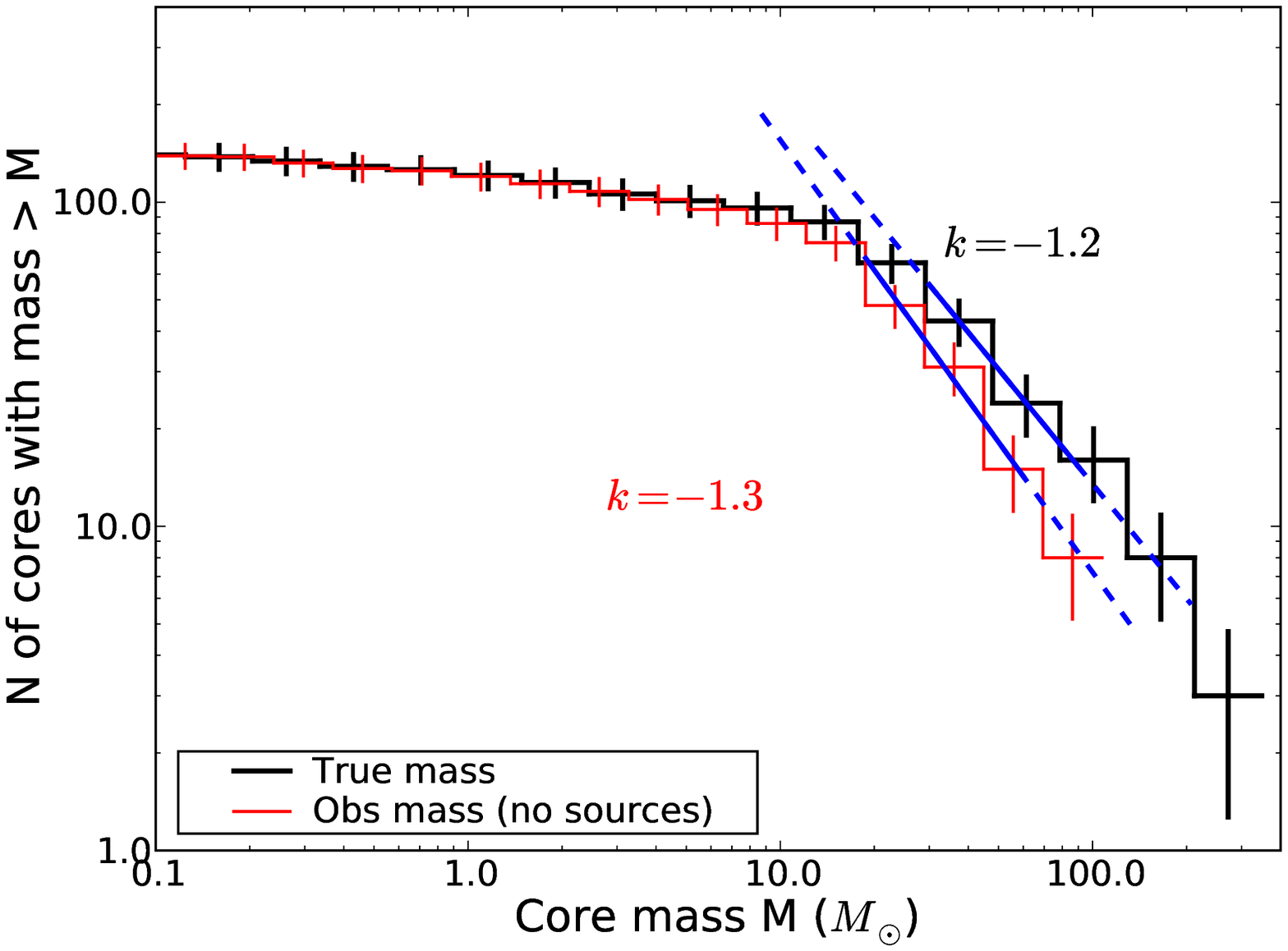}
\caption{
Model III: The effect of lower resolution (caused by longer distance)
on the mass spectra obtained with
Clumpfind clumps using wavelength pair 250/500 $\mu$m and the correct
value of $\beta \sim$ 2.1. Distance to the cloud is 400\,pc (top) and
1000\,pc (bottom). Mass spectra obtained with a distance 100\,pc (without
sources) is shown in Fig.~\ref{fig:CIbo_as}.
} 
\label{fig:CIdist}
\end{figure}

The size of the regions assigned to the cores appears to be important for the
appearance of the mass spectra. Therefore, we tested the effect of different
cloud distances in connection with the Clumpfind clumps of Model III. Because
of the fixed beam sizes, a larger distance means lower linear resolution. In
Fig.~\ref{fig:CIdist} are the results for cloud distances 400\,pc and
1000\,pc.  The slopes for mass spectra are given in the figure, but the values
are highly dependent on the used mass interval. 
The number of clumps identified with Clumpfind in the three
cases, with distance of 100, 400, and 1000\,pc, are 202, 161, and 144,
respectively. The numbers include clumps counted towards all three
directions. 
As the distance increases, the linear size of the identified
clumps increases and some of the clumps are combined. As the average column density
gets smaller, the observed masses approach the true masses. This means that
the difference between the observed and the true mass spectrum decreases when
resolution decreases. This could lead to very different CMFs between low
resolution and high resolution studies of star formation.

We investigated the effect of different wavelength pairs also in Model III and
conclude that the masses obtained with the
wavelength pair 100/350~$\mu$m compared to 250/500~$\mu$m are even more biased than in Model II.

%%%%%%%%%%%%%%%%%%%%%%%%%%%%

\subsection{Estimation of dust spectral indices} \label{sect:specind}

\begin{figure*}
\centering
\includegraphics[width=14cm]{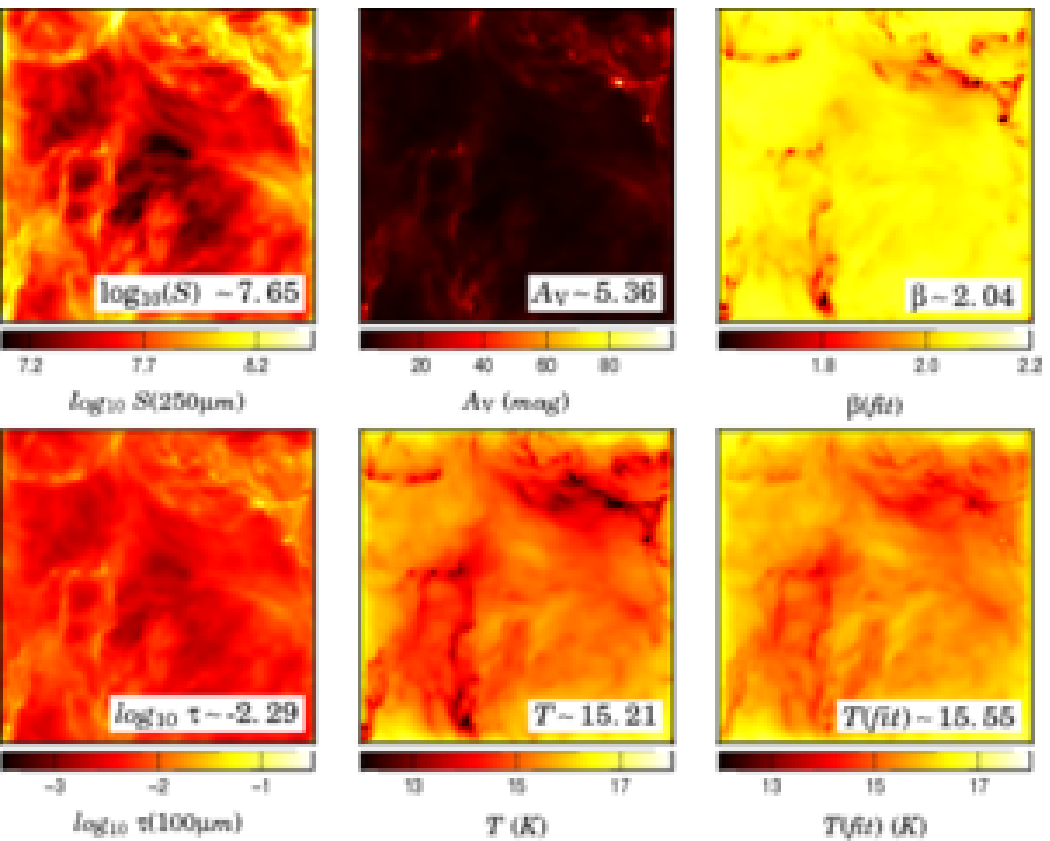}
\caption{
Results of the estimation of dust temperature and spectral index
$\beta$ in Model II. The frames on the top row show the
250\,$\mu$m surface brightness, the visual extinction, and the
estimated spectral indices. On the bottom row are the logarithm of the
100\,$\mu$m optical depth, the real mass averaged dust temperature
along the line-of-sight, and the estimated colour temperature. The
median values of the variables are given in the frames.
}
\label{fig:BETA_TEST_5_BG}
\end{figure*}

\begin{figure}
\centering
\includegraphics[width=7.9cm]{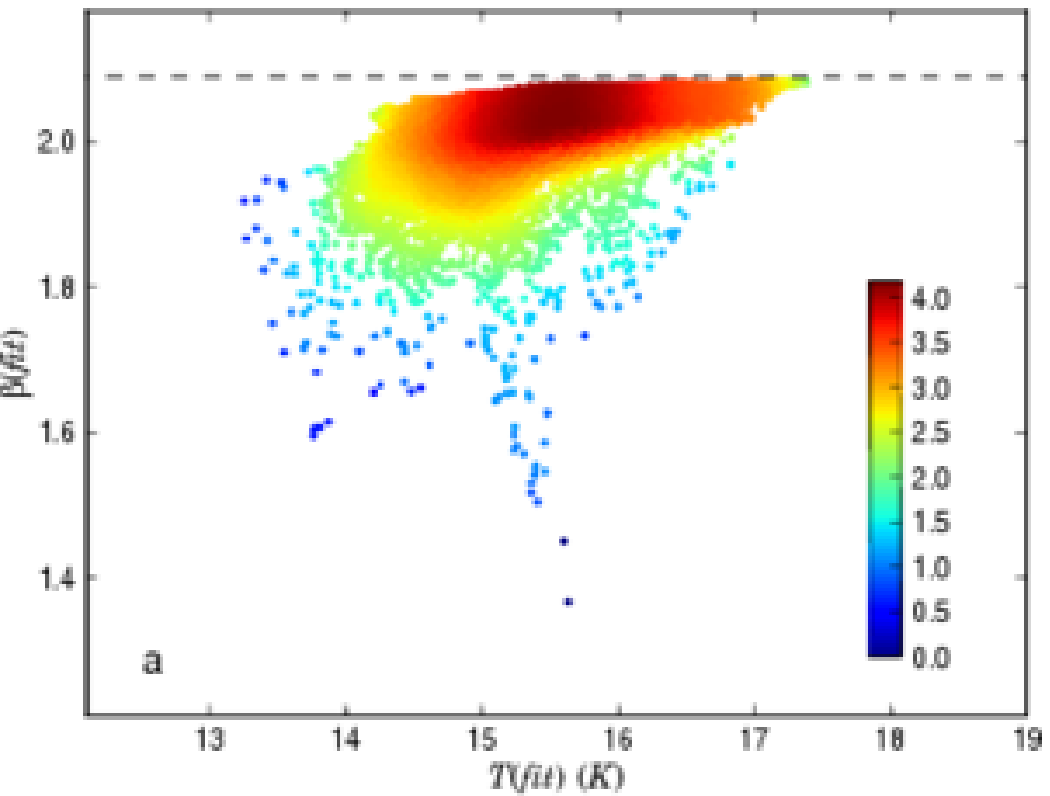}
\includegraphics[width=7.9cm]{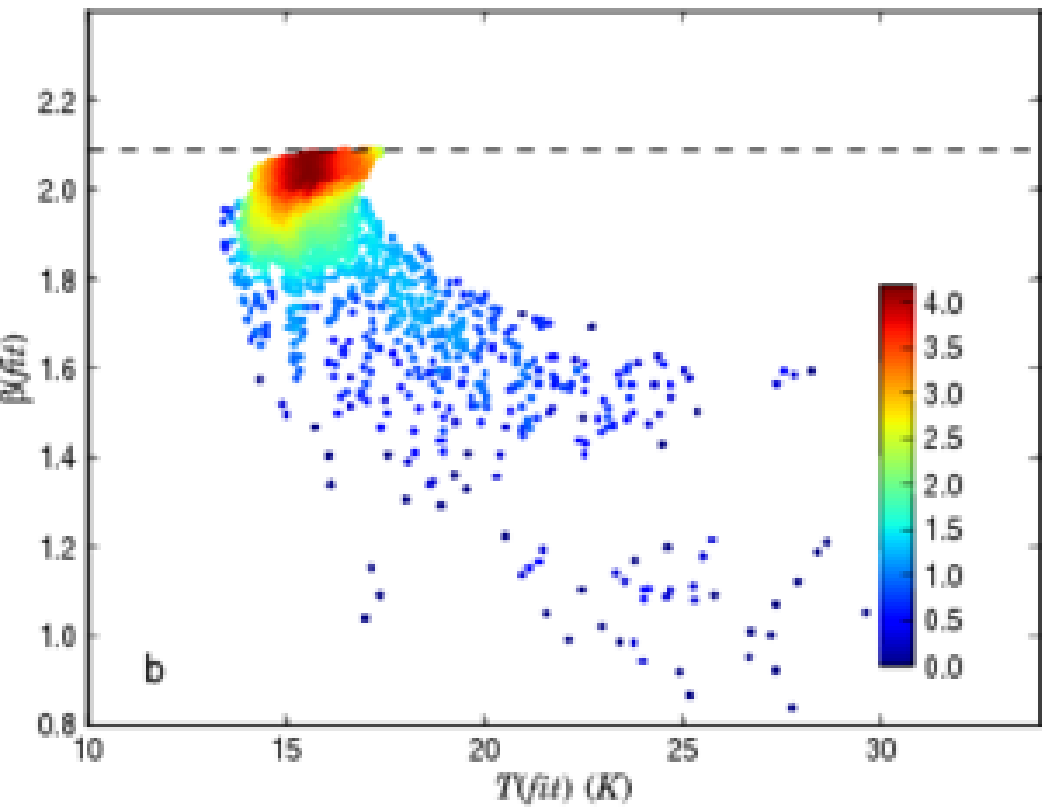}
\caption{
The correlation between the dust colour temperature and the spectral
index as derived from the synthetic observations of Model II without
internal sources ({\em top frame}) and with the internal heating sources
({\em bottom frame}). The colour scale indicates the logarithmic density of
the ($T$, $\beta$) points and the dashed line the average spectral
index in the dust model over the wavelength range used,
100\,$\mu$m--500\,$\mu$m.
}
\label{fig:BETA_TEST_5_BG_plot}
\end{figure}

The spectral index and its variations carry information on the
intrinsic grain properties (see Sect.~\ref{sect:intro}). Because of
the strong anti-correlation between the temperature and the spectral
index, accurate simultaneous determination of $T_{\rm dust}$ and
$\beta$ is difficult in the presence of observational noise, as shown,
e.g. by Shetty et al. (~\cite{shetty09a}). We do not repeat those studies
but concentrate on the importance of the line-of-sight temperature
variations and the radiative transfer effects in our Model II. The
dust colour temperature and the spectral index were obtained by a
simultaneous fit of the simulated observations at 100\,$\mu$m,
160\,$\mu$m, 250\,$\mu$m, 350\,$\mu$m, and 500\,$\mu$m. The analysis
was carried out using data with maps convolved with a FWHM
corresponding to 20 pixels and a pixel scale 0.049\,pc. Noise was not
added to the maps. After the convolution, the noise coming from Monte
Carlo simulation is orders of magnitude below the noise level of
typical observations. The effects discussed below arise from sources
other than the noise.

Fig.~\ref{fig:BETA_TEST_5_BG} shows the results for Model II without
internal sources. The estimated median temperature of 15.55\,K is
close to the real mass weighted average temperature of 15.21\,K.
However, the colour temperature is always higher than the dust
temperature. The difference rises to $\sim$1\,K in the most prominent
filaments and reaches a peak value of $\sim$5\,K towards the densest
core. The spectral indices are correspondingly underestimated. The
observed $\beta$ is close to the true value in diffuse regions, is too
low by at least 0.1 units in the dense regions ($A_{\rm
V}\sim10^{\rm}$ or more), and is underestimated by up to 0.5 units
towards the most opaque cores. This is seen more clearly in 
Fig.\ref{fig:BETA_TEST_5_BG_plot} (top frame) which shows the correlation between
$T_{\rm C}$ and $\beta$. For the main cloud of points, the spectral
index decreases with decreasing temperature. The behaviour is {\em opposite}
to the inverse $T-\beta$ relation that has been detected in
interstellar clouds (e.g. Dupac et al.~\cite{Dupac2003}). The result
suggests that the intrinsic spectral index variations of dust could be
much larger than what is directly observed towards quiescent clouds.
The effect should be taken into account when seeking observational
confirmation for the laboratory results that, after increasing towards
lower temperatures, the sub-millimetre spectral index may again
decrease as temperature falls below $\sim$10\,K (Agladze et al.~\cite{Agladze1996}).

We repeated the ($T_{\rm C}, \beta$) estimation using only wavelengths
100\,$\mu$m, 250\,$\mu$m, and 350\,$\mu$m. The results were
qualitatively similar but the median colour temperature was lower,
14.92\,K, and the median $\beta$ higher, 2.28. For most points, the
spectral index was therefore {\em higher} than the intrinsic spectral
index of the dust grains. In our dust model, the real $\beta$ between
the wavelengths 100\,$\mu$m and 350\,$\mu$m is 2.12 but rises to
$\beta$=2.21 between 250\,$\mu$m and 350\,$\mu$m. The reason for the
high observed $\beta$ was traced back to this frequency dependence. In
tests using a dust model with a constant $\beta$ over all wavelengths,
the estimated $\beta$ remained below the real $\beta$ and approached
that value only in diffuse regions. However, we could confirm with
direct calculations, without the use of radiative transfer models,
that when the spectral index varies with wavelength the fit of
$B_{\nu}(T) \nu^{\beta}$ can result in $\beta$ values that are higher
than the real beta anywhere in that wavelength range. The result
suggests that such wavelength dependence can lead to significant error
in the estimates of the spectral index and, therefore, in the dust
temperature.

\begin{figure}
\centering
\includegraphics[width=7.7cm]{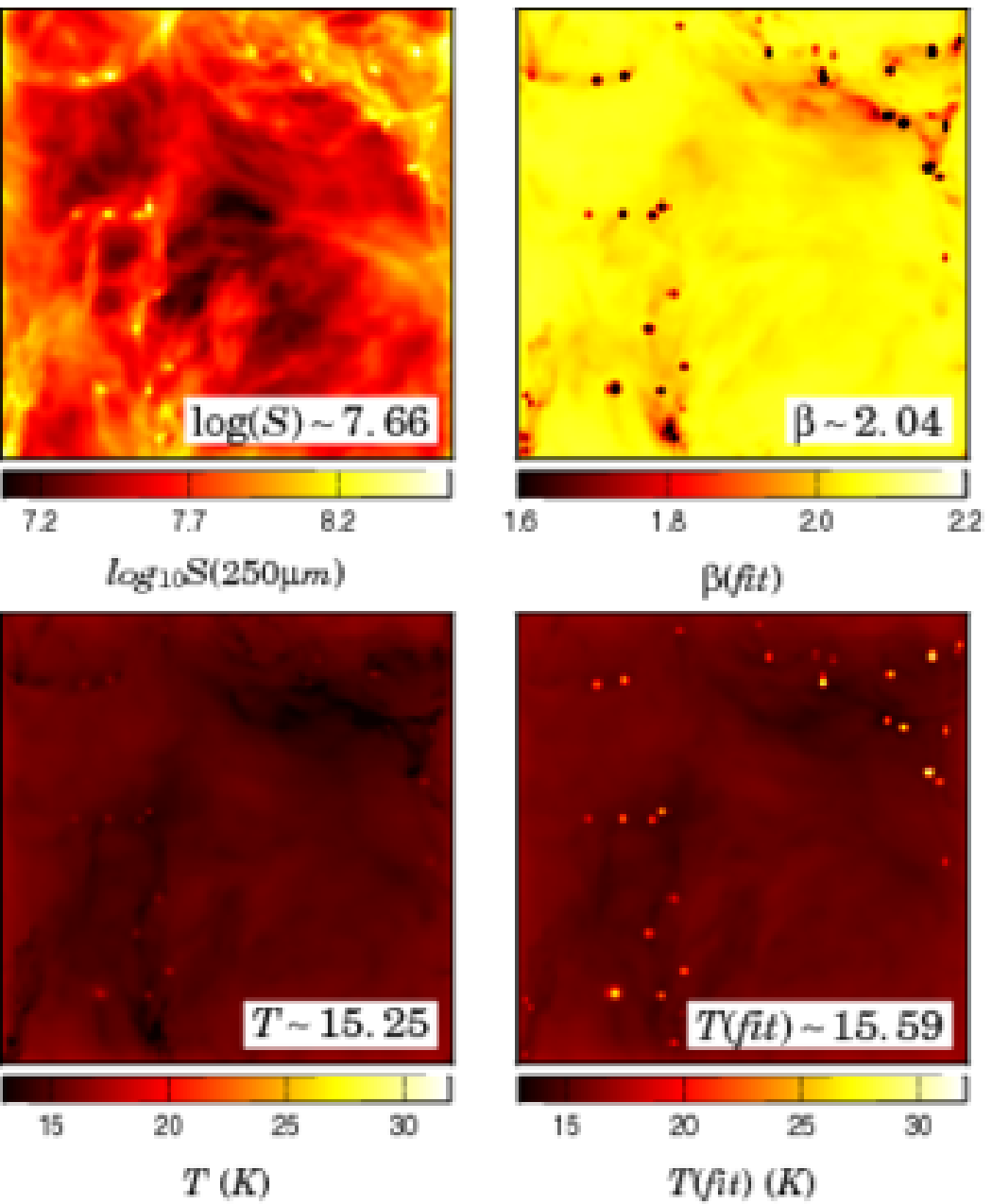}
\caption{
Results of the ($T_{\rm dust}$, $\beta$)- fits for Model II with
internal sources. The frames on the top row show the 250\,$\mu$m
surface brightness and the derived spectral indices. The lower frames
show the mass averaged line-of-sight dust temperature and the
estimated colour temperature. The median value of the variables are
quoted in the frames.
}
\label{fig:BETA_TEST_5_SOU}
\end{figure}

Fig.~\ref{fig:BETA_TEST_5_SOU} and
Fig.\ref{fig:BETA_TEST_5_BG_plot} (bottom frame) show the results for Model II
with the internal radiation sources. Because the sources have only a
very local influence, the median value of $\beta$ is the same as
before and the increase in the average temperature is small.  However,
because of the very strong temperature variations, the bias in the
parameter values is now much larger towards the source regions. 
Furthermore, the internal heating increases the contribution of the
densest cores which, at the full resolution of the cloud model, can
reach optical depths $\tau(100\mu{\rm m})\sim$ 1. The opacity effects
are no longer negligible while the colour temperature determination
itself still assumes an optically thin emission.  Because the spectral
index is now underestimated towards cores that are warm, there is a
strong anticorrelation between the temperature and spectral index (see
Fig.~\ref{fig:BETA_TEST_5_BG_plot} (bottom frame)) This effect is similar to
the inverse relation observed in real clouds. This highlights the
difficulty of separating the intrinsic dust properties from the
effects produced not only by noise but also by radiative transfer
effects. In particular, one must exercise caution when interpreting
such observations when the samples include star forming clouds.

%%%%%%%%%%%%%%%%%%%%%%%%%%%%

\section{Discussion} \label{sect:discussion}

The study showed that the shape of the mass spectrum is quite robust
against errors arising from radiative transfer effects or spatial
variations in dust properties. 
According to observations, the dust opacity varies from cloud to
cloud. In particular, the difference between the diffuse and dense
regions can be a factor of a few (e.g., Stepnik~\cite{Stepnik2003}).
Therefore, dust opacity remains a major factor of uncertainty in the
estimation of cloud masses. We have excluded these errors by always
using the correct dust opacity as obtained from the dust model. More
generally, the mass estimates would of course be directly proportional
to $\kappa^{-1}$. In our case the mass errors are caused by errors in
the derived dust temperatures which, in turn, depend on the assumed
spectral index or the procedure used to obtain $\beta$ from
observations.
For the masses of individual cores, the
main errors result from the uncertain values of the dust opacity $\kappa$ and, to a lesser degree, the spectral index
$\beta$. As long as these values are constant, the mass spectrum may
be shifted but its shape is not much affected, unless the mass spectrum is based on optically thick central parts of the cores.

The errors resulting from radiative transfer effects were mostly smaller and even
systematic and strong variations in the dust properties were not
clearly visible in the shape of the derived mass spectra. Only if the
cores have very high opacity (densest cores in our Model II and most
cores of Model III), the effects resulting from the strong radial
temperature variations become the main source of error, exceeding the
uncertainty in $\kappa$.

It is often assumed that in dense and cold regions the dust sub-mm
emissivity increases and emissivity index $\beta$ changes. If the
change affected only the densest and presumably the most massive
cores, the slope of the mass spectrum should become more shallow.
Although we may have witnessed a very small change (see
Fig.~\ref{fig:msp_mwd_mod}), even the drastic change implemented in
the dust properties of Model I clearly had only a small effect on the
shape of the mass spectrum. We also varied the threshold density at
which the transition from normal dust to the higher emissivity dust
takes place (Fig.~\ref{fig:moddust}). This did shift the mass spectrum
along the mass axis but had no clear effect on the shape of the
spectrum. It seems that, with a constant density threshold, the dust
modification affects all cores almost equally. The denser cores have a
higher abundance of modified dust but its contribution to the observed
signal is small in relative terms because the modified dust resides in
the coldest central parts of the cores and therefore radiates weakly.
However, in denser cores the dust property variations could have a
more important effect than in the more extended low density cores of
Model I. In the AMR models we did not modify the dust. In future work
this effect should be studied further.

The Clumpfind mass spectra obtained with Model II using the correct values of $\kappa$ and $\beta$ and wavelength pair 250 and 500 \,$\mu$m (Fig.~\ref{fig:clumps}) show that the mass estimates are very accurate. We compared also the Clumpfind results to masses within a constant radius of source positions. It is perhaps surprising that the differences between the observed and true masses that were very clear in the case of small radius regions are not seen with Clumpfind clumps. There are probably two reasons for this. Firstly, the Clumpfind clumps tend to be extended which is also visible in their larger masses. This is why Clumpfind is not so sensitive to the effects that are constrained to the densest cores of the clumps. Secondly, some of the compact cores for which the observed masses are the most severely underestimated are no longer found by the Clumpfind routine, depending on how strict criteria are used in finding clumps. The approximated size of the observed cores appears to be an important factor in the obtained mass spectra. As the core sizes increase the observed mass spectra approach the true mass spectra, as demonstrated by Fig.~\ref{fig:clumps}. The mass spectra obtained with Clumpfind clumps and environments of constant radius seem to give similar results, except with the densest cores with small radii. However, we could also limit the results of Clumpfind only to the densest cores. We conclude that when studying the details of mass spectra it is crucial to pay attention to the method of defining the clumps.
The mass spectra obtained with constant radius cores cannot be directly compared to the CMF, as they depict more the column density than mass. 
However, the constant radius areas were a useful tool for studying the bias of mass estimates, independent of the uncertainties in
the definition of the clumps.

We compared the mass estimates obtained using different wavelengths (wavelength pairs 250/500\,$\mu$m, and 100/350\,$\mu$m and five wavelengths (100, 160, 250, 350, 500\,$\mu$m)) to calculate the colour temperature. The observational bias is larger if the shortest wavelengths are used compared to the case with wavelength pair 250 and 500\,$\mu$m. Apparently the massive and dense cores contain significant amount of cold dust that
is no longer seen at 100\,$\mu$m. In our simulations only large grain particles are included. In real
observations the intensity has a significant contribution from
stochastically heated small particles at least up to wavelengths
$\sim$100\,$\mu$m. If no correction is made for this component, the masses
derived from such observations will be even more strongly underestimated.

The SEDs in Fig.~\ref{fig:CIImultipleT_SED} show that for a high density core with large observational mass bias the observed SED differs clearly from the 'true' SED obtained with true values of grain temperature, column density and $\beta$. Temperature variations on the line-of-sight lead to overestimation of temperature and through that to underestimation of mass.

The results in Fig.~\ref{fig:CIdist} for Model III showed that as
the distance to the cloud increases and resolution decreases, the core
sizes increase as smaller clumps are combined together and the mass
estimates get more accurate. 
The mass estimates are worst for the densest regions. Therefore, by
including more of the surrounding areas into the total mass, the mass
will be larger but the relative error will be smaller.

In Model III with cores of high opacity the core masses are
strongly underestimated even with correct values of $\kappa$ and $\beta$. This difference between Models II and III appears to be caused mainly
by the difference in column density. In Model III the densest cores may have 
optical depths higher than for normal stable cores. Our tests showed that the errors in mass estimates become significant when the optical depths of the cores are one or several orders of magnitude higher than for normal Bonnor-Ebert spheres.
This could be the 
case when strong turbulent motions, rotation, or magnetic fields are 
involved. Also for unstable cores, the optical depths will increase 
during the collapse. This means that before the internal heating 
becomes important, observations might underestimate the mass severely.

%The estimates of the column density $N$ were based on an approximation of an optically thin emission
%\begin{equation}
%I_{\nu} = B_{\nu}(T)(1-e^{-\tau}) \approx B_{\nu}(T) \tau = B_{\nu}(T) \kappa N.
%\label{eq:2}
%\end{equation}
%In Model III, the maximum optical depth at 250\,$\mu$m is close to ten and the approximation is not valid everywhere in the cores. In a homogeneous medium, the optical depth reduces the intensity more at the shorter wavelength leading to lower temperature estimates and, therefore, {\em larger} values of the column density.  It would be possible to take this effect into account but the correction would have to be done iteratively, because it depends on the estimated column density. In a real cloud the effect depends critically on the temperature structure along the line of sight. We tested this with a simple two component model where 90\% of the mass was at 7\,K behind a foreground dust screen at 15\,K. With $\tau$ values relevant for Model III we found that an increasing $\tau$ caused the ratio of observed and true masses to {\em decrease} but by less than $\sim$30\%. Therefore, it is very difficult to correct for the effects of high optical depth but these do not appear to be a major factor even in the case of Model III.

\begin{figure}
\includegraphics[width=8cm]{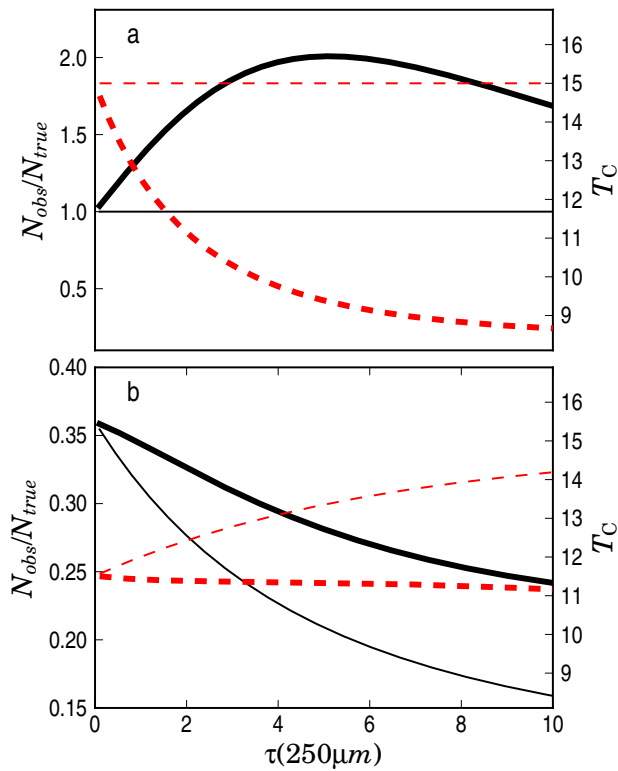}
\caption{
Comparison of colour temperature ($T_{\rm C}$, dashed lines) and the
ratio of estimated and true column densities (solid lines) obtained
from observations at 250$\mu$m and 500$\mu$m. The thick lines
correspond to the optically thin approximation (cf. Eq.~\ref{eq:beta})
and the thin lines to the values obtained by fitting
$I_{\nu}=B_{\nu}(T_{\rm C})(1-exp(-\kappa N))$. Frame $a$ shows the results for a   %$I=B_{\nu}(T)*(1-exp(-\kappa*N))$
homogeneous model with a dust temperature of 15\,K. In frame $b$ the
model consists of two homogeneous slabs, with temperatures of 7\,K and 15\,K (see text
for details). On the horizontal axis is the total opacity of the model
cloud at 250\,$\mu$m. The figure shows that while the exact formula
gives correct values for a homogeneous model, it results in larger
underestimation of column densities in case of the two layer model.
}
\label{fig:hitau_test}
\end{figure}

The estimates of the column density $N$ were based on the
approximation of optically thin emission (see Eq.~\ref{eq:beta}).
In Model II, the maximum optical depth at 100\,$\mu$m is about five
and, even after the convolution with the beam (distance=100pc, 37"
beam), can be up to one.
In Model III, the optical depth at 100\,$\mu$m can reach values of $\sim$5 even
after convolution.
Therefore, the approximation $\tau \ll 1$ is not valid everywhere in the
cores. At least in a homogeneous medium, an increasing optical depth
reduces short wavelength intensity relative to longer wavelengths.
With the approximation of Eq.~\ref{eq:beta} this leads to lower colour
temperature values. The effect counteracts the usual tendency to
underestimate the column densities. This is illustrated in
Fig.~\ref{fig:hitau_test} for a case where observations consist of
intensity measurements at 250\,$\mu$m and 500\,$\mu$m. The upper frame   %tarkista kaavat!!!!!!!!!
shows the situation for a homogeneous, 15\,K slab as a function of the
optical depth. Our use of the approximation $I_{\nu} = B_{\nu}(T_{\rm C}) \nu^{\beta}$    %$I_{\nu} = B_{\nu}(T_{\rm C} nu^{beta}$
results in the column density being overestimated. By fitting
the observations with $I_{\nu}=B_{\nu}(T_{\rm C})(1-exp(-\kappa_{\nu} N))$,   %$I_{\nu}=B_{\nu}(T_{\rm C}(1-exp(-\kappa_{\nu} N$
the correct value is recovered for all optical depths. The lower frame
of Fig.~\ref{fig:hitau_test} shows the situation when the model
consists of two layers at different temperatures. A homogeneous slab
at 7\,K is placed behind another homogeneous layer at 15\,K, the
previous standing for 90\% of the total optical depth that is shown on
the horizontal axis. 
The line-of-sight temperature variation leads to significant
underestimation of column density and the error is larger when we use
the formula that is exact for homogeneous medium. This suggests that
also in our 3D models the errors would have been {\em larger} if the
analysis would have been completed without the assumption of optically
thin emission. In Fig.~\ref{fig:hitau_test} the difference between the
two approaches
is 20--30\% when $\tau(250\,\mu{\rm m})$ is $\sim$1. However, the %$\tau(250\,\mum)$
difference critically depends on the temperature structure of the
cloud examined.

When there are internal heating sources that start to warm up the cores, 
the dust becomes easier to observe and the mass estimates get better. 
In Model II the sources do not change the mass spectrum notably, as it was nearly unbiased to begin with. The importance of the internal source varies, however, from core to core and the largest errors are still of the order of two. The effect of the radiation sources is very local so that a 20 pixel radius can encompass some very dense regions that remain unaffected by the central heating source. In Model III most of the strongly underestimated masses are corrected when internal sources start to make the densest cores visible. However, also in this model some of the core masses are still underestimated by a factor of 2--3. We have used internal sources that are strong enough to correct the mass estimates in most of the cores, but are also clearly visible in surface brightness maps. In future work it could be useful to study also weaker sources and how mass estimates depend on the characteristics of the sources and cores.

The spectral index $\beta$ and its temperature dependency can give information of the properties of interstellar dust grains. The observations (e.g. Dupac et al.\cite{Dupac2003}) show an inverse $T-\beta$ relation in interstellar clouds. However, it is difficult to estimate the reliability of this relation as noise can cause a similar anticorrelation (Shetty~\cite{shetty09a}). The results of Sect.~\ref{sect:specind} showed that temperature variations along the line-of-sight have a strong impact on the observed dust spectral index $\beta$. The value of $\beta$ was strongly underestimated in the direction of dense, inactive cores. This effect is opposite to the observed inverse $T-\beta$ relation. This could mean that in inactive clouds the spectral index variations are much larger than what is seen in the observations. However, in Model II the 100\,$\mu$m optical depth approaches $\tau$=1 in some cores and this could also affect the derived values of $T$ and $\beta$.
Shetty et al.~(\cite{shetty09b}) studied the effects of noise and temperature variations on the estimation of dust properties. They found that estimated colour temperature values could be higher than the physical temperature of either of the two temperature components. We also found that the change of the spectral index as a function of wavelength can cause the value of the estimated spectral index to be higher than the true $\beta$ in our dust model. 
However, the effect was seen only when
using three measurements in a wavelength range where the $\beta$ variations were strong. In SED fits employing five points spread
over a wider wavelength range, the effect disappeared. This may indicate that such high $\beta$ values could appear only when three
frequency points are fitted with a three parameter model that does not exactly fit the observations (i.e., assumes $\beta$
independent of wavelength). The effect depends on the actual shape of the observed spectrum which itself is affected not only by
dust properties but also by the mixture of dust temperatures and other radiative transfer effects.
Our results showed that protostellar sources in the cores can further increase the underestimation of the $\beta$ values. The cores are in this case warm and therefore this can produce a $\beta-T$ anticorrelation that is difficult to distinguish from any inherent $\beta(T)$ relation the dust grains may have. An error in $\beta$ causes an error in the estimated temperature and therefore also in masses.

\section{Conclusions} \label{sect:conclusions}

We have examined the systematic errors in the analysis of
sub-millimetre dust emission observations with the help of MHD model
clouds and radiative transfer modelling. 
We have used different wavelength pairs to compare how the results depend on the used
wavelengths. We have also compared these results to the case when all five Herschel wavelenghts are available.
With three different models,
we have determined the errors in the mass estimates of dense cores and how these are reflected in the shape of the core mass
spectrum. Based on the results we draw the following conclusions:
\begin{itemize}
\item Because of line-of-sight temperature variations the core masses
are usually underestimated. The effect varies strongly depending on
the densities and possible internal heating of the cores.
\item
For normal cores, the largest uncertainties are still caused by the
unknown values of the dust opacity $\kappa$ and, to a
smaller extent, the spectral index $\beta$. In first approximation,
both are multiplicative errors that leave the shape of the mass
spectrum nearly unchanged.
The opacity $\kappa$ is believed to vary by a factor of a few
between diffuse and dense regions. According to our models, the error
resulting from the bias in the colour temperatures is smaller.
\item
With the correct values of $\kappa$ and $\beta$, the mass estimates of
normal cores in hydrostatic equilibrium are precise to some tens of
percent. Although the systematic underestimation of mass is clear for
individual dense cores and can shift the observed mass spectrum along the
mass axis, the errors are not likely to be visible in the slope of the mass spectrum, 
unless the mass spectrum is based on optically thick central parts of the cores.
\item
If the cores have optical depths that are one or several orders of
magnitude higher than expected for Bonnor-Ebert spheres, the errors in
the mass estimates become significant. In our models, the real core
masses were underestimated by up to a factor of ten. However, such
dense cores are also unlikely to be detected even at sub-millimetre
wavelengths, especially in regions of strong background emission.
\item
When observing cold dust, the temperature obtained with wavelength pair 250/500\,$\mu$m gives more accurate mass estimates than the temperature obtained with a fit to 5 wavelengths, if shorter wavelengths ($\sim$100\,$\mu$m) are included.
\item
The presence of internal heating sources can correct the mass
estimates by making the dust in the core centres again visible. Even
in the case of cores with very high opacity, the presence of a typical
protostellar source reduces the bias in mass estimates to some tens of
per cent.
\item
The observed dust spectral index $\beta$ was found to be affected
strongly by the temperature variations along the line-of-sight (and
within the detector beam). The $\beta$ value is strongly
underestimated towards dense, quiescent cores. In SED fits using
three wavelengths the spectral index was seen to be strongly
overestimated because $\beta$ varied over the wavelength range in
question. 
However, the effect was no longer seen in SED fits employing five
frequency points.
\item
When the cores have internal radiation sources, the $\beta$ values are
still strongly underestimated. Because the cores are in this case
warm, this results in a $\beta-T$ anticorrelation that will be difficult to
separate from any intrinsic $\beta(T)$ relation of the dust grains.
\end{itemize}

\acknowledgements 
J.M. and M.J. acknowledge financial support from Academy of Finland
grants 124620 and 127015. J.M. also acknowledges a grant from
The Finnish Academy of Science and Letters, V\"ais\"al\"a Foundation.
T.L. acknowledges a grant from the Magnus Ehrnrooth foundation.

D.C. acknowledges financial support from NSF grants AST0808184 and
AST0908740, and computational resources provided by the National
Institute for Computational Sciences under LRAC allocation MCA98N020
and TRAC allocation TG-AST090110.

P.P. acknowledges support by the National Science Foundation under grant AST0908740 and computer resources provided by
the NASA High End Computing Program.

\end{document}